\UseRawInputEncoding
\documentclass[aps,prd,onecolumn,superscriptaddress,floatfix,nofootinbib]{revtex4-2}
\usepackage{graphicx}% Include figure files
\usepackage{dcolumn}% Align table columns on decimal point
\usepackage{bm}% bold math
\usepackage{amsmath}
\usepackage{hyperref}
\usepackage{lineno}
\usepackage{appendix}
\usepackage{subfigure}
\usepackage{booktabs}

\def\mnras{\rm{MNRAS}}

\def\prd{\rm{Phys.~Rev.~D}}
\def\aap{\rm{A\&A}}

\def\physrep{\rm{Phys.~Rep.}}
\def\apj{\rm{ApJ}}                 % Astrophysical Journal
\def\jcap{\rm{JCAP}}

\usepackage{float}
\begin{document}

\preprint{APS/123-QED}
\title{\boldmath From Theory to Forecast: Neutrino Mass Effects on Mode-Coupling Kernels and Their Observational Implications}

\author{Farshad Kamalinejad} 
\affiliation{Department of Physics, University of Florida, 2001 Museum Road, Gainesville, FL 32611, USA}%Lines break automatically or can be forced with \\
\email{f.kamalinejad@ufl.edu}

\author{Zachary Slepian} 
\affiliation{Department of Astronomy, University of Florida, 211 Bryant Space Science Center, Gainesville, FL 32611, USA}%Lines break automatically or can be forced with \\
\email{zslepian@ufl.eduu}

\date{\today}% It is always \today, today,

\begin{abstract}
We present an analytical expression that gives both the matter and tracer (halo or galaxy) power spectrum with 1-loop corrections that include the neutrino effects on the mode coupling kernels. We use the FFTLog algorithm to accelerate calculating the higher-order corrections to the power spectrum. We then use our power spectrum and bispectrum models to pursue two main goals. First, we examine the impact of neutrino mass on cosmological parameter estimation from both the power spectrum and bispectrum in real space. We create 1-loop power spectrum and bispectrum templates in real-space and fit to the \texttt{Quijote} simulation suite, including the cross-covariance between the power spectrum and the bispectrum. We show the neutrino signature kernels estimate the same cosmological parameters as the model with the SPT (Standard Perturbation Theory) kernels, even for DESI Year 5 volume, except for the galaxy bias parameters inferred from the bispectrum. Second, we investigate to what extent the bispectrum can improve parameter constraints. We perform a Fisher forecast using the power spectrum, the tree-level bispectrum, and a joint analysis that includes the cross-covariance between them. We show that including the bispectrum can substantially reduce the error bars on key parameters. For the neutrino mass in particular, the uncertainty is reduced by $\sim 20\%$.
\end{abstract}
\maketitle
\flushbottom
\section{Introduction}

The precise measurement of neutrino mass is a critical aspect of modern cosmology and fundamental physics. Terrestrial experiments \cite{Fukuda_1999, Eguchi_2003, Ahmad_2002}, such as neutrino oscillations \citep{Pontecorvo:1957cp, Pontecorvo:1957qd}, only reveal the differences in the squares of the masses for the three neutrino flavors. The latest measurements confirm that at least two of the three flavors are massive \cite{abe2024first, 2004, abe2021t2k, esteban2020fate}. Experiments targeting the sum of the masses, such as KATRIN \cite{aker2022katrin}, do not yet have the precision needed to distinguish the neutrino mass ordering. In its most recent measurement, KATRIN reports an upper bound on the electron neutrino mass as $m_{\beta} < 0.8 \; \mathrm{eV}$. This result then implies an upper bound on the mass sum of $\sum m_{\nu} < 2.4 \; \mathrm{eV}$, assuming three degenerate neutrino mass states (\textit{i.e}., all mass states having the same mass) \cite{adame2024desi}.

The neutrino mass affects cosmological observations in several ways. First, since they behave as radiation in the early Universe, they modify the early Integrated Sachs-Wolfe (ISW) effect \cite{ISWeffect}. This can influence the low-$\ell$ multipoles of the Cosmic Microwave Background (CMB) power spectrum. More importantly, neutrinos modify the expansion history of the Universe, shifting the location of the acoustic peaks in the CMB spectra at intermediate multipole moments, $30 < \ell \leq 2000$. Due to the suppression of structure on small scales, they also affect the lensing potential, leaving an imprint on the high-$\ell$ moments of the CMB lensing spectra. Unfortunately, the neutrino mass is degenerate with the Hubble parameter and therefore, precise measurement of the two simultaneously is not possible \cite{ichikawa2008neutrino}. 

Galaxy surveys of Large-Scale Structure (LSS) offer a powerful approach to neutrino mass measurement. Completed (BOSS) \footnote{Baryon Oscillation Spectroscopic Survey} \cite{alam_2017_boss_2ps, alam2017clustering, alam2021completed}, ongoing (DESI\footnote{Dark Energy Spectroscopic Instrument} \cite{DESI:2016,adame2024early, adame2024desi}, Euclid \cite{euclid2021euclid}), and upcoming (Roman \cite{wang2022high}, MegaMapper \cite{schlegel2022megamapper}, SPHEREx \cite{dore2014cosmology}) surveys observe vast numbers of galaxies across a wide redshift range, enabling precise measurements of the Baryon Acoustic Oscillation (BAO) peaks. By probing the late-time Universe, these surveys also provide independent constraints on the reduced Hubble parameter, $h$, separate from the CMB. As a result, BAO distance measurements help break degeneracies in CMB data, yielding the most precise upper bound on the neutrino mass to date: $\sum m_{\nu} < 0.072 \; \mathrm{eV}$ from DESI \cite{adame2024desi}.

The most interesting imprint of neutrino mass occurs on smaller scales, where we observe a constant suppression of the power spectrum due to neutrino free-streaming \cite{kamalinejad2024simpleanalytictreatmentneutrino}. The free-streaming scale is the length scale over which neutrinos free-stream during a Hubble time \cite{lesgourgues2006massive, lesgourgues2012neutrino}:
\begin{align}
    k_{\rm FS}(z) = \sqrt{\frac{3\,\Omega_{\rm m}(z)}{2}}\frac{\mathcal{H}(z)}{c_{\rm s}(z)}\;[{\rm Mpc}^{-1}]
\end{align}
where $\Omega_{\rm m}(z)$\footnote{In this paper, we do not repeat the redshift (or time) dependence. All parameters are given at their present values unless the redshift (time) dependence is explicitly stated.} is the matter density at redshift $z$, $\mathcal{H}(z)$ is the Hubble parameter with respect to the conformal time ($\mathcal{H} \equiv a H$, where $H$ is the Hubble parameter with respect to the time, $t$), and $c_{\rm s}(z)$ is the neutrino sound speed, which can be calculated from the Fermi-Dirac distribution function \cite{fermi1926sulla, Dirac1926_FD_Dist}. The free-streaming scale at the transition redshift, $z_{\rm T}$ is defined as the \textit{non-relativistic} scale, $k_{\rm NR} \equiv k_{\rm FS}(z_{\rm T})$ and we define the neutrino mass fraction as $f_{\nu} \equiv \Omega_{\nu}/\Omega_{\rm m}$ as the ratio of the neutrino density ($\Omega_{\nu}$) to the matter density, $\Omega_{\rm m}$. 

In reality, neutrinos are not described by the fluid equations since they are collisionless. However, \cite{masatoshi2010massive} tested this approximation and found that it holds if neutrinos are non-relativistic when a density perturbation of a given wavenumber \( k \) enters the horizon. This condition corresponds to a total neutrino mass in the range $0.05\;{\rm eV} < \sum m_{\nu} < 0.5\;{\rm eV}$ for modes with  
$k < 0.4\;h/{\rm Mpc}$ at redshifts \( z < 10 \).

To fully exploit LSS redshift surveys, the ideal approach is full-shape power spectrum fitting across all scales, similar to CMB analysis (\textit{e.g.} the latest \textit{Planck} measurement \cite{aghanim2020planck}). However, linear perturbation theory breaks down on scales smaller than $k > 0.1 \; h/\mathrm{Mpc}$, limiting the number of modes we can use. Standard Perturbation Theory (SPT) treats the Cold Dark Matter (CDM) as a pressure-less fluid that can be characterized by only its density and velocity \cite{Bernardeau_2002}. To extend the validity of the scales we may probe, SPT expands the density and velocity perturbation, $\delta$ and $\theta$ around  their linear solutions using SPT kernels, $F_n$ and $G_n$, which are deterministic functions of $n$ Fourier modes \cite{goroff1986coupling}:
\begin{align}
    \delta^{(n)}(k) = \int \prod_{i=1}^{i=n}\frac{d^3\mathbf{q}_i}{(2\pi)^3}F_n(\mathbf{q}_1,\mathbf{q}_2,\cdots,\mathbf{q}_n)\delta_{\rm D}^{[3]}\left(\mathbf{k}-\sum_i \mathbf{q}_i\right) \delta^{(1)}(\mathbf{q}_1)\delta^{(1)}(\mathbf{q}_2)\cdots \delta^{(1)}(\mathbf{q}_n),\label{Eq:delta_n}\\
    \theta^{(n)}(k) = \int \prod_{i=1}^{i=n}\frac{d^3\mathbf{q}_i}{(2\pi)^3}G_n(\mathbf{q}_1,\mathbf{q}_2,\cdots,\mathbf{q}_n)\delta_{\rm D}^{[3]}\left(\mathbf{k}-\sum_i \mathbf{q}_i\right) \delta^{(1)}(\mathbf{q}_1)\delta^{(1)}(\mathbf{q}_2)\cdots \delta^{(1)}(\mathbf{q}_n).\label{Eq:theta_n}
\end{align}
The $F_n$ and $G_n$ kernels are derived under the assumption of a matter-dominated (Einstein-de Sitter, EdS) universe. The kernels at first order are $F_1=G_1=1$\footnote{We will discuss the higher-order kernels later on.}. These kernels allow one to compute higher-order corrections to the matter power spectrum. The structure of these higher-order corrections is very similar to loop corrections in particle physics, and thus, in cosmology, we refer to these corrections as \textit{loop} integrals. The 1-loop corrections to the matter power spectrum involves convolution integrals containing $F_2$ and $F_3$ kernels. However, in redshift space, the 1-loop corrections also include the $G_2$ and $G_3$ kernels, as redshift space mixes the velocity and density perturbations \cite{Bernardeau_2002}. 

In reality, neutrinos (as well as dark energy \cite{Koyama_kernels, tudes2024non,bose2018modelling,upadhye2014large}) modify these kernels. Neutrinos affect the kernels due to their time- and scale-dependent free-streaming \cite{wong2008higher, garny2021loop,saito2009nonlinear}. A complete perturbation theory should account for massive neutrinos in the fluid description of cold dark matter (CDM) and baryons as well. However, solving such a theory analytically is not feasible, and most attempts to find a solution rely on numerical methods \cite{Blas_2014}.

However, it is possible to obtain a simple analytical solution under the following assumptions \cite{wong2008higher}. First, assume that the Universe is EdS with massive neutrinos, such that the expansion around the linear solution can still be applied. Second, assume that neutrino density perturbations are kept only up to linear order and that their velocity perturbations are zero. These assumptions are not entirely valid within the perturbation scheme and break momentum conservation \cite{Blas_2014,garny2021loop}. On the positive side, they allow for an analytical calculation of the modifications to the kernels in the presence of massive neutrinos, allowing one to calculate the loop corrections to the linear power spectrum. 

The first attempt to include massive neutrinos in non-linear perturbation theory was made by \cite{saito2008impact}, who assumed that neutrino perturbations remain in the linear regime. \cite{wong2008higher} later introduced a fully perturbative framework to compute the effects of neutrinos to the SPT kernels for the first time, also assuming linear neutrino perturbations, similar to \cite{saito2008impact}. As discussed earlier, this assumption leads to a violation of momentum conservation. \cite{kamalinejad2020non} employed this framework to derive analytical expressions for the kernels, referred to as signature kernels, and subsequently used them to compute the redshift-space bispectrum. \cite{Blas_2014} numerically solved a fully non-linear two-fluid system but did not provide analytical formulas for higher-order effects. In contrast, \cite{aviles2020lagrangian} developed a Lagrangian approach to describe neutrino kernels without assuming their linearity, while \cite{aviles2021clustering} extended this analysis to the Eulerian framework.

It is important to note that these convolution integrals appearing in the loop corrections are difficult to calculate numerically because the integration runs from zero to infinity, requiring the sampling of the linear power spectrum over a vast range of scales and making the computation costly. One might suggest imposing a cutoff as an upper limit on the integral to address this, as the linear power spectrum is not valid up to infinity. However, this introduces dependence on the choice of the cutoff, which is undesirable. To overcome this, we employ the logarithmic Fast Fourier Transform (FFTLog) method, discussed in detail in \cite{mcewen2016fast, simonovic2018cosmological,schmittfull2016fast}. The FFTLog method samples the power spectrum on a logarithmic grid in $k$, ensuring proper sampling across all scales. Furthermore, because FFTLog uses dimensional regularization in the evaluation of the integrals, any dependence on the cutoff is effectively removed.

In this paper, we utilize the signature kernels introduced in \cite{kamalinejad2020non} to obtain the 1-loop matter power spectrum and the 1-loop galaxy power spectrum in real space. We employ the FFTLog formalism to evaluate the loop integrals. Finally, we investigate whether it is possible to observe these signature terms from neutrinos in upcoming surveys by a Fisher forecast analysis using the covariance matrix estimates from the \texttt{Quijote} N-body simulation suite \cite{villaescusa2020quijote}. If we do not observe any deviation from the SPT kernels, we can conclude that the use of SPT kernels will be entirely sufficient in the next data release of galaxy surveys.

\section{FFTLog Formalism}

In this section, we quickly review the basics of the FFTLog formalism (\cite{simonovic2018cosmological} for a more detailed and thorough discussion). In SPT, the linear power spectrum (tree-level) is corrected by loop integrals. The lowest-order corrections have two contributions: the first is obtained by contracting two second-order density fields together \( (n=2) \) in Eq. (\ref{Eq:delta_n}), while the second involves a linear order \( (n=1) \) combined with a third-order \( (n=3) \) density field. The former is given by \( P^{(22)}(k) \), and the latter by \( P^{(13)}(k) \):
\begin{align}
    P^{(22)}(k) &=2 \int_{\mathbf{q}}F_2^{({\rm s})}(\mathbf{k}-\mathbf{q},\mathbf{q})^2 P_{\rm lin}(|\mathbf{k}-\mathbf{q}|)P_{\rm lin}(q)\label{Eq:P22}\\
    P^{(13)}(k) &= 6 P_{\rm lin}(k)\int_\mathbf{q}F_3^{({\rm s})}(\mathbf{k},\mathbf{q},-\mathbf{q}) P_{\rm lin}(q).\label{Eq:P13}
\end{align}
Here, the superscript $({\rm s})$ on the kernels means symmetrizing the kernels with respect to its arguments. The 1-loop power spectrum is then:
\begin{align}
    P^{1-\rm loop}(k, z) = D(z)^2 P_{\rm lin}(k)+ D(z)^4\left[P^{(13)}(k)+P^{(22)}(k)\right],
\end{align}
where $D(z)$ is the growth rate of matter, which scales as $D(z)\propto 1/(1+z)$ in a matter-dominated universe. The integrals in equations (\ref{Eq:P22})and (\ref{Eq:P13}) are convolution integrals over $\mathbf{q}$ from $q=0$ to $q\rightarrow\infty$. As already stated, numerical evaluation of these integrals is cut-off dependent, and computationally costly. The approach of this paper is to expand the linear power spectrum as a sum of self-similar power law cosmologies in the complex domain as:
\begin{align}
    P_{\rm lin}(k) = \sum_{m=-N/2}^{m=N/2}c_m k^{\nu+i\eta_m}.
\end{align}
Essentially, we have sampled the power spectrum at \( N \) points on a logarithmic axis in \( k \). The coefficients \( c_m \) are then given by:

\begin{align}
    c_m = \frac{1}{N}\sum_{\ell=0}^{\ell=N-1} P_{\rm lin}(k_{\ell})\, k_{\ell}^{-\nu}\, k_{\rm min}^{-i\eta_m}\,e^{2\pi i \eta_m \ell/N},\;\;
    \eta_m = \frac{2\pi m}{\ln{(k_{\rm max}/k_{\rm min})}}.
\end{align}
In this sampling, \( \nu \) is called the bias and is a real number. $\nu$ should be chosen such that the integral we are trying to solve is convergent. This constraint usually defines a range for $\nu$ that depends on the integral, as we will see later. The FFTLog expansion of the linear power spectrum can be handled very quickly by only sampling a little more than \( 100 \) points in Fourier space. In this paper, we always sample the power spectrum at \( N=300 \) points \cite{class_pt, linde2024class}.

Now, with the sampling of the linear power spectrum in a complex power law expansion, the convolution integrals can be performed by utilizing dimensional regularization. Rewriting the SPT kernel in powers of \( q \), \( k \), and \( |\mathbf{k} - \mathbf{q}| \) we obtain \cite{scoccimarro1997cosmological,pajer2013renormalization}:
\begin{align}
    \int_{\mathbf{q}}\frac{1}{q^{2\nu_1}|\mathbf{k}-\mathbf{q}|^{2\nu_2}} = k^{3-2\nu_{12}}I(\nu_1, \nu_2)
    \label{Eq:dim_reg_int}
\end{align}
with $I(\nu_1, \nu_2)$:
\begin{align}
    I(\nu_1, \nu_2) = \frac{1}{8\pi^{3/2}}\frac{\Gamma(3/2-\nu_1)\Gamma(3/2-\nu_2)\Gamma(\nu_{12}-3/2)}{\Gamma(\nu_1)\Gamma(\nu_2)\Gamma(3-\nu_{12})}.
\end{align}
$\nu_{12} = \nu_1+\nu_2$ and $\Gamma$ is the Gamma function. The loop integrals then can be written as integrals like Eq. (\ref{Eq:dim_reg_int}). For instance, the $P^{(22)}(k)$ contribution becomes:
\begin{align}
    P^{(22)}(k) = 2\sum_{m_1, m_2}c_{m_1}c_{m_2}\sum_{n_1, n_2}f_{22}(n_1, n_2) k^{-2(n_1+n_2)}\int_{\mathbf{q}}\frac{1}{q^{2\nu_1-2n_1}|\mathbf{k}-\mathbf{q}|^{2\nu_2-2n_2}},
\end{align}
where $\nu_{1} = -\left( \nu + i \eta_{m_{1}} \right)/2$, $\nu_{2} = - \left( \nu + i \eta_{m_{2}} \right)/2$ and $n_1$ and $n_2$ are integer powers of $q^2$ and $|\mathbf{k}-\mathbf{q}|^{2}$ in the expansion of the $F_2$ kernel. The rational coefficients are brought together in the form of the matrix $f_{22}(n_1, n_2)$ which is an $n_1\times n_2$ dimensional matrix. A similar approach can be applied to $P^{(13)}(k)$ which we do not show here (see \cite{simonovic2018cosmological}). Performing the summation over $n_1$ and $n_2$ and evaluating the integrals using Eq. (\ref{Eq:dim_reg_int}) gives:
\begin{align}
    P^{(22)}(k) = k^3 \sum_{m_1, m_2}c_{m_1}k^{-2\nu_1} M_{22}(\nu_1, \nu_2)c_{m_2}k^{-2\nu_2},
\end{align}
where 
\begin{align}
    M_{22}(\nu_1, \nu_2) = \frac{\left(\frac{3}{2} - \nu_{12}\right)\left(\frac{1}{2} - \nu_{12}\right) \left[\nu_1 \nu_2 \left(98 \nu_{12}^2 - 14 \nu_{12} + 36\right) - 91 \nu_{12}^2 + 3 \nu_{12} + 58\right]}{196 \nu_1 \left(1 + \nu_1\right) \left(\frac{1}{2} - \nu_1\right) \nu_2 \left(1 + \nu_2\right) \left(\frac{1}{2} - \nu_2\right)} I\left(\nu_1, \nu_2\right).
\end{align}
The strength of the FFTLog method lies in the fact that the \( M_{22} \) matrix (and the analogous matrix for the \( P^{(13)} \) term, \( M_{13} \)) is completely independent of cosmology. The cosmology is encoded solely in the \( c_m \) coefficients. As a result, these matrices are computed only once throughout the entire process, significantly speeding up the computation.

The neutrinos introduce additional terms, referred to as signature terms \cite{kamalinejad2020non}, to the kernels. These signature terms have a specific scale dependence, which can be incorporated into the FFTLog algorithm. In what follows, we first derive the neutrino signatures in the kernels and use them to construct the matter and halo power spectra and bispectra.

\section{Neutrino Signature Kernels}
\label{sec:kernels}
\subsection{Second-Order Kernels}
Due to their free-streaming, neutrinos modify the growth rate of structure, introducing both scale- and time-dependent effects. This, along with the free-streaming length, alters SPT considerably.

% We can approximate the SPT kernels in the presence of massive neutrinos with the assumption that neutrino density perturbations are retained only up to linear order, and velocity perturbations are negligible on scales smaller than the free-streaming scale. The validity of this assumption can be supported by the fluid equations for the neutrinos:
% \begin{align}
%     &\dot{\delta}_{\nu}+\theta_{\nu}=0.
% \end{align}
% \cite{kamalinejad2024simpleanalytictreatmentneutrino} explores the two-fluid equations in the linear regime, yielding an analytic solution for the density perturbations of cold dark matter (CDM) and neutrinos. 

% Assuming an EdS Universe, changing the time variable from \( \tau \) to \( s \equiv \ln{a(\tau)} \), and looking for the asymptotic solution on small scales, the neutrino density perturbation is \cite{kamalinejad2024simpleanalytictreatmentneutrino}:
% \begin{align}
%     \delta_{\nu}(k, s)=\delta_{\nu}^0+\frac{1}{2}\delta_{\rm cb}(k, a) \frac{a}{a_{\rm T}}\left(\frac{k_{\rm NR}}{k}\right)^2,
% \end{align}
% which leads to a simple solution, \( \theta_{\nu} \propto a^{3/2} \). Therefore, $\theta_{\rm \nu}$ does not grow as fast as the density perturbations, $\delta_{\rm \nu}$, so we can neglect it. Although this result arises from linear theory, we extend it to higher-order corrections by assuming that neutrino velocity perturbations remain negligible at all orders.

Based on \cite{wong2008higher}, one can obtain correction terms to the SPT kernels. A complete and detailed derivation is given in \cite{kamalinejad2020non} where these corrected kernels are termed ``neutrino signature'' kernels. We have the changes to the kernels as:
\begin{align}
    \Delta F_2(\mathbf{k},\mathbf{q}) =& \frac{6}{245}f_{\nu}\left[1-\left(\frac{\mathbf{k} \cdot \mathbf{q}}{k q}\right)^2 \right],\label{Eq:F2_sig}\\
    \Delta G_2(\mathbf{k},\mathbf{q}) =& -f_{\nu} \left[\frac{51}{245} + \frac{3}{10} \frac{\mathbf{k} \cdot \mathbf{q}}{k q} \left(\frac{k}{q} + \frac{q}{k}\right) + \frac{96}{245} \left(\frac{\mathbf{k} \cdot \mathbf{q}}{k q}\right)^2\right].\label{Eq:G2_sig}
\end{align}
Upon inspecting these new corrections, we see that the corrections depend on the scales and also depend on the relative angle between $\mathbf{k}$ and $\mathbf{q}$. Also, we notice that the $F_2$ kernel does not receive any contributions to the dipole (\textit{i.e.} terms like $\mathbf{k}\cdot \mathbf{q}$) but the $G_2$ kernel gets this correction. 

Although there are now successful perturbative frameworks that predict the impact of neutrino mass on LSS, such as \cite{aviles2020lagrangian, aviles2021clustering, Blas_2014, senatore2017effective}, none of these theories provide a simple analytic description of neutrinos; they all depend on numerical implementations which can often be computationally expensive. To speed up computations, simplified approximations are often made, which reduce the accuracy of the resulting models. For instance, \cite{kamalinejad2020non} shows that the neutrino signature kernels in equations (\ref{Eq:F2_sig}) and (\ref{Eq:G2_sig}) are very similar to those derived by \cite{Avilesfolpnu}, which are based on a more accurate perturbation theory that includes neutrinos. Therefore, our approach can still provide a relatively accurate description of the impact of neutrinos on the PT kernels.

One of the caveats of this approach is the issue of momentum conservation (\cite{Blas_2014} for a detailed discussion). The SPT kernels respect momentum conservation because the kernels at any order $F_n(k_1, k_2, \cdots, k_n)$ scale as $k^2/q^2$ where we have defined $\mathbf{k}= \mathbf{k}_1+\mathbf{k}_2+\cdots+\mathbf{k}_n$. This scaling assumes that each of the momenta $k_i\sim q$ and the total sum is $k$$\ll$$q$. However, for the neutrino signature kernels in equations (\ref{Eq:F2_sig}) and (\ref{Eq:G2_sig}) this scaling also receives a linear contribution, $k/q$. 

% It is more straightforward to see why this term violates momentum conservation in one dimension, as the coefficient of \( k \) corresponds to a spatial derivative, \( d/dx \), in configuration space. Since momentum conservation is given by $\dot{\delta} + dv/dx$, the presence of this additional \( d/dx \) term prevents the equation from summing to zero. This is more of an issue when one extends perturbation theory to higher loop corrections \cite{Blas_2014}.

\subsection{Third-Order Kernels}
We also need to calculate the neutrino mass correction to the third-order kernel, $F_3$, if we want to find the 1-loop power spectrum. \cite{wong2008higher} also provides the framework in which we can calculate the third-order kernels. 
The 1-loop correction to the power spectrum contains a convolutional integral on $F_3$ in the $P^{(13)}(k)$ contribution. We can write the $F_3^{({\rm s})}(\mathbf{k}, \mathbf{q}, -\mathbf{q})$ as:
\begin{align}
    F_3^{({\rm s})}(\mathbf{k}, \mathbf{q}, -\mathbf{q}) = \frac{1}{3}&\Bigg[\mathcal{A}(\mathbf{k}, -\mathbf{q})F_2(\mathbf{k}, \mathbf{q}) +\mathcal{A}(\mathbf{k},\mathbf{q})F_2(\mathbf{k}, -\mathbf{q})\nonumber\\&+\mathcal{B}(\mathbf{k}, -\mathbf{q})G_2(\mathbf{k}, \mathbf{q}) +\mathcal{B}(\mathbf{k},\mathbf{q})G_2(\mathbf{k}, -\mathbf{q})\nonumber\\&+\mathcal{C}(\mathbf{k}, -\mathbf{q})G_2(\mathbf{k}, \mathbf{q}) +\mathcal{C}(\mathbf{k},\mathbf{q})G_2(\mathbf{k}, -\mathbf{q})\Bigg],
\end{align}
where we have grouped different terms based on their similarities. The coefficients $\mathcal{A}(\mathbf{k}, \mathbf{q})$ , $\mathcal{B}(\mathbf{k}, \mathbf{q})$ and  $\mathcal{C}(\mathbf{k}, \mathbf{q})$ describe the mode-coupling in the third order kernel and are defined as:
\begin{align}
    \mathcal{A}(\mathbf{k}, \mathbf{q}) &= \frac{7}{18}X\frac{\mathbf{k}\cdot \mathbf{q}}{q^2},\\
    \mathcal{B}(\mathbf{k}, \mathbf{q}) &=\frac{7}{18}Y\frac{\mathbf{k}\cdot\left(\mathbf{k}-\mathbf{q}\right)}{|\mathbf{k}-\mathbf{q}|^2},\\
    \mathcal{C}(\mathbf{k}, \mathbf{q}) &=\frac{1}{9}Z\frac{k^2\left(\mathbf{k}-\mathbf{q}\right)\cdot\mathbf{q}}{\left|\mathbf{k}-\mathbf{q}\right|^2q^2} .
\end{align}
In an EdS Universe with massless neutrinos, the coefficients \(X\), \(Y\), and \(Z\) are all equal to one. The neutrino mass changes the \(F_3\) kernel by modifying the \(F_2\) and \(G_2\) kernels stated earlier, as well as by altering the \(X\), \(Y\), and \(Z\) coefficients:
\begin{align}
    X = 1+\frac{2}{105}f_{\nu},\;
    Y = 1+\frac{13}{21}f_{\nu},\;
    Z = 1+\frac{8}{15}f_{\nu}.
\end{align}
For a typical neutrino mass, $f_{\nu} \sim 0.01$ which means that the corrections to the kernels do not exceed $2\%$ at best. Therefore, the neutrino signature amplitude is highly suppressed compared to the SPT terms.

\section{Matter Power Spectrum at 1-Loop}

After the non-relativistic transition of massive neutrinos, they will contribute both to the total matter density as $\Omega_{\rm m} = \Omega_{\rm c}+\Omega_{\rm b} + \Omega_{\nu}$, as well as to the total matter density perturbation ($\delta_{\rm m}$) as the weighted sum of the CDM+baryons ($\delta_{\rm cb}$) and neutrinos ($\delta_{\nu}$) \cite{lesgourgues2006massive}:
\begin{align}
    \delta_{\rm m} =  (1-f_{\nu})\delta_{\rm cb}+f_{\nu}\delta_{\nu}
    \label{Eq:delta_m}.
\end{align}
The expansion of $\delta_{\rm m}$ beyond linear order should include corrections up to the third-order. However, as previously mentioned, we retain neutrino density perturbations only at linear order. Expanding Eq. (\ref{Eq:delta_m}) up to third order with neutrinos at linear order, and using the definition of the power spectrum along with higher-order kernels, we can derive the 1-loop matter power spectrum as:
\begin{align}
    P^{\rm 1-loop}(k) = P^{\rm tree}(k)+(1-2f_{\nu})P^{(13)}_{\rm cb, cb}(k) \nonumber+ (1-2f_{\nu})P^{(22)}_{\rm cb, cb}(k)+2f_{\nu}P^{(13)}_{\rm cb\nu}(k).
\end{align}
$P^{\rm tree}(k)$ is the tree-level power spectrum, $P^{(13)}_{\rm cb, cb}(k)$ comes from the contraction of the third-order CDM+baryon overdensity with a first-order term, $P^{(22)}_{\rm cb, cb}(k)$ is produced by contracting two second-order CDM+baryon overdensities, and lastly $P^{(13)}_{\rm cb\nu}(k)$ results from the contraction between a linear neutrino overdensity and a third-order CDM+baryon term. The explicit forms of these loop integrals are given by:
\begin{align}
    P^{(13)}_{\rm cb, cb}(k) &= 6 P_{\rm cb}(k)\int_\mathbf{q}F_3^{({\rm s})}(\mathbf{k},\mathbf{q},-\mathbf{q}) P_{\rm cb}(q),\\
    P^{(22)}_{\rm cb, cb}(k) &=2 \int_{\mathbf{q}}F_2^{({\rm s})}(\mathbf{k}-\mathbf{q},\mathbf{q})^2 P_{\rm cb}(|\mathbf{k}-\mathbf{q}|)P_{\rm cb}(q),\\
    P^{(13)}_{\rm cb\nu}(k) &= 3 P_{\rm cb\nu}(k)\int_\mathbf{q}F_3^{({\rm s})}(\mathbf{k},\mathbf{q},-\mathbf{q}) P_{\rm cb}(q).
\end{align}
We note that the $P^{(22)}$ contribution does not receive any corrections from neutrinos in the integrated power spectra. However, changes in the kernels still contribute. On the other hand, $P^{(13)}$ does receive a correction in the form of the cross-power spectrum, $P_{\rm cb\nu}$, as well as from the kernel modifications.

\subsection{Tree-Level Effects}
The neutrino effects in the tree-level power spectrum are well-known. On very small scales, the neutrinos cause a constant, scale-independent suppression equal to $-8f_{\nu}$ \cite{hu1998weighing}. On large scales, the neutrinos do not change the matter power spectrum because their free-streaming scale is much smaller than the scales of interest ($k\ll k_{\rm FS}$). The interesting feature of the neutrinos in the matter power spectrum is in the intermediate regimes where the suppression is scale-dependent. \cite{kamalinejad2024simpleanalytictreatmentneutrino} shows that this suppression in intermediate regimes is proportional to $1/k^2$ by solving the two-fluid equations iteratively. They also offer a closed-form formula for the matter power spectrum that is valid on all scales, assuming linear perturbation theory. This is important here because we need the cross-power spectrum of matter and neutrinos $P_{\rm cb\nu}$ in order to compute loop corrections. Their formula will enable calculating these spectra analytically. However, in this paper, we use \texttt{class} \cite{Diego_Blas_2011,lesgourgues2011cosmic} to calculate all the spectra including $P_{\rm cb\nu}$.

\subsection{\texorpdfstring{$P^{(22)}$}{P(22)} Correction}

The FFTLog method discussed earlier provides a fast approach to include the signature terms in the higher-order loop integrals. The overall strategy is to express the kernels in powers of $k$, $q$, and $|\mathbf{k}-\mathbf{q}|$, and then use the FFTLog algorithm to calculate the integrals.

We denote the SPT kernels from now on as $\Tilde{F}$ and $\Tilde{G}$, and the signature terms as $\Delta F$ and $\Delta G$. Since the loop integrals from the SPT kernels have already been calculated in the literature, we do not reproduce them here. Instead, we focus solely on the signature terms.

The $P^{(22)}(k)$ integral does not include any contribution from the cross-power spectrum. Dropping the ${\rm cb,cb}$ subscript for brevity, as discussed in the previous paragraph, we can write the $P^{(22)}(k)$ contribution as the sum of the neutrino-less term and the neutrino term:
\begin{align}
    \Delta P^{(22)}(k) &= 2 \int_{\mathbf{q}}2 \Tilde{F}_2(\mathbf{k}-\mathbf{q},\mathbf{q})\Delta F_2(\mathbf{k}-\mathbf{q},\mathbf{q})P(|\mathbf{k}-\mathbf{q}|)P(q)\nonumber,\\
    &=-\frac{12f_{\nu}}{245}\int_{\mathbf{q}}2 \sigma^2(\mathbf{k}-\mathbf{q},\mathbf{q})\Tilde{F}_2(\mathbf{k}-\mathbf{q},\mathbf{q})P(|\mathbf{k}-\mathbf{q}|)P(q),
\end{align}
where $\sigma^2\left(\mathbf{k}, \mathbf{q}\right) = \left(\mathbf{k}\cdot\mathbf{q}/k q\right)^2-1$ is the tidal operator.
This integral is calculated in previous work using the FFTLog method \cite{simonovic2018cosmological}. The $M$ matrix corresponding to this integral is given by:\footnote{Here, we used \cite{simonovic2018cosmological} notation.}
\begin{align}
    M_{\mathcal{G}_2}(\nu_1, \nu_2) = \frac{(3-2\nu_{12})(1-\nu_{12})(6+7\nu_{12})}{28\nu_1(1+\nu_1)\nu_2(1+\nu_2)}I(\nu_1, \nu_2).
\end{align}
With this matrix, the neutrino signature correction in the $(22)$ term will be:
\begin{align}
    \Delta P^{(22)}(k) = \frac{12f_{\nu}}{245}k^3\sum_{m_1, m_2}c_{m_1}k^{-2\nu_1} M_{\mathcal{G}_2}(\nu_1, \nu_2) c_{m_2}k^{-2\nu_2}.
\end{align}
This integral is convergent for a wide range of biases, $-3<\nu<1/2$. We take the bias parameters to be equal to $\nu = -0.3$ for the $(22)$ contribution which guarantees convergence \cite{simonovic2018cosmological}.
\subsection{\texorpdfstring{$P^{(13)}$}{P(13)} Correction}

The $F^{({\rm s})}_3(\mathbf{k},
    \mathbf{q}, -\mathbf{q})$ kernel
    given in the previous section can be expanded into SPT terms and neutrino signature terms from the $F_2$ and $G_2$ kernels as well as the coefficients explained earlier.  We can write the terms in the third-order kernel as:
\begin{align}
    F^{({\rm s})}_3 = \Tilde{F}^{({\rm s})}_3+\frac{1}{3}&\Bigg[\Delta \mathcal{A}(\mathbf{k},
    -\mathbf{q})\,\Tilde{F}_2(\mathbf{k},
    \mathbf{q})+\Delta \mathcal{A} (\mathbf{k},
    \mathbf{q})\,\Tilde{F}_2(\mathbf{k},
    -\mathbf{q})\nonumber\\&+ \mathcal{A}(\mathbf{k},
    -\mathbf{q}) \,\Delta F_2(\mathbf{k},
    \mathbf{q}) + \mathcal{A}(\mathbf{k},
    \mathbf{q}) \,\Delta F_2(\mathbf{k},
    -\mathbf{q})\nonumber\\&+\Delta \mathcal{B} (\mathbf{k},
    -\mathbf{q})\,\Tilde{G}_2(\mathbf{k},
    \mathbf{q})+\Delta \mathcal{B} (\mathbf{k},
    \mathbf{q})\,\Tilde{G}_2(\mathbf{k},
    -\mathbf{q})\nonumber\\&+ \mathcal{B}(\mathbf{k},
    -\mathbf{q}) \,\Delta G_2(\mathbf{k},
    \mathbf{q}) + \mathcal{B}(\mathbf{k},
    \mathbf{q}) \,\Delta G_2(\mathbf{k},
    -\mathbf{q})\nonumber\\&+\Delta \mathcal{C} (\mathbf{k},
    -\mathbf{q})\,\Tilde{G}_2(\mathbf{k},
    \mathbf{q})+\Delta \mathcal{C} (\mathbf{k},
    \mathbf{q})\,\Tilde{G}_2(\mathbf{k},
    -\mathbf{q})\nonumber\\&+ \mathcal{C}(\mathbf{k},
    -\mathbf{q})\, \Delta G_2(\mathbf{k},
    \mathbf{q}) + \mathcal{C}(\mathbf{k},
    \mathbf{q}) \,\Delta G_2(\mathbf{k},
    -\mathbf{q})\Bigg].
    \label{Eq:F_3}
\end{align}
We group the terms that are similar in their form. The first two lines are similar since they contain the changes in $\mathcal{A}$ and the $F_2$ kernel. In fact, the second line is zero due to the symmetries of $\Delta F_2$ under the transformation $\mathbf{q} \rightarrow -\mathbf{q}$. We then denote the first line of Eq. (\ref{Eq:F_3}) as $\mathcal{F}_{\mathcal{A}}$. The terms that contain $|\mathbf{k} - \mathbf{q}|$ arise from functions with an argument of $(\mathbf{k}, \mathbf{q})$ in the kernels such as $\Delta \mathcal{B}(\mathbf{k}, -\mathbf{q})\,\Tilde{G}_2(\mathbf{k}, \mathbf{q})$ and $\mathcal{B}(\mathbf{k}, -\mathbf{q})\,\Delta \Tilde{G}_2(\mathbf{k}, \mathbf{q})$. We denote them as $\mathcal{F}_{-}$. Lastly, the terms involving $|\mathbf{k} + \mathbf{q}|$ originate from contributions with $(\mathbf{k}, -\mathbf{q})$ as the argument of the kernels such as $\Delta \mathcal{B}(\mathbf{k},\mathbf{q})\,\Tilde{G}_2(\mathbf{k}, -\mathbf{q})$ and $\mathcal{B}(\mathbf{k},\mathbf{q})\,\Delta \Tilde{G}_2(\mathbf{k}, -\mathbf{q})$. We denote them as $\mathcal{F}_{+}$.  $\mathcal{F}_{\mathcal{A}}$, $\mathcal{F}_{-}$ and $\mathcal{F}_{+}$ are:
\begin{align}
    \mathcal{F}_{\mathcal{A}} &= \frac{1}{3}\Bigg[\Delta \mathcal{A}(\mathbf{k},
    -\mathbf{q})\,\Tilde{F}_2(\mathbf{k},
    \mathbf{q})+\Delta \mathcal{A} (\mathbf{k},
    \mathbf{q})\,\Tilde{F}_2(\mathbf{k},
    -\mathbf{q})\nonumber\\&\quad\qquad+ \mathcal{A}(\mathbf{k},
    -\mathbf{q}) \,\Delta F_2(\mathbf{k},
    \mathbf{q}) + \mathcal{A}(\mathbf{k},
    \mathbf{q}) \,\Delta F_2(\mathbf{k},
    -\mathbf{q})\Bigg],\\
    \nonumber\\
    \mathcal{F}_{-} &= \frac{1}{3}\Bigg[\Delta \mathcal{B} (\mathbf{k},
    -\mathbf{q})\,\Tilde{G}_2(\mathbf{k},
    \mathbf{q})+ \mathcal{B}(\mathbf{k},
    -\mathbf{q}) \,\Delta G_2(\mathbf{k},
    \mathbf{q}) \nonumber\\&\quad\qquad+\Delta \mathcal{C} (\mathbf{k},
    -\mathbf{q})\,\Tilde{G}_2(\mathbf{k},
    \mathbf{q})+ \mathcal{C}(\mathbf{k},
    -\mathbf{q})\, \Delta G_2(\mathbf{k},
    \mathbf{q})\Bigg],\\
    \nonumber\\
    \mathcal{F}_{+} &= \frac{1}{3}\Bigg[\Delta \mathcal{B} (\mathbf{k},
    \mathbf{q})\,\Tilde{G}_2(\mathbf{k},
    -\mathbf{q})+ \mathcal{B}(\mathbf{k},
    \mathbf{q}) \,\Delta G_2(\mathbf{k},
    -\mathbf{q}) \nonumber\\&\quad\qquad+\Delta \mathcal{C} (\mathbf{k},
    \mathbf{q})\,\Tilde{G}_2(\mathbf{k},
    -\mathbf{q})+ \mathcal{C}(\mathbf{k},
    \mathbf{q})\, \Delta G_2(\mathbf{k},
    -\mathbf{q})\Bigg].
\end{align}

Since we are integrating over the entire range of $\mathbf{q}$, we can make a transformation from $\mathbf{q} \rightarrow -\mathbf{q}$. This transformation will make the $\mathcal{F}_{+}$ and $\mathcal{F}_{-}$ integrals exactly the same. Therefore:
\begin{align}
    F^{({\rm s})}_3 = \Tilde{F}^{({\rm s})}_3+\Bigg[\mathcal{F}_{\mathcal{A}} +  2\mathcal{F}_{-}\Bigg]
    \label{Eq:F(s)_3}.
\end{align}
In what follows, we calculate the  neutrino-induced terms (those in the square bracket).

\subsubsection{$\mathcal{F}_{\mathcal{A}}$ Term}
The first line in Eq. (\ref{Eq:F_3}) is denoted by $\mathcal{F}_{\mathcal{A}}$. We now use the law of cosines to rewrite this term in terms of $k$, $q$ and $|\mathbf{k}-\mathbf{q}|$ so that we can use the FFTLog method. Using this substitution we find:
\begin{align}
    \mathcal{F}_{\mathcal{A}} = \frac{1}{3}\frac{f_{\nu}}{540}\Bigg[&\frac{2 k^2 |\mathbf{k}-\mathbf{q}|^2}{q^4}-\frac{|\mathbf{k}-\mathbf{q}|^4}{k^2 q^2}+\frac{2 |\mathbf{k}-\mathbf{q}|^2}{k^2}-\frac{k^4}{q^4}\nonumber\\&-\frac{3 k^2}{q^2}-\frac{q^2}{k^2}+\frac{4 |\mathbf{k}-\mathbf{q}|^2}{q^2}-\frac{|\mathbf{k}-\mathbf{q}|^4}{q^4}-3\Bigg].
\end{align}
% This term will have the following contribution to the $P^{(13)}$:
% \begin{align}
%     \Delta P^{(13)}_{\mathcal{A}}(k) = 6 P_{\rm comp.}(k) \sum_{m}c_{m}\sum_{n_1, n_2}f_{\mathcal{A}}(n_1, n_2) k^{-2(n_1+n_2)}\int_{\mathbf{q}}\frac{1}{q^{2\nu_1-2n_1}|\mathbf{k}-\mathbf{q}|^{-2n_2}}
% \end{align}
% % where $f_{\mathcal{A}}(n_1, n_2)$ is a matrix given by:
% \begin{align}
%     f_{\mathcal{A}}(n_1, n_2) = \frac{1}{3}\frac{f_{\nu}}{540}\begin{pmatrix}
% -1 & -3 & -3 & -1 \\
% 2 & 4 & 2 & 0 \\
% -1 & -1 & 0 & 0
% \end{pmatrix},
% \:n_1 = -2, -1, 0, 1,\: n_2 = 0, 1, 2
% \end{align}
% performing the sum on $n_1$ and $n_2$ we find:
The $M$ matrix corresponding to this kernel can be found as:
\begin{align}
    M_{\mathcal{A}} = -\frac{1}{3}\frac{f_{\nu}}{540}\frac{\nu^4+\nu^3-3 \nu^2-\nu+1}{2 \pi  \nu \left(\nu^2-1\right)}\tan (\pi  \nu).
\end{align}
\subsubsection{\texorpdfstring{$\mathcal{F}_{-}$ Term}{F- Term}}

$\mathcal{F}_{-}$ terms are the contributions in Eq. (\ref{Eq:F_3}) that include terms with $(\mathbf{k}, \mathbf{q})$ as the arguments of their kernels such as $\Delta \mathcal{B}(\mathbf{k}, -\mathbf{q})\,\Tilde{G}_2(\mathbf{k}, \mathbf{q})$ and $\mathcal{B}(\mathbf{k}, -\mathbf{q})\,\Delta \Tilde{G}_2(\mathbf{k}, \mathbf{q})$. Similarly to $\mathcal{F}_{\mathcal{A}}$, using the law of cosines we can write $\mathcal{F}_{-}$ as:
\begin{align}
    \mathcal{F}_{-} = \frac{1}{3}\frac{f_{\nu}}{540}\Bigg[&-\frac{15 k^6}{98 |\mathbf{k}-\mathbf{q}|^2 q^4}-\frac{51 k^4}{49 |\mathbf{k}-\mathbf{q}|^2 q^2}-\frac{143 k^2 |\mathbf{k}-\mathbf{q}|^2}{98 q^4}+\frac{3 q^4}{2 k^2 |\mathbf{k}-\mathbf{q}|^2}\nonumber\\&-\frac{|\mathbf{k}-\mathbf{q}|^4}{k^2 q^2}+\frac{198 k^2}{49 |\mathbf{k}-\mathbf{q}|^2}+\frac{7 |\mathbf{k}-\mathbf{q}|^2}{2 k^2}+\frac{47 k^4}{49 q^4}+\frac{34 k^2}{49 q^2}\nonumber\\&-\frac{4 q^2}{k^2}-\frac{213 q^2}{49 |\mathbf{k}-\mathbf{q}|^2}+\frac{66 |\mathbf{k}-\mathbf{q}|^2}{49 q^2}+\frac{32 |\mathbf{k}-\mathbf{q}|^4}{49 q^4}+\frac{115}{49} \Bigg],
\end{align}
with $M$ matrix:
\begin{align}
    M_{-}(\nu) = -\frac{1}{3}\frac{f_{\nu}}{540}\frac{34 \nu^4-47 \nu^3-200 \nu^2-196 \nu+66}{392 \pi  \nu \left(\nu^2-1\right)} \tan (\pi  \nu).
\end{align}
Now that we have obtained the required matrices, we can write the neutrino corrections to $P^{(13)}$:
\begin{align}
    \Delta P^{(13)}_{\rm comp.}(k) = 6k^3P_{\rm comp.}(k)\sum_m c_m k^{-2\nu}\left(M_{\mathcal{A}}+2 M_{-} \right).
\end{align}
where $P_{\rm comp.}$ is either $P_{\rm cb}$ or $P_{\rm cb\nu}$.
Finally, the total one-loop matter power spectrum is:
\begin{align}
    P^{\rm 1-loop}(k) = P^{\rm tree}(k)&+\left(1-2f_{\nu}\right)\left[P^{(13)}_{\rm cb, cb}(k) +\Delta P^{(13)}_{\rm cb, cb}(k)\right]\nonumber\\&+ (1-2f_{\nu})\left[P^{(22)}_{\rm cb, cb}(k)+\Delta P^{(22)}_{\rm cb, cb}(k)\right]\nonumber\\&+2f_{\nu}\left[P^{(13)}_{\rm cb\nu}(k)+\Delta P^{(13)}_{\rm cb\nu}(k)\right].
\end{align}

\subsection{IR Resummation}

Accurately capturing the BAO features in the 1-loop power spectrum is crucial. The 1-loop SPT power spectrum fails to describe the BAO feature correctly because nonlinear evolution induces bulk flows that wash out the oscillatory signal. This effect is not properly accounted for in the SPT framework. To address this issue, the IR resummation method has been developed \cite{blas2016time, senatore2018ir, ivanov2018infrared, lewandowski2020analytic}. In our analysis, we follow the implementation of \cite{class_pt} for IR resummation.

\subsection{Results for the Matter Field}
\label{sec:main_res}
\begin{figure}[ht]
    \centering
    {\includegraphics[width=0.7\textwidth]{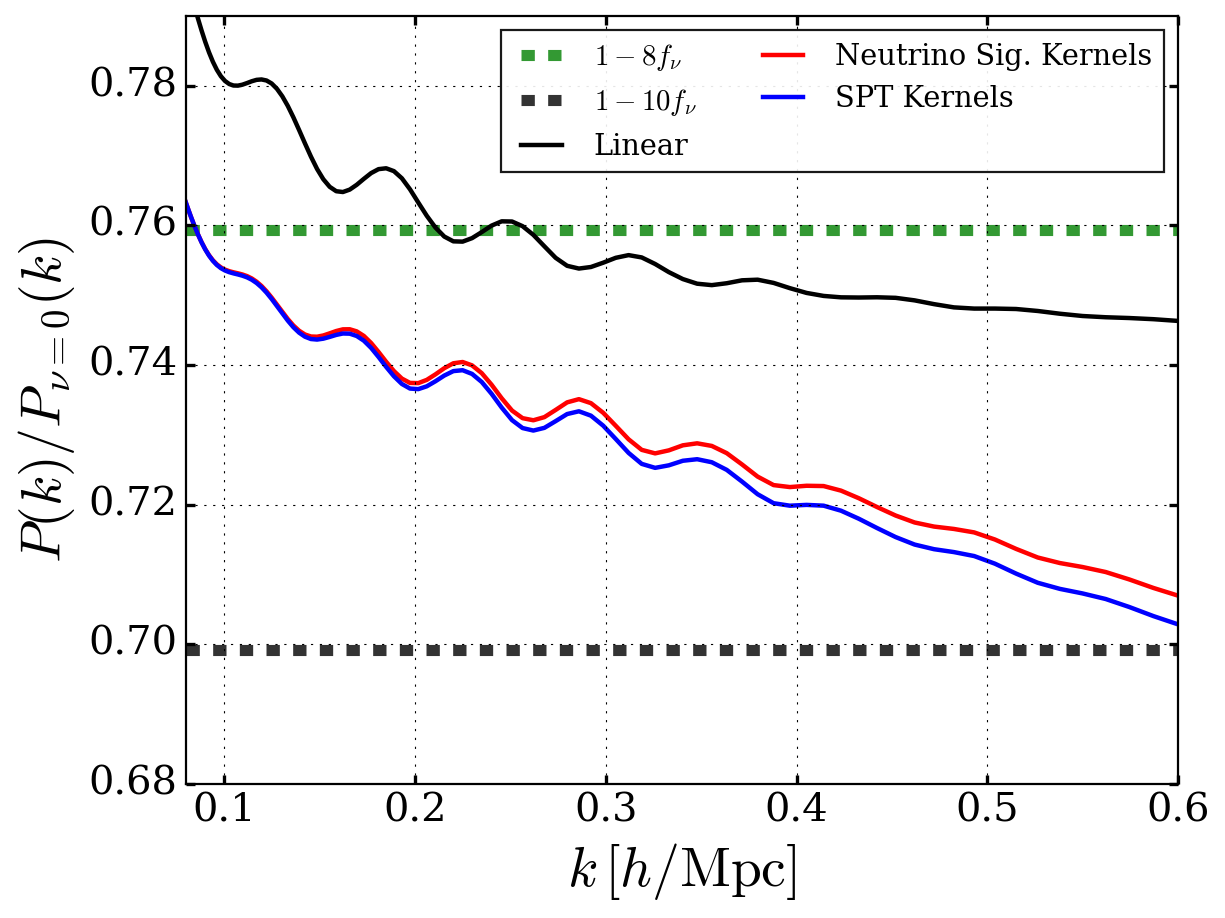} }%
    \caption{The linear (black) and 1-loop matter power spectrum obtained from the SPT kernels (blue) and neutrino signature kernels (red), normalized by their respective massive neutrino-less power spectra. The linear prediction is roughly equal to $1-8f_{\nu}$ \cite{hu1998weighing} and the 1-loop ratio reaches $1-10f_{\nu}$ as was predicted by previous work \cite{wong2008higher, levi2016massive}. The 1-loop power spectrum obtained from the neutrino signature kernels adds a subtle amount of power on small scales relative to the SPT kernels, and this addition depends on $\sum m_{\nu}$, enabling this latter to be probed by redshift surveys. The fiducial cosmology is that of the \texttt{Quijote} simulation \cite{villaescusa2020quijote} listed in the main text in \S\ref{sec:main_res}.}%
    \label{fig:two_images}%
\end{figure}

These extra signature terms will change the matter power spectrum amplitude on small scales. Fig. \ref{fig:two_images} shows the non-linear matter power spectrum with SPT kernels (blue curve), 1-loop matter power spectrum with signature terms (red curve), and linear power spectrum (black curve) all normalized by the power spectrum in the absence of massive neutrinos. The 1-loop spectra are also normalized in such a way that they give the same value for the Root Mean Square (RMS) of the amplitude of the fluctuations on $8\;h/{\rm Mpc}$ scale, $\sigma_8$. The cosmology we considered in making this plot is standard $\Lambda$CDM with three degenerate neutrino masses $\sum m_{\nu} = 0.4\;{\rm eV}$, $\Omega_{\rm m} = 0.3175$, $\Omega_{\rm b} = 0.049$, $\sigma_8 = 0.834$, $h = 0.6711$ and $n_{\rm s} = 0.9624$.\footnote{These parameters are the fiducial parameters of the \texttt{Quijote} N-body simulation suite \cite{villaescusa2020quijote}.} All the linear power spectra are computed using \texttt{class} \cite{Diego_Blas_2011}.

As seen from Fig. \ref{fig:two_images}, the linear power spectrum limits to $P\sim (1-8f_{\nu})P_{\nu = 0}$ as expected from the linear theory \cite{kamalinejad2024simpleanalytictreatmentneutrino, hu1998weighing}. The 1-loop power spectrum also goes to $P\sim (1-10f_{\nu})P_{\nu = 0}$ as was previously discussed \cite{Blas_2014, wong2008higher, saito2008impact, ruggeri2018demnuni,hannestad2020spoon, levi2016massive}. We can observe that the neutrinos will increase the power slightly on small scales. This is similar to the result of \cite{Blas_2014} and is due to the fact the neutrinos are enhancing the $F^{({\rm s})}_2$ and $F^{({\rm s})}_3$ kernels, which will create an enhancement in the 1-loop results.

% It is well-known that the SPT 1-loop power spectrum cannot accurately describe the matter clustering on smaller scales than $k_{\rm max}\sim 0.15\;h/{\rm Mpc}$. This is the primary reason why the EFTofLSS was developed. Therefore, 

When would this difference be observable? Similar to any measurement we carry out, the power spectrum measurements also include error bars and noises. The error bars are characterized by the covariance matrix of the power spectrum, often measured from N-body simulations. The noise originates from the fact that we observe tracers (halos and galaxies) and these objects are discrete, rather than continuous and therefore, there is going to be shot-noise (Poisson noise) present in our measurements. For the matter power spectrum, however, the shot-noise is zero because the matter field from N-body simulations is continuous. We can introduce the theoretical Gaussian covariance matrix of the power spectrum as \cite{biagetti2022covariance}:
\begin{align}
    \mathbf{C}^{\rm PP}(k) = \frac{2(2\pi)^3}{4\pi k^2 \Delta k V_{\rm eff}}P(k)^2
    \label{Eq:Theo_Cov},
\end{align}
where $V_{\rm eff}$ is the effective volume of the redshift survey, $\Delta k$ is the bin-width in Fourier space and $P(k)$ is the linear matter power spectrum. We note that this is the diagonal component of the total matter power spectrum covariance matrix, which is valid in the linear regime \cite{biagetti2022covariance}. On smaller scales, there will be an additional term  from the matter trispectrum which we neglect \cite{mohamed,novell2024approximations,biagetti2022covariance,hamilton2006measuring,wadekar2020galaxy}. 

We also construct the power spectrum covariance matrix from the 500 realizations of the \texttt{Quijote} N-body simulation suite \cite{villaescusa2020quijote} with the fiducial cosmology explained two paragraphs ago. We will discuss the covariance matrix in more depth in \S\ref{sec:Fisher}. 

We note that the covariance matrix estimated from the 500 realizations is noisy and is not reliable for more precise analysis. It should also be corrected by the Hartlap factor \cite{hartlap} which in our case is about 0.9.
\begin{figure}[h]
    \centering
    {\includegraphics[width=0.7\textwidth]{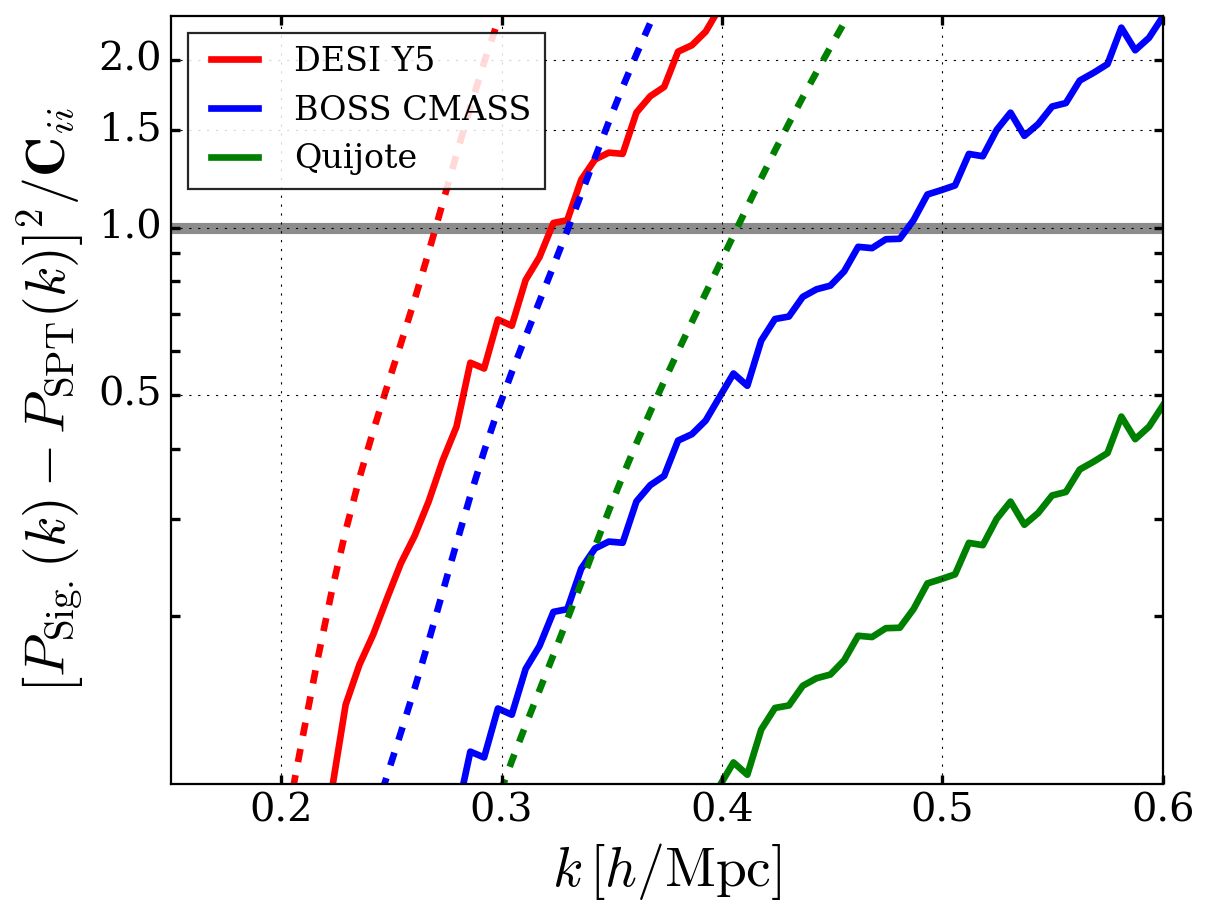} }%
    \caption{The square of the difference between the matter power spectrum at 1-loop obtained from the neutrino signature kernels and from the SPT kernels, normalized by the diagonal of the matter covariance matrix obtained from 500 realizations of the \texttt{Quijote} simulations (solid curves) with the Hartlap factor taken into account. The dashed curves are the same but the covariance matrix is the theoretical covariance matrix from Eq. (\ref{Eq:Theo_Cov}). (\ref{Eq:Theo_Cov}). We rescale the covariance matrix to match different survey volumes. $V_{\rm DESI} = 25\;[{\rm Gpc}/h]^3$, $V_{\rm BOSS} = 5\;[{\rm Gpc}/h]^3$ and $V_{\texttt{Quijote}} = 1\;[{\rm Gpc}/h]^3$. The horizontal thick gray line indicates when the difference between the matter power spectra from the two models equals the error bar on the power spectrum given by the diagonal of the covariance matrix. At the scale ($k$) where the curves cross this line, the difference between the models becomes statistically distinguishable. For a DESI Y5-like survey (red curve), it is at $k_{\rm max}$ $\sim$ $0.3\;h/{\rm Mpc}$.}%
    \label{fig:chi2_for_volumes}%
\end{figure}
Now, once the covariance matrix is constructed either from Eq. (\ref{Eq:Theo_Cov}) or from N-body simulations, we proceed to define the quantity $\chi^2(k)$. This is given by the square of the difference between the 1-loop power spectra obtained using the SPT kernels and the neutrino signature kernels, normalized by the diagonal of the covariance matrix:
\begin{align}
    \chi ^2(k) = \frac{\left[P_{\rm Sig.}(k)-P_{\rm SPT}(k)\right]^2}{\mathbf{C}^{{\rm PP}}_{ii}(k)}.
    \label{eq:chi2}
\end{align}
We note that the $\chi^2(k)$ we defined here is not the same as the measure of the goodness-of-fit which is most commonly defined as $\chi^2 = (\mathbf{D}-\mathbf{M})^{\rm T}\mathbf{C}^{-1}(\mathbf{D}-\mathbf{M})$. Here we are expressing the difference between the models in terms of the standard deviation obtained from the square-root of the diagonal of the covariance matrix.

The covariance matrix is inversely proportional to the survey volume $V_{\rm eff}$ as seen from Eq. (\ref{Eq:Theo_Cov}). Therefore, in order to see the impact of the survey volume we can rescale the covariance matrix to the desired volume. Let us consider \texttt{Quijote} volume with $V_{\texttt{Quijote}} = 1\;[{\rm Gpc}/h]^3$ \cite{villaescusa2020quijote}, BOSS CMASS volume with $V_{\rm BOSS} = 5\;[{\rm Gpc}/h]^3$ \cite{alam2017clustering} and DESI Y5 volume with $V_{\rm DESI} = 25\;[{\rm Gpc}/h]^3$ \cite{DESI:2016, grove2022desi}.

Fig. \ref{fig:chi2_for_volumes} shows the $\chi^2(k)$ as a function of $k$ for different survey volumes. The scale where $\chi^2(k)$ reaches 1 is where the difference between the models is smaller than the error bar on the power spectrum, meaning that the survey can distinguish between the two models for a fixed cosmology. This scale for \texttt{Quijote} volume is around $k_{\rm max}$$\sim 1$$\;h/{\rm Mpc}$, for BOSS CMASS volume is $k_{\rm max}$$\sim 0.5$$\;h/{\rm Mpc}$ and for DESI Y5 volume is $k_{\rm max}$$\sim 0.3$$\;h/{\rm Mpc}$. The difference between the dashed curves (where the theoretical covariance matrix from Eq. (\ref{Eq:Theo_Cov}) is used) and the solid curves (where the covariance matrix is estimated from 500 \texttt{Quijote} realizations) in Fig. \ref{fig:chi2_for_volumes} is due to the fact that Eq. (\ref{Eq:Theo_Cov}) only includes the Gaussian contribution and ignores the trispectrum term \cite{biagetti2022covariance}. The purpose of the plot is to show that for the matter density field where the cosmology is completely known the difference between the models can be detected at the scale where the curves cross the gray horizontal line. This line indicates when  the difference between the models is the same as the error bar on the power spectrum.

% We note that we have used the 1-loop power spectra instead of the linear power spectrum in Eq. (\ref{Eq:Theo_Cov}). Although this approach is not mathematically rigorous, it provides a rough estimate of the covariance matrix on small scales. This approximation matches the covariance from simulations to an acceptable degree. The dashed curves in Fig. \ref{fig:chi2_for_volumes} correspond to using the non-linear power spectrum instead of the linear power in Eq. (\ref{eq:chi2}), while the solid curves correspond to the use of the \texttt{Quijote} covariance matrix.

However, in reality, the situation is more complicated because we do not know the cosmological parameters and should fit for the parameters. We also need to take the galaxy biasing \cite{desjacques} into account, which will increase the size of the parameter space. A more realistic discussion on the impact of kernels in the parameter estimation comes in \S\ref{sec:Fisher}.

\section{Tracer Power Spectrum at 1-loop}

So far we have discussed the matter field and the matter power spectrum. However, we do not observe the matter field directly. We observe galaxies in a redshift survey and halos in N-body simulations. Halos are gravitationally bound structures that are formed by merging of smaller objects. However, the theoretical predictions are about the matter field, and the galaxies and halos are only biased tracers of it. We expand the halo (or galaxy) overdensity in configuration space in terms of the underlying matter density up to third order as \cite{desjacques, mcdonald2009clustering, leonardo2015bias,assassi2014renormalized,mirbabayi2015biased}:
\begin{align}
    \delta_{\rm h} = b_1 \delta_{\rm m}  + \frac{b_2}{2}\delta_{\rm m} ^2 + b_{\mathcal{G}_2} \mathcal{G}_2 +\frac{b_3}{6}\delta_{\rm m}^3 + b_{\Gamma_3}\Gamma_3+ b_{\delta\mathcal{G}_2}\delta \mathcal{G}_2 + b_{\mathcal{G}_3} \mathcal{G}_3 + R_*^2 \partial^2 \delta_{\rm m}+\epsilon
    \label{Eq:halo_expansion},
\end{align}
where $b_1$ is the linear bias, $b_2$ is the non-linear (quadratic) bias, $b_{\mathcal{G}_2}$ is the tidal tensor bias and $b_3$ is the cubic bias. $b_{\Gamma_3}$, $b_{\delta \mathcal{G}_2}$, $b_{\mathcal{G}_3}$ and $R_*$ are higher-order biases. $\epsilon$ is the stochastic terms.  ${\mathcal{G}_2}$ is the tidal tensor operator. In Fourier space we have:
\begin{align}
    {\mathcal{G}_2} = \int_\mathbf{p}\frac{d^3 \mathbf{p}}{(2\pi)^3}\sigma^2(\mathbf{p}, \mathbf{k}-\mathbf{p})\delta(\mathbf{p})\delta(\mathbf{k}-\mathbf{p}).
\end{align}
All higher-order biases and operators such as $\Gamma_3$ and $\mathcal{G}_3$ are defined in \cite{assassi2014renormalized,class_pt, simonovic2018cosmological, Ortol_Leonard_2025}. 

In our work, we only consider biases up to second order. This choice is made because the 1-loop power spectrum does not include contributions from \( b_3 \), \( b_{\delta \mathcal{G}_2} \), \( b_{\mathcal{G}_3} \), \( R_{*} \), and \( \epsilon \), as discussed in the renormalized theory of galaxy biasing \cite{assassi2014renormalized}. It does, however, contain a correction from \( b_{\Gamma_3} \), which we deliberately set to zero since it has been shown in the literature that this bias is highly degenerate with other parameters, and current data cannot give reliable measurements of it \cite{ivanov2020cosmological}. Therefore, our bias expansion only includes \( b_1 \), \( b_2 \), and \( b_{\mathcal{G}_2} \), which offers enough freedom in parameter estimation to extend to smaller scales.

Although the terms proportional to $b_2^2$, $b_{\mathcal{G}_2}^2$ and $b_{\mathcal{G}_2} b_2$ do not explicitly depend on the kernels through their loop integrals (and are therefore insensitive to the new neutrino signatures as we will see shortly), we have to include them in our model since they still contribute to the total amplitude of the power spectrum and covariance matrix. We also include a counter-term equal to $c_{\rm ctr} \,k^2 P_{\rm lin}(k)$ that re-sums the small-scale behavior of the matter power spectrum as described in the EFT framework \cite{pajer2013renormalization, carrasco2012effective,senatore2017effective, senatore2015bias}. Therefore, our bias model has 4 nuisance parameters $(b_1, b_2, b_{\mathcal{G}_2}, c_{\rm ctr})$ and 6  cosmological parameters $(\sum m_{\nu}, \Omega_{\rm m}, \Omega_{\rm b}, \sigma_8, h, n_{\rm s})$.

Using the bias expansion we described earlier, we obtain the power spectrum model at 1-loop for the galaxies in real space \cite{simonovic2018cosmological, class_pt}:
\begin{align}
     P_{\rm gg}(k) &= b_1^2 P^{\rm 1-loop}(k)+ b_1 b_2 P_{b_1 b_2}(k)+2 b_1 b_{\mathcal{G}_2} P_{b_1 b_{\mathcal{G}_2}}(k)\nonumber\\&+\frac{1}{4}b_2^2 P_{b_2^2}(k) + b_{\mathcal{G}_2}^2P_{b_{\mathcal{G}_2}^2}(k) + \frac{1}{2}b_2 b_{\mathcal{G}_2} P_{ b_2 b_{\mathcal{G}_2}}(k) \nonumber\\&+ c_{\rm ctr}k^2 P_{\rm lin}(k).
     \label{Eq:1loop}
\end{align}
We need to expand the density fields as in Eq. (\ref{Eq:halo_expansion}) and apply the kernel to the higher-order CDM+baryon overdensity fields as described in the previous section. It is straightforward to show that the expressions for $P_{b_1 b_2}(k)$ and $P_{b_1 b_{\mathcal{G}_2}}(k)$ are given by:
\begin{align}
    P_{b_1 b_2}(k) &= 2(1-3f_{\nu})\int_{\mathbf{q}}F_2(\mathbf{k}-\mathbf{q}, \mathbf{q})P_{\rm cb}(|\mathbf{k}-\mathbf{q}|)P_{\rm cb}(q)\nonumber\\&+4f_{\nu}\int_{\mathbf{q}}F_2(\mathbf{k}-\mathbf{q}, \mathbf{q})P_{\rm cb\nu}(|\mathbf{k}-\mathbf{q}|)P_{\rm cb}(q)
    \label{Eq:Pb1b2}
\end{align}
and
\begin{align}
    P_{b_1 b_{\mathcal{G}_2}}(k) &= 4(1-f_{\nu})\int_{\mathbf{q}}\sigma^2(\mathbf{q}, \mathbf{k}-\mathbf{q})F_2(\mathbf{k}, -\mathbf{q})P_{\rm cbm}(k)P_{\rm cbm}(q)\nonumber\\&+2(1-f_{\nu})\int_{\mathbf{q}}\sigma^2(\mathbf{q}, \mathbf{k}-\mathbf{q})F_2(\mathbf{q},\mathbf{k} -\mathbf{q})P_{\rm cbm}(|\mathbf{k}-\mathbf{q}|)P_{\rm cbm}(q),
    \label{Eq:Pb1bG2}
\end{align}
where $\sigma^2\left(\mathbf{k}, \mathbf{q}\right) = \left(\mathbf{k}\cdot\mathbf{q}/k q\right)^2-1$. The contributions that do not depend on the signature term, and therefore do not contribute to the difference between the SPT model and neutrino signature model ($\chi^2(k)$) are:
\begin{align}
    &P_{b_2^2}(k) = 2\int_{\mathbf{q}}P_{\rm lin}(q)P_{\rm lin}(|\mathbf{k}-\mathbf{q}|),\\
    &P_{b_{\mathcal{G}_2}^2}(k) = 2\int_{\mathbf{q}} \sigma^4(\mathbf{q},\mathbf{k}-\mathbf{q})P_{\rm lin}(q)P_{\rm lin}(|\mathbf{k}-\mathbf{q}|),\\
    &P_{ b_2 b_{\mathcal{G}_2}}(k) = 2\int_{\mathbf{q}} \sigma^2(\mathbf{q},\mathbf{k}-\mathbf{q})P_{\rm lin}(q)P_{\rm lin}(|\mathbf{k}-\mathbf{q}|),
\end{align}
where we have contracted a matter density field with the CDM+baryon field as $P_{\rm cbm} = \langle\delta_{\rm cb} \delta_{\rm m}\rangle$. Now, as before, we express the kernel as the sum of the SPT kernel and the neutrino signature correction, then compute the integrals using the FFTLog method. Fortunately, all of these integrals have already been calculated previously \cite{simonovic2018cosmological, mcewen2016fast}, so we do not repeat them here.

It is important to mention that all the integrals in \( P_{b_1 b_2}(k) \) (Eq. (\ref{Eq:Pb1b2})) \( P_{b_1 b_{\mathcal{G}_2}}(k) \) (Eq. (\ref{Eq:Pb1bG2})) and  are divergent for biases outside of the range \( -3 < \nu < -1/2 \) and \( -3 < \nu < 1/2 \), respectively. For these terms, we use a bias parameter that prevent the divergence. Specifically, we set \( \nu = -1.75 \) for both integrals.  We fix $\nu = -0.3$ for all other contributions in Eq. (\ref{Eq:1loop}) \cite{simonovic2018cosmological}.

\section{Tracer Real-Space Bispectrum}

The first measurement of the galaxy bispectrum was obtained from the IRAS PCSz galaxy redshift survey by \cite{feldman2001constraints}. In the following decades the bispectrum grew in popularity and now it is one of the most important observables. The bispectrum is a direct probe of Primordial Non-Gaussianities (PNGs) \cite{Guhaprimordial,bisp_png,emiliano2006cosmology, scoccimarro2000bispectrum, dizgah2021primordial,baldauf2011primordial,LewandowskiLimits,ShirasakiConstraining} and it has been proven in multiple studies that the bispectrum helps reduce the error bars on the parameters \cite{GualdiEnhancing,IvanovPrecision,PhilcoxCosmology,senatoreforecast,simbigchang, hahn2023rmsscriptsizeimbigcosmological} by breaking the degeneracies in the power spectrum parameter space \cite{fry_1993, yankelevich2019cosmological, Hahn_2020,rezaei_mahdi_BAO}. There have been numerous BOSS, eBOSS and DESI bispectrum analyses which can be found in \cite{Gilclustering, Gilinterpretation,ChildBispectrum,PearsonDetection,SlepianDetection,CabassConstraints,SpaarCosmological,gil2016clustering, D_Amico_2024, Ivanov_2022,philcox2022boss,LuPreference,Ivanovoptimal,PearsonErratum}.

The effect of neutrinos on the kernels also extends to the tree-level bispectrum, as studied for the first time in \cite{kamalinejad2020non}. As it turns out, the neutrinos affect the estimation of the halo or galaxy biases. The tracer bispectrum can be defined as \cite{Bernardeau_2002, scoccimarro1998nonlinear, scoccimarro1999bispectrum}:
\begin{align}
    B(\mathbf{k}_1, \mathbf{k}_2, \mathbf{k}_3)(2\pi)^3 \delta_{\rm D}^{[3]}(\mathbf{k}_1+ \mathbf{k}_2+ \mathbf{k}_3) = \langle \delta_{\rm h}(\mathbf{k}_1)\delta_{\rm h}(\mathbf{k}_2)\delta_{\rm h}(\mathbf{k}_3) \rangle
    \label{Eq:B_def},
\end{align}
where $\delta_{\rm h}$ can be obtained from Eq. (\ref{Eq:halo_expansion}) and the Dirac delta function $\delta_{\rm D}^{[3]}$ ensures that $\mathbf{k}_1$,  $\mathbf{k}_2$, and $\mathbf{k}_3$ form a closed triangle in Fourier space.

Similar to the tracers power spectrum we studied earlier, we need to write the tracers bispectrum in terms of the CDM+baryon density $(\delta_{\rm cb})$ and the neutrino density $(\delta_{\nu})$ since the kernels only apply to $\delta_{\rm cb}$, and not the neutrinos. Using the PT expansion of equations (\ref{Eq:delta_n}) and (\ref{Eq:theta_n}) along with the contractions of the density fields we can obtain the real-space galaxy bispectrum as \cite{Bernardeau_2002}:
\begin{align}
    B_h(k_1, k_2, k_3) &= b_1^3\Bigg[2\big(1-3f_{\nu}\big) \bigg(F_2(\mathbf{k}_1, \mathbf{k}_2)+\Delta F_2(\mathbf{k}_1, \mathbf{k}_2)\bigg) P_{\rm cb}(k_1) P_{\rm cb}(k_2) \\&\qquad\quad+ 3 f_{\nu} {F}_2(\mathbf{k}_1, \mathbf{k}_2) P_{\rm cb\nu}(k_1) P_{\rm cb}(k_2) \nonumber\\  &\qquad\quad+3 f_{\nu} {F}_2(\mathbf{k}_1, \mathbf{k}_2) P_{\rm cb\nu}(k_2) P_{\rm cb}(k_1)\Bigg]\nonumber\\&+b_1^2b_{\mathcal{G}_2}\Bigg[2\;\,\sigma(\mathbf{k}_1, \mathbf{k}_2)^2\,P_{\rm lin}(k_1) P_{\rm lin}(k_2) \Bigg]\nonumber\\
    &+ \frac{b_1^2 b_2}{2} \Bigg[2 P_{\rm lin}(k_1)P_{\rm lin}(k_2)\Bigg]+{\rm cyc.}\nonumber\,,
\end{align}
where $\Delta F_2$ is the neutrino signature effects in the kernels (Eq. (\ref{Eq:F2_sig})) and cyc. means that the bispectrum needs to be cyclically summed over $\mathbf{k}_1$, $\mathbf{k}_2$ and $\mathbf{k}_3$. Since the neutrino effects are only observable at wave-numbers larger than their non-relativistic scale (\( k_{\rm NR} \)), we assume that the bispectrum is only affected by the signature kernels at \( k \) values larger than \( k_{\rm NR} \). This is achieved by using a Heaviside function to filter out the neutrino effects at smaller wave-numbers, similar to the approach adopted in \cite{kamalinejad2020non}. The tree-level bispectrum then also depends on the biases ($b_1, b_2, b_{\mathcal{G}_2}$) and six cosmological parameters $(\sum m_{\nu}, \Omega_{\rm m}, \Omega_{\rm b}, \sigma_8, h, n_{\rm s})$.

In practice, both the theoretical power spectrum and bispectrum should be binned in order to be compared to data. For the power spectrum, this means that we need to average over spherical shells with width $\Delta k$ in Fourier space as:
\begin{align}
    \bar{P}(k) = \frac{\sum_{\mathbf{q}\in\mathbf{k}}VP(q)}{4\pi k^2 \Delta k V} = \frac{1}{k^2 \Delta k}\int_{k-\Delta k/2}^{k+\Delta k/2}q^2P(q) dq,
    \label{Eq:binn_PS}
\end{align}
with $\bar{P}$ denoting the binned power spectrum.

For the bispectrum, the binning is more complicated because different binning schemes have been shown to create systematic issues \cite{D_Amico_2024}. We calculate the binned bispectrum $\bar{B}$ according to \cite{D_Amico_2024}:
\begin{align}
    \bar{B}(k_1, k_2, k_3) = \frac{1}{V_{\rm T}}\int_{q_i\in k_i} q_1\, q_2\, q_3\, B(q_1, q_2, q_3) \,dq_1 \,dq_2\, dq_3
    \label{Eq:Binn_Bis},
\end{align}
where $V_{\rm T}$ is the normalization factor and is given by:
\begin{align}
    V_{\rm T} = \int q_1\, q_2\, q_3 \,dq_1 \,dq_2\, dq_3
    \label{Eq:Vt}.
\end{align}
We only consider bin centers $k_1, k_2, k_3$ that form a closed triangle, satisfying the triangle condition ($|k_1 - k_2| < k_3 < k_1 + k_2$ for all permutations), as well as the condition $k_1 \geq k_2 \geq k_3$. As regards our numerical implementation, we divide each $k$ bin into 5 smaller intervals, creating 125 triples $(q_1, q_2, q_3)$ overall. We then evaluate the bispectrum at each triple and approximate the integrals in equations  (\ref{Eq:Binn_Bis}) and (\ref{Eq:Vt}) with a sum over all $q_i$ that respect the $q_1 \geq q_2 \geq q_3$ condition. In our binning scheme, we do not consider open triangles formed by $q_1, q_2, q_3$ triplets. We also exclude folded triangles satisfying $q_1 + q_2 = q_3$ (or similar configurations).

\section{Fisher Forecast \& Observability of the Signature Kernels}
\label{sec:Fisher} 
The Fisher forecast formalism provides an estimate of the error bars and parameter correlations around a fiducial value, given a survey volume, redshift, and other related survey specifications \cite{Tegmark_1998}. In this work, we apply this formalism to assess whether the kernels produce a statistically meaningful difference (\textit{i.e.} there is at least $1\sigma$ difference between the two models) in the estimation of cosmological parameters in future surveys \cite{noriega2022fast, kamalinejad2020non}. The approach is to fit two different models (one with SPT kernels and the other with neutrino signature kernels) to N-body simulation data to obtain the central values. 
We then perform a Fisher forecast  around those values to determine the error bars, as the kernels can affect both the central values and the error bars. This method is computationally cheaper than performing a Markov Chain Monte Carlo (MCMC) analysis and should be adequate for our purposes.

\subsection{Covariance Matrix Estimation \& Maximum Likelihood}
\label{sec:Cov}
Using the \texttt{Quijote} simulation suite \cite{villaescusa2020quijote} we estimate the covariance matrix of the power spectrum, bispectrum and a joint covariance of the power spectrum+bispectrum, which is almost noise-less since \texttt{Quijote } has 15,000 realizations with fiducial cosmology given by the following parameters: $\sum m_{\nu}=0.0\;{\rm eV}$, $ \Omega_{\rm m} = 0.3175$, $\Omega_{\rm b} = 0.049$, $\sigma_8 = 0.834$, $h = 0.6711$, and $n_{\rm s} = 0.9624$ in a box with volume $V_{\texttt{Quijote}} = 1 \;[{\rm Gpc}/h]^3$ at redshift $z=0$. We can represent the joint covariance matrix of the power spectrum+bispectrum as \cite{biagetti2022covariance,FumagalliFitting, sugi_covar}:
\begin{align}
    \mathbf{C} = \begin{pmatrix}
    \mathbf{C}^{\rm P} & \mathbf{C}^{\rm PB}    \\
    \mathbf{C}^{\rm PB} & \mathbf{C}^{\rm B}
\end{pmatrix},
\label{Eq:JointCov}
\end{align}
$\mathbf{C}^{\rm P} \equiv \langle \delta\mathbf{P} \delta\mathbf{P}\rangle$ represents the power spectrum covariance matrix (we have defined $\delta\mathbf{P}\equiv\mathbf{P}-\langle \mathbf{P}\rangle$), $\mathbf{C}^{\rm PB} \equiv \langle \delta\mathbf{P}\delta\mathbf{B}\rangle$ is the cross-covariance between the power spectrum and bispectrum and lastly, $\mathbf{C}^{\rm B} \equiv \langle \delta\mathbf{B}\delta\mathbf{B}\rangle$ is the bispectrum covariance matrix.

We also use 500 realizations with neutrino mass of $\sum m_{\nu} = 0.4\;{\rm eV}$ (which is the most massive case) to estimate the signal corresponding to the power spectrum and bispectrum in real space \cite{Oddolikelihood}. We then fit the power spectrum, bispectrum and power spectrum+bispectrum templates obtained by assuming the neutrino signature model and by assuming the SPT model to find the recovered central values. One might think that the separate fit of the power spectrum and bispectrum is not different from the joint fit of the power spectrum+bispectrum but it is not the case, due to the cross-covariance between the power spectrum and bispectrum, $\mathbf{C}^{\rm PB}$ \cite{biagetti2022covariance,novell2024approximations,Salvalaggiocovariance, gualdi2020galaxy}. The fit we mentioned earlier is done by maximizing the likelihood function:
\begin{align}
    \ln{\mathcal{L}}(\mathbf{D}|\mathbf{M}) = -\frac{1}{2}(\mathbf{D}-\mathbf{M})\mathbf{C}^{-1}(\mathbf{D}-\mathbf{M})^{\rm T},
\end{align}
where $\mathbf{D}$ is the data vector corresponding to the power spectrum $\mathbf{P}$, bispectrum $\mathbf{B}$, or the joint power spectrum and bispectrum. This last is $\mathbf{P}$ and $\mathbf{B}$ stacked together as a vector: \textit{i.e.} $\mathbf{D} = (\mathbf{P}, \mathbf{B})$. The model vector $\mathbf{M}$ is the model and $\mathbf{C}^{-1}$ is the inverse of the covariance matrix. We note that we use $\mathbf{C}^{\rm P}$ for fitting the power spectrum, $\mathbf{C}^{\rm B}$ for the bispectrum and $\mathbf{C}$ (Eq. (\ref{Eq:JointCov})) for the joint power spectrum+bispectrum. It is straightforward to show that in the absence of cross-covariance between $\mathbf{P}$ and $\mathbf{B}$ the covariance matrix becomes block-diagonal, and the inverse of a block-diagonal matrix is equal to the inverse of blocks, separately. This shows that a joint fit of the power spectrum and bispectrum is different from two separate fits of the power spectrum and bispectrum.

To extend the analysis to future surveys, we use $V_{\rm DESI}^{\rm P} = 25\; [{\rm Gpc}/h]^3$ \cite{adame2024desi,noriega2022fast} and $V_{\rm DESI}^{\rm B} = 14\; [{\rm Gpc}/h]^3$. \footnote{The effective volume of the power spectrum is not necessarily equal to the bispectrum} The fitting process of the power spectrum and bispectrum separately is not sensitive to the survey volume since we rescale the $\mathbf{C}^{\rm P}$ for the power spectrum and $\mathbf{C}^{\rm B}$ for the bispectrum, which does not change the location of the maximum likelihood. For the joint analysis, since the covariance matrix is represented as blocks ($\mathbf{C}^{\rm P}$, $\mathbf{C}^{\rm PB}$ and $\mathbf{C}^{\rm B}$ in Eq. (\ref{Eq:JointCov})) and each of the blocks are rescaled differently, the survey volume affects the best-fit. Therefore, we multiply the cross covariance by the geometric mean of $V_{\rm DESI}^{\rm P}$ and $V_{\rm DESI}^{\rm B}$ which is $V_{\rm DESI}^{\rm PB} = 18.7\; [{\rm Gpc}/h]^3$.

It is crucial to mention the range of wave numbers we use. For the power spectrum our model with four nuisance parameters $(b_1, b_2, b_{\mathcal{G}_2}, c_{\rm ctr})$ can describe the \texttt{Quijote} power spectrum data to $k_{\rm max} = 0.3\;h/{\rm Mpc}$. For the bispectrum, we follow the recommendation made by \cite{Ivanov_2022} and choose $k_{\rm max} = 0.08\;h/{\rm Mpc}$. Their analysis is at $\bar{z}=0.61$ and our analysis is at $\bar{z}=0$. We found that the model still works well at this $k_{\max}$. On this range, the tree-level bispectrum can describe the data well enough. We noticed that increasing the $k_{\rm max}$ leads to a poorer fit. For the power spectrum, the bin width is equal to $\Delta k = 0.00628\;h/{\rm Mpc}$, leading to 47 bins, and for the bispectrum the bin width is $\Delta k = 0.01881\;h/{\rm Mpc}$, producing 13 triangle bins.
% \begin{table}[h]
%     \centering
%     \label{tab:empty}
%     \begin{tabular}{c|ccc|ccc}
%         \toprule
%         & \multicolumn{3}{c|}{Neutrino Signature} & \multicolumn{3}{c}{SPT} \\
%          Parameters& P & B & P+B (DESI Y5) & P & B & P+B (DESI Y5) \\
%         \midrule
%         $b_1$  & 1.703 & 1.766 & 1.795 & 1.703 & 1.721 & 1.795 \\
%         $b_2$  & -1.676 & -0.033 & -0.052 & -1.676 & -0.092 & -0.053 \\
%         $b_{\mathcal{G}_2}$  & 0.038 & -0.106 & -0.089 & 0.039 & -0.219 & -0.088 \\
%         $c_{\rm ctr}$  & 63.420 & - & 36.220 & 63.420 & - & 36.210 \\
%         $\sum m_{\nu}$  & 0.400 & 0.395 & 0.400 & 0.400 & 0.388 & 0.400 \\
%         $\Omega_{\rm m}$  & 0.318 & 0.317 & 0.318 & 0.318 & 0.318 & 0.318 \\
%         $\sigma_8$  & 0.834 & - & 0.834 & 0.834 & - & 0.834 \\
%         $h$  & 0.671 & 0.668 & 0.671 & 0.671 & 0.664 & 0.671 \\
%         \bottomrule
%     \end{tabular}
%     \caption{The best-fit values of the halo biases, counter-term, and cosmological parameters obtained from maximizing the likelihood function for the neutrino signature model vs. the SPT model. We used $k_{\rm max} = 0.3\;h/{\rm Mpc}^{-1}$ for the power spectrum and $k_{\rm max} = 0.1\;h/{\rm Mpc}^{-1}$ for the bispectrum. Comparing the recovered values between the neutrino signature model and the SPT model reveals that the difference between the two models is really small. However, the real difference will be revealed later in the triangle plots of the Fisher forecast.}
%     \label{Tab:bestfit}
% \end{table}

Here are the parameters we consider in our fitting procedure. In all our fits, we fix the values of $\Omega_{\rm b}$ and $n_{\rm s}$ to the fiducial values given by the \texttt{Quijote} simulation \cite{villaescusa2020quijote} since the LSS does not give very good estimates of these parameters. For the power spectrum, we find the best-fit values of the $b_1, b_2, b_{\mathcal{G}_2}, c_{\rm ctr}, \sum m_{\nu}, \Omega_{\rm m}, \sigma_8$ and $h$. We also fix the value of $\sigma_8$ for the bispectrum due to the degeneracy between $b_1$ and $\sigma_8$ in the tree-level bispectrum \cite{fry_1993, yankelevich2019cosmological}. Adding the higher-order biases will alleviate this degeneracy but $b_1$ and $\sigma_8$ remain highly correlated. For the joint analysis of the power spectrum+bispectrum, we fit all the parameters except for $\Omega_{\rm b}$ and $n_{\rm s}$. We consider the biases and counter-term as free parameters \cite{McDonald_2009, Saito_2014}.

We also note that the initialization of the fitting process (the initial guess of the best-fit parameters) is the same when fitting both models. Therefore, initialization is not a factor that affects the difference in the best-fit values. The way we initialize the best-fit parameter estimation is as follows. Since we know the cosmological parameters with which \texttt{Quijote} simulation is run, we only fit for the biases and the counter-term. Next, we use these biases and the counter-term to initialize the maximization of the likelihood function for all parameters.

\subsection{Effect of Neutrino Signature Kernels on Parameter Estimation}
Once the central values are recovered, we perform a Fisher forecast around those central-values to obtain an estimation of the error bars on the parameters and also to see how the kernels affect the correlations between parameters. The Fisher information matrix is obtained from the derivative of the model vector, $\mathbf{M}$ and the inverse of the covariance matrix as \cite{Tegmark_1998}:
\begin{align}
    \mathbf{F}_{\rm ij} = \frac{\partial \mathbf{M}}{\partial \theta_i}\mathbf{C}^{-1}\frac{\partial \mathbf{M}}{\partial \theta_j},
\end{align}
where $i$ and $j$ represent the $i^{\rm th}$ and $j^{\rm th}$ element of the Fisher matrix. For the power spectrum, the parameter space we consider is $\theta:\{b_1, b_2, b_{\mathcal{G}_2}, c_{\rm ctr}, \sum m_{\nu}, \Omega_{\rm m}, \Omega_{\rm b}, \sigma_8, h, n_{\rm s}\}$, for the bispectrum, $\theta:\{b_1, b_2, b_{\mathcal{G}_2}, \Omega_{\rm m}, h, n_{\rm s}\}$. For the joint forecast we consider the same parameters as the power spectrum. We remind the reader that we did not fit for $\Omega_{\rm b}$ and $n_{\rm s}$ in the power spectrum (and the bispectrum), but we considered them in the Fisher analysis. This approach was taken to assess how the neutrino signature kernels affect parameter estimation. Since the kernel effects are small, we do not expect them to significantly shift the parameter values. Moreover, because $\Omega_{\rm b}$ and $n_{\rm s}$ are not well-constrained, any potential shift would not be meaningful.

Once we obtain the Fisher matrix, we compute its inverse and use a multivariate Gaussian distribution to generate 10,000 synthetic samples. Finally, we employ \texttt{mcsamples} from the \texttt{GetDist} package \cite{lewis2019getdistpythonpackageanalysing} to obtain the samples needed for the triangle plot (Fig. \ref{fig:QT_triangle_plot}), which are produced using the same package.

\begin{table}[h]
    \centering
    \renewcommand{\arraystretch}{1.2}
    \setlength{\arrayrulewidth}{1pt} % Thicker lines
    \begin{tabular}{|c||c c c|}
        \hline
        & \multicolumn{3}{c|}{\((\theta_{\rm Sig.} - \theta_{\rm SPT})/\sigma_{\theta}\times 100\)} \\
        Parameters & \textbf{P} & \textbf{B} & \textbf{P+B} \\
        \hline
        $b_1$               &  0.29  &  35.00 & 0.48 \\
        $b_2$               &  0.02  &  14.70 & 1.44 \\
        $b_{\mathcal{G}_2}$ & 0.34   &  44.28 & --1.14 \\
        $c_{\rm ctr}$       & --0.18 &   --   & 0.04 \\
        $\sum m_{\nu}$      & --0.05 &   --   &  0.00 \\
        $\Omega_{\rm m}$    & --0.25 & --1.12 & 0.00 \\
        $\sigma_8$          & --0.07 &   --   & 0.01 \\
        $h$                 & 0.05   &  0.00  &  0.01 \\
        $n_{\rm s}$         & 0.00   & --     & 0.00 \\
        \hline
    \end{tabular}
    \caption{Percent shift of best-fit parameters between the two models ($\theta_{\rm Sig.}$ which is the best-fit of the neutrino signature model vs. $\theta_{\rm SPT}$, which is the best-fit of the SPT model), normalized by the error bars obtained from the Fisher forecast, $\sigma_{\theta}$, shown for a DESI Y5-like survey. The signature-kernel-induced shifts remain small and generally below the detection threshold, except in the bispectrum for galaxy biases $b_1$, $b_2$ and $b_{\mathcal{G}_2}$, similarly to the results in \cite{kamalinejad2020non}. The shift of the linear bias $b_1$ in the bispectrum when using the SPT model is $0.35\sigma$. For the higher-order biases $b_2$ and $b_{\mathcal{G}_2}$ the shifts are $0.14\sigma$ and $0.44\sigma$, respectively. The negative shifts indicate that the SPT model prefers larger values.}
    \label{tab:galaxy_bias_desi}
\end{table}

Table \ref{tab:galaxy_bias_desi} shows the percent shift of the estimated parameters. These shifts are the differences in the best-fit values obtained from our two models, normalized by the error bar derived from the Fisher matrix. The shifts are shown when the neutrino signature kernels are used compared to the SPT kernels for the power spectrum, bispectrum, and joint fit. As we can see, for the power spectrum and power spectrum+bispectrum, the shift is less than 2\%, even for the DESI Y5 volume. This indicates that kernel effects in the higher-order power spectrum are not observable, even in future surveys. For the bispectrum, however, the linear bias $b_1$ and the tidal tensor bias $b_{\mathcal{G}_2}$ shift by $0.35\sigma$ $0.45\sigma$ respectively, and the non-linear bias shifts by about $0.15\sigma$ for the DESI Y5 volume.

\subsection{Error Forecasts and Bispectrum Gains for DESI Y5}
\label{subsec: Triplot}

\begin{figure}[h]
    \centering
    {\includegraphics[width=1.\textwidth]{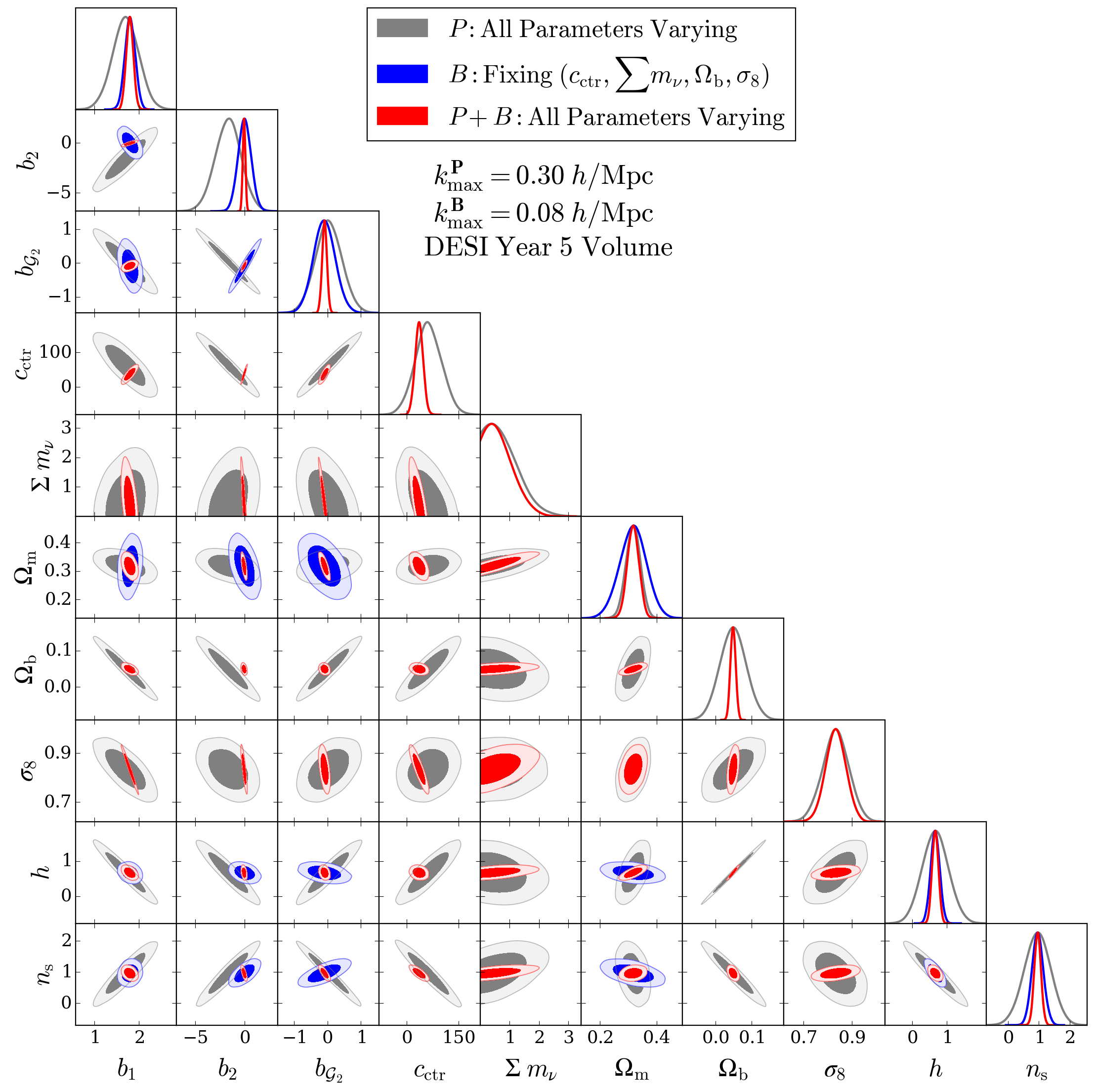} }%
    \caption{The $1\sigma$ and $2\sigma$ error bars obtained from the Fisher matrix analysis of the real-space power spectrum (gray), real-space bispectrum (blue), and the joint power spectrum + bispectrum (red) are shown. We use a covariance matrix estimated from 15,000 realizations of the \texttt{Quijote} simulation suite  and rescaled to match the volume of DESI Y5, and we include the cross-covariance between the power spectrum and bispectrum. The central values are obtained from fitting our neutrino signature model to the 500 \texttt{Quijote} simulation suite realizations with $\sum m_{\nu} = 0.4\,{\rm eV}$, discussed in more details in \S\ref{sec:Cov}. We use a scale cut of $k_{\rm max} = 0.3\;h/{\rm Mpc}$ for the power spectrum and $k_{\rm max} = 0.08\;h/{\rm Mpc}$ for the bispectrum. As the figure shows, the bispectrum helps break degeneracies present in the power spectrum, leading to a significant reduction in the error bars for all parameters. In the bispectrum-only forecast, we fix the value of $\sum m_{\nu}$, as the chosen bispectrum $k_{\rm max}$ does not provide sufficient information on it. We also fix $\Omega_{\rm b}$ and $\sigma_8$, since the bispectrum alone does not constrain them tightly at this scale cut.}%
    \label{fig:QT_triangle_plot}%
\end{figure}

Now that we have established that the impact of the neutrino signature kernels on the recovered parameters is minimal, primarily affecting only the biases derived from the bispectrum. Let us examine the potential of a DESI Y5-like survey for constraining cosmological parameters. Fig.~\ref{fig:QT_triangle_plot} shows the triangle plot obtained from the Fisher matrix analysis, as described above, for a DESI Y5-like sample. The gray ellipses indicate the $1\sigma$ and $2\sigma$ constraints from the power spectrum, the blue ellipses correspond to the bispectrum, and the red ellipses represent the joint analysis, which includes the rescaling of the cross-covariance matrix as described above in \S\ref{sec:Cov}.

\begin{figure}[h]
    \centering
    {\includegraphics[width=0.6\textwidth]{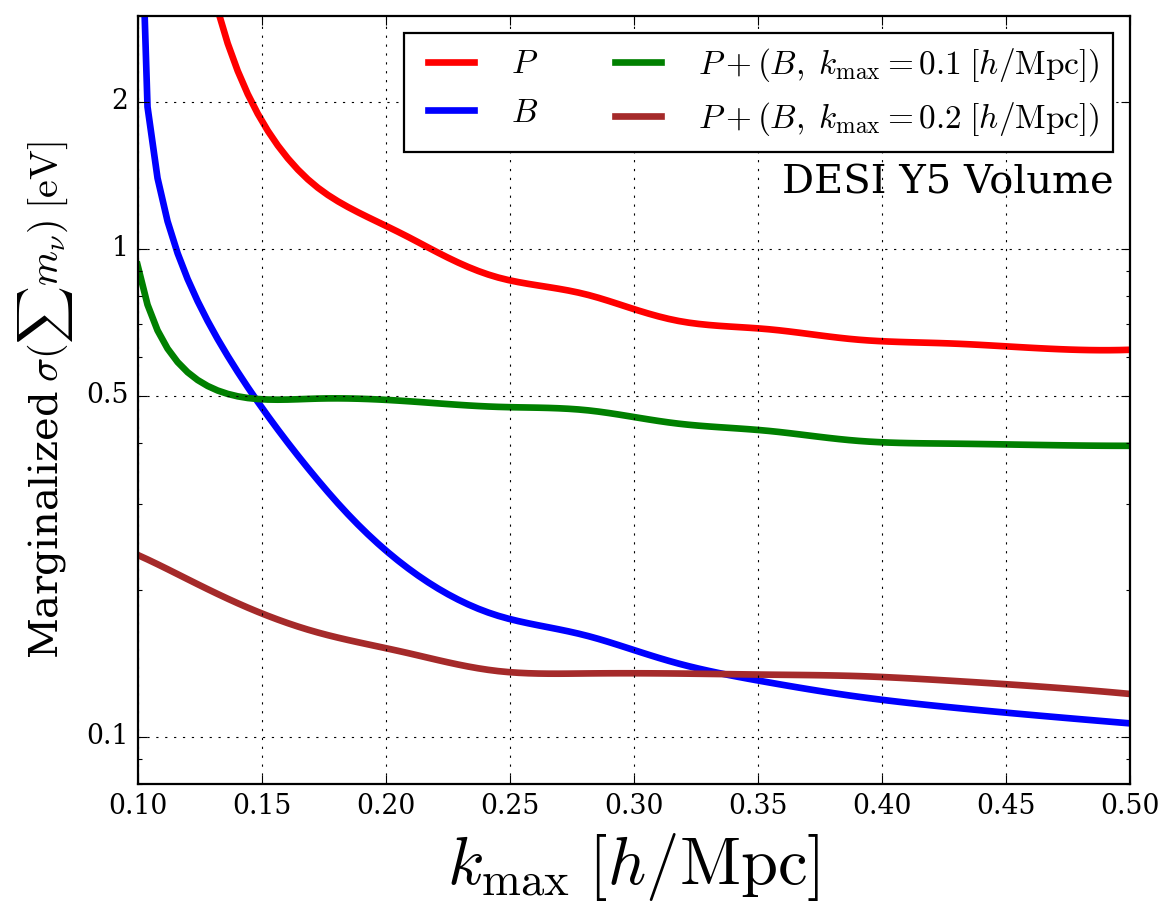} }%
    \caption{The marginalized $1\sigma$ error bar on the neutrino mass, obtained from forecasts using the power spectrum (red), bispectrum (blue), and the joint power spectrum + bispectrum with a scale cut for the bispectrum of $k_{\rm max} = 0.1$ (green) and $0.2\;h/{\rm Mpc}$ (red). The uncertainty on the neutrino mass from the power spectrum alone saturates at $k_{\rm max} = 0.3\;h/{\rm Mpc}$. However, the bispectrum continues to reduce the error bars on the neutrino mass as $k_{\rm max}$ increases. This improvement arises from the rapidly increasing number of triangle configurations available as we go to higher $k_{\rm max}$. The joint power spectrum + bispectrum forecast also demonstrates a substantial improvement over the power spectrum alone. This improvement is because the bispectrum breaks the degeneracies present in the power spectrum.}%
    \label{fig:sigmavs.k}%
\end{figure}

Fig.~\ref{fig:QT_triangle_plot} reveals that even with a simple bispectrum model and a conservative scale cut of $k_{\rm max} = 0.08\;h/{\rm Mpc}$, the bispectrum helps reduce the overall uncertainties on several parameters. This reduction is particularly evident for the bias parameters, $\Omega_{\rm b}$, $h$, and $\sigma_8$. We quantify this reduction as the “information gain” on each parameter, defined by $(\sigma^{\mathbf{P}}_{\rm parameter} - \sigma^{\mathbf{P+B}}_{\rm parameter}) / \sigma^{\mathbf{P}}_{\rm parameter}$, where $\sigma^{\mathbf{P}}_{\rm parameter}$ and $\sigma^{\mathbf{P+B}}_{\rm parameter}$ are the error bars from the power spectrum alone and from the joint power spectrum and bispectrum analysis, respectively. The information gain from including the bispectrum for cosmological parameters is substantial. The information gain, in percentage, is as follows. $\Omega_{\rm m}$: 18.5\%, neutrino mass: 18\%, $\Omega_{\rm b}$: 80\%, $\sigma_8$: 23\%, $h$: 75\%, and $n_{\rm s}$: 75\%. This improvement arises because the bispectrum can help break intrinsic degeneracies in the power spectrum, such as between $b_1$ and $\sigma_8$. This is clearly visible in the $b_1$–$\sigma_8$ panel, where the red ellipse (joint power spectrum and bispectrum) is not only smaller than the gray (power spectrum only) but also oriented differently. A similar effect is observed in the ($\Omega_{\rm m}$, $h$) and ($\Omega_{\rm m}$, $n_{\rm s}$) panels.
Thus far, we have used the bispectrum up to $k_{\rm max} = 0.08\,h/{\rm Mpc}$. Although the bispectrum is not valid on larger Fourier wave numbers (smaller scales), let us extend the bispectrum $k_{\rm max}$ to see what happens to the error bar on the neutrino mass. 
Fig. ~\ref{fig:sigmavs.k} shows the $1\sigma$ error bar on the neutrino mass while marginalizing over all other parameters. In this plot, for the power spectrum and the power spectrum + bispectrum, we have fixed $\Omega_{\rm b}$, while for the bispectrum alone we have fixed both $\Omega_{\rm b}$ and $\sigma_8$. As we can see, the power spectrum curve (red) saturates at $k_{\rm max} = 0.3\;h/{\rm Mpc}$, and there is not much additional information to be gained from it on the neutrino mass.

The green and brown curves show the joint power spectrum + bispectrum forecast. To obtain these curves, we fixed the $k_{\rm max}$ of the bispectrum to $k_{\rm max} = 0.1\;h/{\rm Mpc}$ for the green curve and $k_{\rm max} = 0.2\;h/{\rm Mpc}$ for the brown curve. This plot clearly shows that even a simple bispectrum model can reduce the error bar from the red curve down to the green curve, which is a significant improvement, even at $k_{\rm max} = 0.2\;h/{\rm Mpc}$, the error bar is reduced by about a factor of two.

The bispectrum (blue) curve continues to decrease as $k_{\rm max}$ increases. This is because, as $k_{\rm max}$ grows, the number of triangle configurations increases much more rapidly than the number of modes in the power spectrum, leading to a significant gain in information.

Let us now compare our findings with previous work. \cite{kamalinejad2020non} obtains a similar result by creating synthetic data using the theoretical covariance matrix for the redshift-space bispectrum up to $k_{\rm max} = 0.35 \;h/{\rm Mpc}$ and finds that only the biases are significantly shifted. Here, we also observe that the cosmological parameters are not sensitive to the effects of the kernels. They also hypothesize that this is because most of the information on the cosmological parameters comes from the power spectrum, and the kernel effects only add a small fraction to the information obtained from the power spectrum. For the biases, however, there is no additional information coming from the power spectrum; therefore, they are more sensitive to the neutrino effects in the kernels. 

\cite{noriega2022fast} also performs the same test as we did here but for the redshift-space power spectrum created using their own code, \texttt{folps$\nu$}. \cite{NoriegaComparing} also performs a similar test and shows that the recovered parameters for both kernels are within $1\sigma$ of each other. However, for the neutrinos \cite{NoriegaComparing} reports a $14\%$ shift when switching from the SPT kernels to their \texttt{fk} kernels, which is different from our results.

Therefore, from the parameter estimation perspective, it appears that the power spectrum is not sensitive to the effects of the signature kernels. For the bispectrum, however, the biases are the parameters most affected.

\section{Discussion and Conclusions }
The neutrinos modify the growth rate of matter, making it both time- and scale-dependent. This leads to a time- and scale-dependent effect on the higher-order corrections to the power spectrum as well as the tree-level bispectrum. Computing these modifications is challenging but has been successfully addressed in previous works \cite{saito2008impact, wong2008higher, levi2016massive, Blas_2014, garny2021loop, aviles2020lagrangian, aviles2021clustering, kamalinejad2020non}.

Although the approach in \cite{wong2008higher} faces fundamental challenges, such as violating momentum conservation, it provides a simple framework in which an analytical formula describing the modifications of neutrinos to the SPT kernels can be obtained. This serves as a computationally inexpensive approximation to the neutrino effects on the power spectrum and bispectrum, first calculated in \cite{kamalinejad2020non}. In this paper, we built on their work by deriving the new 
$F_3^{({\rm s})}$ kernel and computing the loop corrections to the real-space power spectrum using the FFTLog method.

We then investigated whether these kernel effects are observable in the summary statistics of halos in large-scale structure. To assess this, we used the \texttt{Quijote} simulation suite to estimate the covariance matrix and signal. We fit our model of the power spectrum and bispectrum to obtain the central values, which we then used as the fiducial cosmological parameters and halo biases for performing a Fisher forecast. Unlike \cite{noriega2022fast}, we included the cross-covariance between the power spectrum and bispectrum. Our results indicate that the power spectrum is insensitive to the kernel modifications, whereas the bispectrum exhibits a $\sim$ $0.5\sigma$ shift in the bias parameters, consistent with the findings of \cite{kamalinejad2020non}. This suggests that the bispectrum is sensitive to the neutrino-modified kernels, particularly in the estimation of bias parameters.

We then performed a Fisher forecast for the power spectrum, bispectrum, and the  joint power spectrum + bispectrum and found that the bispectrum can significantly reduce the error bars on the parameters, even at for $k_{\rm max}<0.1\;h/{\rm Mpc}$ for the bispectrum. This is because the bispectrum breaks degeneracies in the power spectrum parameter space, as shown in \cite{fry_1993, yankelevich2019cosmological, Hahn_2020}. We also extended the forecast to higher values of $k_{\rm max}$ for both the power spectrum and the bispectrum. We found that the error bars on all cosmological parameters are significantly reduced when additional bispectrum modes are included. This improvement is driven by the rapid increase in the number of triangle configurations with increasing $k_{\rm max}$. Although the tree-level approximation breaks down on scales smaller than $k_{\rm max}$$\sim$$ 0.1\,h/{\rm Mpc}$, our forecast demonstrates that the bispectrum provides sufficient additional information to warrant the inclusion of loop corrections in future analyses, particularly for upcoming datasets such as DESI Y5.
\clearpage
\section*{Acknowledgments}

FK acknowledges valuable discussions with members of the Slepian Research Group. ZS acknowledges support from NASA grant 80NSSC24M0021. Additionally, the authors express their appreciation to William Ortolá Leonard, Matthew Reinhard, Jessica Chellino, Alex Krolewski, and Simon May for their extensive comments on the manuscript. FK also thanks H.J. Seo, J. Hou, and A. Greco for insightful discussions.

\bibliography{apssamp}% Produces the bibliography via BibTeX.

\begin{thebibliography}{125}%
\makeatletter
\providecommand \@ifxundefined [1]{%
 \@ifx{#1\undefined}
}%
\providecommand \@ifnum [1]{%
 \ifnum #1\expandafter \@firstoftwo
 \else \expandafter \@secondoftwo
 \fi
}%
\providecommand \@ifx [1]{%
 \ifx #1\expandafter \@firstoftwo
 \else \expandafter \@secondoftwo
 \fi
}%
\providecommand \natexlab [1]{#1}%
\providecommand \enquote  [1]{``#1''}%
\providecommand \bibnamefont  [1]{#1}%
\providecommand \bibfnamefont [1]{#1}%
\providecommand \citenamefont [1]{#1}%
\providecommand \href@noop [0]{\@secondoftwo}%
\providecommand \href [0]{\begingroup \@sanitize@url \@href}%
\providecommand \@href[1]{\@@startlink{#1}\@@href}%
\providecommand \@@href[1]{\endgroup#1\@@endlink}%
\providecommand \@sanitize@url [0]{\catcode `\\12\catcode `\$12\catcode `\&12\catcode `\#12\catcode `\^12\catcode `\_12\catcode `\%12\relax}%
\providecommand \@@startlink[1]{}%
\providecommand \@@endlink[0]{}%
\providecommand \url  [0]{\begingroup\@sanitize@url \@url }%
\providecommand \@url [1]{\endgroup\@href {#1}{\urlprefix }}%
\providecommand \urlprefix  [0]{URL }%
\providecommand \Eprint [0]{\href }%
\providecommand \doibase [0]{https://doi.org/}%
\providecommand \selectlanguage [0]{\@gobble}%
\providecommand \bibinfo  [0]{\@secondoftwo}%
\providecommand \bibfield  [0]{\@secondoftwo}%
\providecommand \translation [1]{[#1]}%
\providecommand \BibitemOpen [0]{}%
\providecommand \bibitemStop [0]{}%
\providecommand \bibitemNoStop [0]{.\EOS\space}%
\providecommand \EOS [0]{\spacefactor3000\relax}%
\providecommand \BibitemShut  [1]{\csname bibitem#1\endcsname}%
\let\auto@bib@innerbib\@empty
%</preamble>
\bibitem [{\citenamefont {Fukuda}\ \emph {et~al.}(1999)\citenamefont {Fukuda} \emph {et~al.}}]{Fukuda_1999}%
  \BibitemOpen
  \bibfield  {author} {\bibinfo {author} {\bibfnamefont {Y.}~\bibnamefont {Fukuda}} \emph {et~al.},\ }\bibfield  {title} {\bibinfo {title} {Measurement of the flux and zenith-angle distribution of upward throughgoing muons by super-kamiokande},\ }\href {https://doi.org/10.1103/physrevlett.82.2644} {\bibfield  {journal} {\bibinfo  {journal} {Physical Review Letters}\ }\textbf {\bibinfo {volume} {82}},\ \bibinfo {pages} {2644–2648} (\bibinfo {year} {1999})}\BibitemShut {NoStop}%
\bibitem [{\citenamefont {Eguchi}\ \emph {et~al.}(2003)\citenamefont {Eguchi} \emph {et~al.}}]{Eguchi_2003}%
  \BibitemOpen
  \bibfield  {author} {\bibinfo {author} {\bibfnamefont {K.}~\bibnamefont {Eguchi}} \emph {et~al.},\ }\bibfield  {title} {\bibinfo {title} {First results from kamland: Evidence for reactor antineutrino disappearance},\ }\bibfield  {journal} {\bibinfo  {journal} {Physical Review Letters}\ }\textbf {\bibinfo {volume} {90}},\ \href {https://doi.org/10.1103/physrevlett.90.021802} {10.1103/physrevlett.90.021802} (\bibinfo {year} {2003})\BibitemShut {NoStop}%
\bibitem [{\citenamefont {Ahmad}\ \emph {et~al.}(2002)\citenamefont {Ahmad} \emph {et~al.}}]{Ahmad_2002}%
  \BibitemOpen
  \bibfield  {author} {\bibinfo {author} {\bibfnamefont {Q.~R.}\ \bibnamefont {Ahmad}} \emph {et~al.},\ }\bibfield  {title} {\bibinfo {title} {Direct evidence for neutrino flavor transformation from neutral-current interactions in the sudbury neutrino observatory},\ }\bibfield  {journal} {\bibinfo  {journal} {Physical Review Letters}\ }\textbf {\bibinfo {volume} {89}},\ \href {https://doi.org/10.1103/physrevlett.89.011301} {10.1103/physrevlett.89.011301} (\bibinfo {year} {2002})\BibitemShut {NoStop}%
\bibitem [{\citenamefont {Pontecorvo}(1957)}]{Pontecorvo:1957cp}%
  \BibitemOpen
  \bibfield  {author} {\bibinfo {author} {\bibfnamefont {B.}~\bibnamefont {Pontecorvo}},\ }\bibfield  {title} {\bibinfo {title} {{Mesonium and anti-mesonium}},\ }\href@noop {} {\bibfield  {journal} {\bibinfo  {journal} {Sov. Phys. JETP}\ }\textbf {\bibinfo {volume} {6}},\ \bibinfo {pages} {429} (\bibinfo {year} {1957})}\BibitemShut {NoStop}%
\bibitem [{\citenamefont {Pontecorvo}(1958)}]{Pontecorvo:1957qd}%
  \BibitemOpen
  \bibfield  {author} {\bibinfo {author} {\bibfnamefont {B.}~\bibnamefont {Pontecorvo}},\ }\bibfield  {title} {\bibinfo {title} {Inverse $ beta $ processes and nonconservation of lepton charge},\ }\href@noop {} {\bibfield  {journal} {\bibinfo  {journal} {Zhur. Eksptl'. i Teoret. Fiz.}\ }\textbf {\bibinfo {volume} {34}} (\bibinfo {year} {1958})}\BibitemShut {NoStop}%
\bibitem [{\citenamefont {Abe}\ \emph {et~al.}(2024)\citenamefont {Abe} \emph {et~al.}}]{abe2024first}%
  \BibitemOpen
  \bibfield  {author} {\bibinfo {author} {\bibfnamefont {S.}~\bibnamefont {Abe}} \emph {et~al.},\ }\bibfield  {title} {\bibinfo {title} {First joint oscillation analysis of super-kamiokande atmospheric and t2k accelerator neutrino data},\ }\href@noop {} {\bibfield  {journal} {\bibinfo  {journal} {arXiv preprint arXiv:2405.12488}\ } (\bibinfo {year} {2024})}\BibitemShut {NoStop}%
\bibitem [{\citenamefont {Maltoni}\ \emph {et~al.}(2004)\citenamefont {Maltoni} \emph {et~al.}}]{2004}%
  \BibitemOpen
  \bibfield  {author} {\bibinfo {author} {\bibfnamefont {M.}~\bibnamefont {Maltoni}} \emph {et~al.},\ }\bibfield  {title} {\bibinfo {title} {Status of global fits to neutrino oscillations},\ }\href {https://doi.org/10.1088/1367-2630/6/1/122} {\bibfield  {journal} {\bibinfo  {journal} {New Journal of Physics}\ }\textbf {\bibinfo {volume} {6}},\ \bibinfo {pages} {122–122} (\bibinfo {year} {2004})}\BibitemShut {NoStop}%
\bibitem [{\citenamefont {Abe}\ \emph {et~al.}(2021)\citenamefont {Abe} \emph {et~al.}}]{abe2021t2k}%
  \BibitemOpen
  \bibfield  {author} {\bibinfo {author} {\bibfnamefont {K.}~\bibnamefont {Abe}} \emph {et~al.},\ }\bibfield  {title} {\bibinfo {title} {T2k measurements of muon neutrino and antineutrino disappearance using 3.13$\times$ 10 21 protons on target},\ }\href@noop {} {\bibfield  {journal} {\bibinfo  {journal} {Physical Review D}\ }\textbf {\bibinfo {volume} {103}},\ \bibinfo {pages} {L011101} (\bibinfo {year} {2021})}\BibitemShut {NoStop}%
\bibitem [{\citenamefont {Esteban}\ \emph {et~al.}(2020)\citenamefont {Esteban} \emph {et~al.}}]{esteban2020fate}%
  \BibitemOpen
  \bibfield  {author} {\bibinfo {author} {\bibfnamefont {I.}~\bibnamefont {Esteban}} \emph {et~al.},\ }\bibfield  {title} {\bibinfo {title} {The fate of hints: updated global analysis of three-flavor neutrino oscillations},\ }\href@noop {} {\bibfield  {journal} {\bibinfo  {journal} {Journal of High Energy Physics}\ }\textbf {\bibinfo {volume} {2020}},\ \bibinfo {pages} {1} (\bibinfo {year} {2020})}\BibitemShut {NoStop}%
\bibitem [{\citenamefont {Aker}\ \emph {et~al.}(2022)\citenamefont {Aker} \emph {et~al.}}]{aker2022katrin}%
  \BibitemOpen
  \bibfield  {author} {\bibinfo {author} {\bibfnamefont {M.}~\bibnamefont {Aker}} \emph {et~al.},\ }\bibfield  {title} {\bibinfo {title} {Katrin: status and prospects for the neutrino mass and beyond},\ }\href@noop {} {\bibfield  {journal} {\bibinfo  {journal} {Journal of Physics G: Nuclear and Particle Physics}\ }\textbf {\bibinfo {volume} {49}},\ \bibinfo {pages} {100501} (\bibinfo {year} {2022})}\BibitemShut {NoStop}%
\bibitem [{\citenamefont {Adame}\ \emph {et~al.}(2024{\natexlab{a}})\citenamefont {Adame} \emph {et~al.}}]{adame2024desi}%
  \BibitemOpen
  \bibfield  {author} {\bibinfo {author} {\bibfnamefont {A.}~\bibnamefont {Adame}} \emph {et~al.},\ }\bibfield  {title} {\bibinfo {title} {Desi 2024 vii: Cosmological constraints from the full-shape modeling of clustering measurements},\ }\href@noop {} {\bibfield  {journal} {\bibinfo  {journal} {arXiv preprint arXiv:2411.12022}\ } (\bibinfo {year} {2024}{\natexlab{a}})}\BibitemShut {NoStop}%
\bibitem [{\citenamefont {{Sachs}}\ and\ \citenamefont {{Wolfe}}(1967)}]{ISWeffect}%
  \BibitemOpen
  \bibfield  {author} {\bibinfo {author} {\bibfnamefont {R.~K.}\ \bibnamefont {{Sachs}}}\ and\ \bibinfo {author} {\bibfnamefont {A.~M.}\ \bibnamefont {{Wolfe}}},\ }\bibfield  {title} {\bibinfo {title} {{Perturbations of a Cosmological Model and Angular Variations of the Microwave Background}},\ }\href {https://doi.org/10.1086/148982} {\bibfield  {journal} {\bibinfo  {journal} {\apj}\ }\textbf {\bibinfo {volume} {147}},\ \bibinfo {pages} {73} (\bibinfo {year} {1967})}\BibitemShut {NoStop}%
\bibitem [{\citenamefont {Ichikawa}(2008)}]{ichikawa2008neutrino}%
  \BibitemOpen
  \bibfield  {author} {\bibinfo {author} {\bibfnamefont {K.}~\bibnamefont {Ichikawa}},\ }\bibfield  {title} {\bibinfo {title} {Neutrino mass constraint from cmb and its degeneracy with other cosmological parameters},\ }in\ \href@noop {} {\emph {\bibinfo {booktitle} {Journal of Physics: Conference Series}}},\ Vol.\ \bibinfo {volume} {120}\ (\bibinfo {organization} {IOP Publishing},\ \bibinfo {year} {2008})\ p.\ \bibinfo {pages} {022004}\BibitemShut {NoStop}%
\bibitem [{\citenamefont {{Alam}}\ \emph {et~al.}(2017)\citenamefont {{Alam}} \emph {et~al.}}]{alam_2017_boss_2ps}%
  \BibitemOpen
  \bibfield  {author} {\bibinfo {author} {\bibfnamefont {S.}~\bibnamefont {{Alam}}} \emph {et~al.},\ }\bibfield  {title} {\bibinfo {title} {{The clustering of galaxies in the completed SDSS-III Baryon Oscillation Spectroscopic Survey: cosmological analysis of the DR12 galaxy sample}},\ }\href {https://doi.org/10.1093/mnras/stx721} {\bibfield  {journal} {\bibinfo  {journal} {\mnras}\ }\textbf {\bibinfo {volume} {470}},\ \bibinfo {pages} {2617} (\bibinfo {year} {2017})},\ \Eprint {https://arxiv.org/abs/1607.03155} {arXiv:1607.03155 [astro-ph.CO]} \BibitemShut {NoStop}%
\bibitem [{\citenamefont {Alam}\ \emph {et~al.}(2017)\citenamefont {Alam} \emph {et~al.}}]{alam2017clustering}%
  \BibitemOpen
  \bibfield  {author} {\bibinfo {author} {\bibfnamefont {S.}~\bibnamefont {Alam}} \emph {et~al.},\ }\bibfield  {title} {\bibinfo {title} {The clustering of galaxies in the completed sdss-iii baryon oscillation spectroscopic survey: cosmological analysis of the dr12 galaxy sample},\ }\href@noop {} {\bibfield  {journal} {\bibinfo  {journal} {Monthly Notices of the Royal Astronomical Society}\ }\textbf {\bibinfo {volume} {470}},\ \bibinfo {pages} {2617} (\bibinfo {year} {2017})}\BibitemShut {NoStop}%
\bibitem [{\citenamefont {Alam}\ \emph {et~al.}(2021)\citenamefont {Alam} \emph {et~al.}}]{alam2021completed}%
  \BibitemOpen
  \bibfield  {author} {\bibinfo {author} {\bibfnamefont {S.}~\bibnamefont {Alam}} \emph {et~al.},\ }\bibfield  {title} {\bibinfo {title} {Completed sdss-iv extended baryon oscillation spectroscopic survey: Cosmological implications from two decades of spectroscopic surveys at the apache point observatory},\ }\href@noop {} {\bibfield  {journal} {\bibinfo  {journal} {Physical Review D}\ }\textbf {\bibinfo {volume} {103}},\ \bibinfo {pages} {083533} (\bibinfo {year} {2021})}\BibitemShut {NoStop}%
\bibitem [{\citenamefont {{DESI Collaboration}}\ \emph {et~al.}(2016)\citenamefont {{DESI Collaboration}}, \citenamefont {{Aghamousa}} \emph {et~al.}}]{DESI:2016}%
  \BibitemOpen
  \bibfield  {author} {\bibinfo {author} {\bibnamefont {{DESI Collaboration}}}, \bibinfo {author} {\bibfnamefont {A.}~\bibnamefont {{Aghamousa}}}, \emph {et~al.},\ }\bibfield  {title} {\bibinfo {title} {{The DESI Experiment Part I: Science,Targeting, and Survey Design}},\ }\href@noop {} {\bibfield  {journal} {\bibinfo  {journal} {ArXiv e-prints}\ } (\bibinfo {year} {2016})},\ \Eprint {https://arxiv.org/abs/1611.00036} {arXiv:1611.00036 [astro-ph.IM]} \BibitemShut {NoStop}%
\bibitem [{\citenamefont {Adame}\ \emph {et~al.}(2024{\natexlab{b}})\citenamefont {Adame} \emph {et~al.}}]{adame2024early}%
  \BibitemOpen
  \bibfield  {author} {\bibinfo {author} {\bibfnamefont {A.}~\bibnamefont {Adame}} \emph {et~al.},\ }\bibfield  {title} {\bibinfo {title} {The early data release of the dark energy spectroscopic instrument},\ }\href@noop {} {\bibfield  {journal} {\bibinfo  {journal} {The Astronomical Journal}\ }\textbf {\bibinfo {volume} {168}},\ \bibinfo {pages} {58} (\bibinfo {year} {2024}{\natexlab{b}})}\BibitemShut {NoStop}%
\bibitem [{\citenamefont {Collaboration}\ \emph {et~al.}(2021)\citenamefont {Collaboration} \emph {et~al.}}]{euclid2021euclid}%
  \BibitemOpen
  \bibfield  {author} {\bibinfo {author} {\bibfnamefont {E.}~\bibnamefont {Collaboration}} \emph {et~al.},\ }\bibfield  {title} {\bibinfo {title} {Euclid preparation: Ix. euclidemulator2--power spectrum emulation with massive neutrinos and self-consistent dark energy perturbations},\ }\href@noop {} {\bibfield  {journal} {\bibinfo  {journal} {Monthly Notices of the Royal Astronomical Society}\ }\textbf {\bibinfo {volume} {505}},\ \bibinfo {pages} {2840} (\bibinfo {year} {2021})}\BibitemShut {NoStop}%
\bibitem [{\citenamefont {Wang}\ \emph {et~al.}(2022)\citenamefont {Wang} \emph {et~al.}}]{wang2022high}%
  \BibitemOpen
  \bibfield  {author} {\bibinfo {author} {\bibfnamefont {Y.}~\bibnamefont {Wang}} \emph {et~al.},\ }\bibfield  {title} {\bibinfo {title} {The high latitude spectroscopic survey on the nancy grace roman space telescope},\ }\href@noop {} {\bibfield  {journal} {\bibinfo  {journal} {The Astrophysical Journal}\ }\textbf {\bibinfo {volume} {928}},\ \bibinfo {pages} {1} (\bibinfo {year} {2022})}\BibitemShut {NoStop}%
\bibitem [{\citenamefont {Schlegel}\ \emph {et~al.}(2022)\citenamefont {Schlegel} \emph {et~al.}}]{schlegel2022megamapper}%
  \BibitemOpen
  \bibfield  {author} {\bibinfo {author} {\bibfnamefont {D.~J.}\ \bibnamefont {Schlegel}} \emph {et~al.},\ }\bibfield  {title} {\bibinfo {title} {The megamapper: A stage-5 spectroscopic instrument concept for the study of inflation and dark energy},\ }\href@noop {} {\bibfield  {journal} {\bibinfo  {journal} {arXiv preprint arXiv:2209.04322}\ } (\bibinfo {year} {2022})}\BibitemShut {NoStop}%
\bibitem [{\citenamefont {Dor{\'e}}\ \emph {et~al.}(2014)\citenamefont {Dor{\'e}} \emph {et~al.}}]{dore2014cosmology}%
  \BibitemOpen
  \bibfield  {author} {\bibinfo {author} {\bibfnamefont {O.}~\bibnamefont {Dor{\'e}}} \emph {et~al.} (\bibinfo {collaboration} {SPHEREx}),\ }\bibfield  {title} {\bibinfo {title} {{Cosmology with the SPHEREX All-Sky Spectral Survey}},\ }\href@noop {} {\bibfield  {journal} {\bibinfo  {journal} {arXiv preprint arXiv:1412.4872}\ } (\bibinfo {year} {2014})},\ \Eprint {https://arxiv.org/abs/1412.4872} {arXiv:1412.4872 [astro-ph.CO]} \BibitemShut {NoStop}%
\bibitem [{\citenamefont {Kamalinejad}\ and\ \citenamefont {Slepian}(2024)}]{kamalinejad2024simpleanalytictreatmentneutrino}%
  \BibitemOpen
  \bibfield  {author} {\bibinfo {author} {\bibfnamefont {F.}~\bibnamefont {Kamalinejad}}\ and\ \bibinfo {author} {\bibfnamefont {Z.}~\bibnamefont {Slepian}},\ }\href {https://arxiv.org/abs/2203.13103} {\bibinfo {title} {A simple analytic treatment of neutrino mass impact on the full power spectrum shape via a two-fluid approximation}} (\bibinfo {year} {2024}),\ \Eprint {https://arxiv.org/abs/2203.13103} {arXiv:2203.13103 [astro-ph.CO]} \BibitemShut {NoStop}%
\bibitem [{\citenamefont {Lesgourgues}\ and\ \citenamefont {Pastor}(2006)}]{lesgourgues2006massive}%
  \BibitemOpen
  \bibfield  {author} {\bibinfo {author} {\bibfnamefont {J.}~\bibnamefont {Lesgourgues}}\ and\ \bibinfo {author} {\bibfnamefont {S.}~\bibnamefont {Pastor}},\ }\bibfield  {title} {\bibinfo {title} {Massive neutrinos and cosmology},\ }\href@noop {} {\bibfield  {journal} {\bibinfo  {journal} {Physics Reports}\ }\textbf {\bibinfo {volume} {429}},\ \bibinfo {pages} {307} (\bibinfo {year} {2006})}\BibitemShut {NoStop}%
\bibitem [{\citenamefont {Lesgourgues}\ and\ \citenamefont {Pastor}(2012)}]{lesgourgues2012neutrino}%
  \BibitemOpen
  \bibfield  {author} {\bibinfo {author} {\bibfnamefont {J.}~\bibnamefont {Lesgourgues}}\ and\ \bibinfo {author} {\bibfnamefont {S.}~\bibnamefont {Pastor}},\ }\bibfield  {title} {\bibinfo {title} {Neutrino mass from cosmology},\ }\href@noop {} {\bibfield  {journal} {\bibinfo  {journal} {Advances in High Energy Physics}\ }\textbf {\bibinfo {volume} {2012}},\ \bibinfo {pages} {608515} (\bibinfo {year} {2012})}\BibitemShut {NoStop}%
\bibitem [{\citenamefont {Fermi}(1926)}]{fermi1926sulla}%
  \BibitemOpen
  \bibfield  {author} {\bibinfo {author} {\bibfnamefont {E.}~\bibnamefont {Fermi}},\ }\bibfield  {title} {\bibinfo {title} {Sulla quantizzazione del gas perfetto monoatomico},\ }\href@noop {} {\bibfield  {journal} {\bibinfo  {journal} {Rendiconti Lincei}\ }\textbf {\bibinfo {volume} {145}} (\bibinfo {year} {1926})}\BibitemShut {NoStop}%
\bibitem [{\citenamefont {{Dirac}}(1926)}]{Dirac1926_FD_Dist}%
  \BibitemOpen
  \bibfield  {author} {\bibinfo {author} {\bibfnamefont {P.~A.~M.}\ \bibnamefont {{Dirac}}},\ }\bibfield  {title} {\bibinfo {title} {{On the Theory of Quantum Mechanics}},\ }\href {https://doi.org/10.1098/rspa.1926.0133} {\bibfield  {journal} {\bibinfo  {journal} {Proceedings of the Royal Society of London Series A}\ }\textbf {\bibinfo {volume} {112}},\ \bibinfo {pages} {661} (\bibinfo {year} {1926})}\BibitemShut {NoStop}%
\bibitem [{\citenamefont {Masatoshi}\ and\ \citenamefont {Eiichiro}(2010)}]{masatoshi2010massive}%
  \BibitemOpen
  \bibfield  {author} {\bibinfo {author} {\bibfnamefont {S.}~\bibnamefont {Masatoshi}}\ and\ \bibinfo {author} {\bibfnamefont {K.}~\bibnamefont {Eiichiro}},\ }\bibfield  {title} {\bibinfo {title} {Massive neutrinos in cosmology: Analytic solutions and fluid approximation},\ }\href@noop {} {\bibfield  {journal} {\bibinfo  {journal} {Phys. Rev. D}\ }\textbf {\bibinfo {volume} {81}},\ \bibinfo {pages} {123516} (\bibinfo {year} {2010})}\BibitemShut {NoStop}%
\bibitem [{\citenamefont {Aghanim}\ \emph {et~al.}(2020)\citenamefont {Aghanim} \emph {et~al.}}]{aghanim2020planck}%
  \BibitemOpen
  \bibfield  {author} {\bibinfo {author} {\bibfnamefont {N.}~\bibnamefont {Aghanim}} \emph {et~al.},\ }\bibfield  {title} {\bibinfo {title} {Planck 2018 results-vi. cosmological parameters},\ }\href@noop {} {\bibfield  {journal} {\bibinfo  {journal} {Astronomy \& Astrophysics}\ }\textbf {\bibinfo {volume} {641}},\ \bibinfo {pages} {A6} (\bibinfo {year} {2020})}\BibitemShut {NoStop}%
\bibitem [{\citenamefont {Bernardeau}\ \emph {et~al.}(2002)\citenamefont {Bernardeau} \emph {et~al.}}]{Bernardeau_2002}%
  \BibitemOpen
  \bibfield  {author} {\bibinfo {author} {\bibfnamefont {F.}~\bibnamefont {Bernardeau}} \emph {et~al.},\ }\bibfield  {title} {\bibinfo {title} {Large-scale structure of the universe and cosmological perturbation theory},\ }\href {https://doi.org/10.1016/s0370-1573(02)00135-7} {\bibfield  {journal} {\bibinfo  {journal} {Physics Reports}\ }\textbf {\bibinfo {volume} {367}},\ \bibinfo {pages} {1–248} (\bibinfo {year} {2002})}\BibitemShut {NoStop}%
\bibitem [{\citenamefont {Goroff}\ \emph {et~al.}(1986)\citenamefont {Goroff} \emph {et~al.}}]{goroff1986coupling}%
  \BibitemOpen
  \bibfield  {author} {\bibinfo {author} {\bibfnamefont {M.}~\bibnamefont {Goroff}} \emph {et~al.},\ }\bibfield  {title} {\bibinfo {title} {Coupling of modes of cosmological mass density fluctuations},\ }\href@noop {} {\bibfield  {journal} {\bibinfo  {journal} {Astrophysical Journal, Part 1 (ISSN 0004-637X), vol. 311, Dec. 1, 1986, p. 6-14.}\ }\textbf {\bibinfo {volume} {311}},\ \bibinfo {pages} {6} (\bibinfo {year} {1986})}\BibitemShut {NoStop}%
\bibitem [{\citenamefont {{Koyama}}\ \emph {et~al.}(2009)\citenamefont {{Koyama}} \emph {et~al.}}]{Koyama_kernels}%
  \BibitemOpen
  \bibfield  {author} {\bibinfo {author} {\bibfnamefont {K.}~\bibnamefont {{Koyama}}} \emph {et~al.},\ }\bibfield  {title} {\bibinfo {title} {{Nonlinear evolution of the matter power spectrum in modified theories of gravity}},\ }\href {https://doi.org/10.1103/PhysRevD.79.123512} {\bibfield  {journal} {\bibinfo  {journal} {\prd}\ }\textbf {\bibinfo {volume} {79}},\ \bibinfo {eid} {123512} (\bibinfo {year} {2009})},\ \Eprint {https://arxiv.org/abs/0902.0618} {arXiv:0902.0618 [astro-ph.CO]} \BibitemShut {NoStop}%
\bibitem [{\citenamefont {T{\"u}des}\ and\ \citenamefont {Amendola}(2024)}]{tudes2024non}%
  \BibitemOpen
  \bibfield  {author} {\bibinfo {author} {\bibfnamefont {B.}~\bibnamefont {T{\"u}des}}\ and\ \bibinfo {author} {\bibfnamefont {L.}~\bibnamefont {Amendola}},\ }\bibfield  {title} {\bibinfo {title} {Non-linear cosmological perturbations for coupled dark energy},\ }\href@noop {} {\bibfield  {journal} {\bibinfo  {journal} {arXiv preprint arXiv:2411.06014}\ } (\bibinfo {year} {2024})}\BibitemShut {NoStop}%
\bibitem [{\citenamefont {Bose}\ \emph {et~al.}(2018)\citenamefont {Bose}, \citenamefont {Baldi},\ and\ \citenamefont {Pourtsidou}}]{bose2018modelling}%
  \BibitemOpen
  \bibfield  {author} {\bibinfo {author} {\bibfnamefont {B.}~\bibnamefont {Bose}}, \bibinfo {author} {\bibfnamefont {M.}~\bibnamefont {Baldi}},\ and\ \bibinfo {author} {\bibfnamefont {A.}~\bibnamefont {Pourtsidou}},\ }\bibfield  {title} {\bibinfo {title} {Modelling non-linear effects of dark energy},\ }\href@noop {} {\bibfield  {journal} {\bibinfo  {journal} {Journal of Cosmology and Astroparticle Physics}\ }\textbf {\bibinfo {volume} {2018}}\bibinfo  {number} { (04)},\ \bibinfo {pages} {032}}\BibitemShut {NoStop}%
\bibitem [{\citenamefont {Upadhye}\ \emph {et~al.}(2014)\citenamefont {Upadhye} \emph {et~al.}}]{upadhye2014large}%
  \BibitemOpen
\bibfield  {number} {  }\bibfield  {author} {\bibinfo {author} {\bibfnamefont {A.}~\bibnamefont {Upadhye}} \emph {et~al.},\ }\bibfield  {title} {\bibinfo {title} {Large-scale structure formation with massive neutrinos and dynamical dark energy},\ }\href@noop {} {\bibfield  {journal} {\bibinfo  {journal} {Physical Review D}\ }\textbf {\bibinfo {volume} {89}},\ \bibinfo {pages} {103515} (\bibinfo {year} {2014})}\BibitemShut {NoStop}%
\bibitem [{\citenamefont {Wong}(2008)}]{wong2008higher}%
  \BibitemOpen
  \bibfield  {author} {\bibinfo {author} {\bibfnamefont {Y.~Y.}\ \bibnamefont {Wong}},\ }\bibfield  {title} {\bibinfo {title} {Higher order corrections to the large scale matter power spectrum in the presence ofmassive neutrinos},\ }\href@noop {} {\bibfield  {journal} {\bibinfo  {journal} {Journal of Cosmology and Astroparticle Physics}\ }\textbf {\bibinfo {volume} {2008}}\bibinfo  {number} { (10)},\ \bibinfo {pages} {035}}\BibitemShut {NoStop}%
\bibitem [{\citenamefont {Garny}\ and\ \citenamefont {Taule}(2021)}]{garny2021loop}%
  \BibitemOpen
\bibfield  {number} {  }\bibfield  {author} {\bibinfo {author} {\bibfnamefont {M.}~\bibnamefont {Garny}}\ and\ \bibinfo {author} {\bibfnamefont {P.}~\bibnamefont {Taule}},\ }\bibfield  {title} {\bibinfo {title} {Loop corrections to the power spectrum for massive neutrino cosmologies with full time-and scale-dependence},\ }\href@noop {} {\bibfield  {journal} {\bibinfo  {journal} {Journal of Cosmology and Astroparticle Physics}\ }\textbf {\bibinfo {volume} {2021}}\bibinfo  {number} { (01)},\ \bibinfo {pages} {020}}\BibitemShut {NoStop}%
\bibitem [{\citenamefont {Saito}\ \emph {et~al.}(2009)\citenamefont {Saito}, \citenamefont {Takada},\ and\ \citenamefont {Taruya}}]{saito2009nonlinear}%
  \BibitemOpen
\bibfield  {number} {  }\bibfield  {author} {\bibinfo {author} {\bibfnamefont {S.}~\bibnamefont {Saito}}, \bibinfo {author} {\bibfnamefont {M.}~\bibnamefont {Takada}},\ and\ \bibinfo {author} {\bibfnamefont {A.}~\bibnamefont {Taruya}},\ }\bibfield  {title} {\bibinfo {title} {Nonlinear power spectrum in the presence of massive neutrinos: Perturbation theory approach, galaxy bias, and parameter forecasts},\ }\href@noop {} {\bibfield  {journal} {\bibinfo  {journal} {Physical Review D—Particles, Fields, Gravitation, and Cosmology}\ }\textbf {\bibinfo {volume} {80}},\ \bibinfo {pages} {083528} (\bibinfo {year} {2009})}\BibitemShut {NoStop}%
\bibitem [{\citenamefont {Blas}\ \emph {et~al.}(2014)\citenamefont {Blas} \emph {et~al.}}]{Blas_2014}%
  \BibitemOpen
  \bibfield  {author} {\bibinfo {author} {\bibfnamefont {D.}~\bibnamefont {Blas}} \emph {et~al.},\ }\bibfield  {title} {\bibinfo {title} {Structure formation with massive neutrinos: going beyond linear theory},\ }\href {https://doi.org/10.1088/1475-7516/2014/11/039} {\bibfield  {journal} {\bibinfo  {journal} {Journal of Cosmology and Astroparticle Physics}\ }\textbf {\bibinfo {volume} {2014}}\bibinfo  {number} { (11)},\ \bibinfo {pages} {039–039}}\BibitemShut {NoStop}%
\bibitem [{\citenamefont {Saito}\ \emph {et~al.}(2008)\citenamefont {Saito} \emph {et~al.}}]{saito2008impact}%
  \BibitemOpen
\bibfield  {number} {  }\bibfield  {author} {\bibinfo {author} {\bibfnamefont {S.}~\bibnamefont {Saito}} \emph {et~al.},\ }\bibfield  {title} {\bibinfo {title} {Impact of massive neutrinos on the nonlinear matter power spectrum},\ }\href@noop {} {\bibfield  {journal} {\bibinfo  {journal} {Physical Review Letters}\ }\textbf {\bibinfo {volume} {100}},\ \bibinfo {pages} {191301} (\bibinfo {year} {2008})}\BibitemShut {NoStop}%
\bibitem [{\citenamefont {Kamalinejad}\ and\ \citenamefont {Slepian}(2020)}]{kamalinejad2020non}%
  \BibitemOpen
  \bibfield  {author} {\bibinfo {author} {\bibfnamefont {F.}~\bibnamefont {Kamalinejad}}\ and\ \bibinfo {author} {\bibfnamefont {Z.}~\bibnamefont {Slepian}},\ }\bibfield  {title} {\bibinfo {title} {{Neutrino Mass Signatures in the Galaxy Bispectrum}},\ }\href@noop {} {\bibfield  {journal} {\bibinfo  {journal} {arXiv preprint}\ } (\bibinfo {year} {2020})},\ \Eprint {https://arxiv.org/abs/2011.00899} {arXiv:2011.00899 [astro-ph.CO]} \BibitemShut {NoStop}%
\bibitem [{\citenamefont {Aviles}\ and\ \citenamefont {Banerjee}(2020)}]{aviles2020lagrangian}%
  \BibitemOpen
  \bibfield  {author} {\bibinfo {author} {\bibfnamefont {A.}~\bibnamefont {Aviles}}\ and\ \bibinfo {author} {\bibfnamefont {A.}~\bibnamefont {Banerjee}},\ }\bibfield  {title} {\bibinfo {title} {A lagrangian perturbation theory in the presence of massive neutrinos},\ }\href@noop {} {\bibfield  {journal} {\bibinfo  {journal} {Journal of Cosmology and Astroparticle Physics}\ }\textbf {\bibinfo {volume} {2020}}\bibinfo  {number} { (10)},\ \bibinfo {pages} {034}}\BibitemShut {NoStop}%
\bibitem [{\citenamefont {Aviles}\ \emph {et~al.}(2021)\citenamefont {Aviles} \emph {et~al.}}]{aviles2021clustering}%
  \BibitemOpen
\bibfield  {number} {  }\bibfield  {author} {\bibinfo {author} {\bibfnamefont {A.}~\bibnamefont {Aviles}} \emph {et~al.},\ }\bibfield  {title} {\bibinfo {title} {Clustering in massive neutrino cosmologies via eulerian perturbation theory},\ }\href@noop {} {\bibfield  {journal} {\bibinfo  {journal} {Journal of Cosmology and Astroparticle Physics}\ }\textbf {\bibinfo {volume} {2021}}\bibinfo  {number} { (11)},\ \bibinfo {pages} {028}}\BibitemShut {NoStop}%
\bibitem [{\citenamefont {McEwen}\ \emph {et~al.}(2016)\citenamefont {McEwen} \emph {et~al.}}]{mcewen2016fast}%
  \BibitemOpen
\bibfield  {number} {  }\bibfield  {author} {\bibinfo {author} {\bibfnamefont {J.~E.}\ \bibnamefont {McEwen}} \emph {et~al.},\ }\bibfield  {title} {\bibinfo {title} {Fast-pt: a novel algorithm to calculate convolution integrals in cosmological perturbation theory},\ }\href@noop {} {\bibfield  {journal} {\bibinfo  {journal} {Journal of Cosmology and Astroparticle Physics}\ }\textbf {\bibinfo {volume} {2016}}\bibinfo  {number} { (09)},\ \bibinfo {pages} {015}}\BibitemShut {NoStop}%
\bibitem [{\citenamefont {Simonovi{\'c}}\ and\ \citenamefont {Bothers}(2018)}]{simonovic2018cosmological}%
  \BibitemOpen
\bibfield  {number} {  }\bibfield  {author} {\bibinfo {author} {\bibfnamefont {M.}~\bibnamefont {Simonovi{\'c}}}\ and\ \bibinfo {author} {\bibnamefont {Bothers}},\ }\bibfield  {title} {\bibinfo {title} {Cosmological perturbation theory using the fftlog: formalism and connection to qft loop integrals},\ }\href@noop {} {\bibfield  {journal} {\bibinfo  {journal} {Journal of Cosmology and Astroparticle Physics}\ }\textbf {\bibinfo {volume} {2018}}\bibinfo  {number} { (04)},\ \bibinfo {pages} {030}}\BibitemShut {NoStop}%
\bibitem [{\citenamefont {Schmittfull}\ \emph {et~al.}(2016)\citenamefont {Schmittfull} \emph {et~al.}}]{schmittfull2016fast}%
  \BibitemOpen
\bibfield  {number} {  }\bibfield  {author} {\bibinfo {author} {\bibfnamefont {M.}~\bibnamefont {Schmittfull}} \emph {et~al.},\ }\bibfield  {title} {\bibinfo {title} {Fast large scale structure perturbation theory using one-dimensional fast fourier transforms},\ }\href@noop {} {\bibfield  {journal} {\bibinfo  {journal} {Physical Review D}\ }\textbf {\bibinfo {volume} {93}},\ \bibinfo {pages} {103528} (\bibinfo {year} {2016})}\BibitemShut {NoStop}%
\bibitem [{\citenamefont {Villaescusa-Navarro}\ \emph {et~al.}(2020)\citenamefont {Villaescusa-Navarro} \emph {et~al.}}]{villaescusa2020quijote}%
  \BibitemOpen
  \bibfield  {author} {\bibinfo {author} {\bibfnamefont {F.}~\bibnamefont {Villaescusa-Navarro}} \emph {et~al.},\ }\bibfield  {title} {\bibinfo {title} {The quijote simulations},\ }\href@noop {} {\bibfield  {journal} {\bibinfo  {journal} {The Astrophysical Journal Supplement Series}\ }\textbf {\bibinfo {volume} {250}},\ \bibinfo {pages} {2} (\bibinfo {year} {2020})}\BibitemShut {NoStop}%
\bibitem [{\citenamefont {{Chudaykin}}\ \emph {et~al.}(2020)\citenamefont {{Chudaykin}} \emph {et~al.}}]{class_pt}%
  \BibitemOpen
  \bibfield  {author} {\bibinfo {author} {\bibfnamefont {A.}~\bibnamefont {{Chudaykin}}} \emph {et~al.},\ }\bibfield  {title} {\bibinfo {title} {{Nonlinear perturbation theory extension of the Boltzmann code CLASS}},\ }\href {https://doi.org/10.1103/PhysRevD.102.063533} {\bibfield  {journal} {\bibinfo  {journal} {\prd}\ }\textbf {\bibinfo {volume} {102}},\ \bibinfo {eid} {063533} (\bibinfo {year} {2020})},\ \Eprint {https://arxiv.org/abs/2004.10607} {arXiv:2004.10607 [astro-ph.CO]} \BibitemShut {NoStop}%
\bibitem [{\citenamefont {Linde}\ \emph {et~al.}(2024)\citenamefont {Linde} \emph {et~al.}}]{linde2024class}%
  \BibitemOpen
  \bibfield  {author} {\bibinfo {author} {\bibfnamefont {D.}~\bibnamefont {Linde}} \emph {et~al.},\ }\bibfield  {title} {\bibinfo {title} {Class-oneloop: accurate and unbiased inference from spectroscopic galaxy surveys},\ }\href@noop {} {\bibfield  {journal} {\bibinfo  {journal} {Journal of Cosmology and Astroparticle Physics}\ }\textbf {\bibinfo {volume} {2024}}\bibinfo  {number} { (07)},\ \bibinfo {pages} {068}}\BibitemShut {NoStop}%
\bibitem [{\citenamefont {Scoccimarro}(1997)}]{scoccimarro1997cosmological}%
  \BibitemOpen
\bibfield  {number} {  }\bibfield  {author} {\bibinfo {author} {\bibfnamefont {R.}~\bibnamefont {Scoccimarro}},\ }\bibfield  {title} {\bibinfo {title} {Cosmological perturbations: Entering the nonlinear regime},\ }\href@noop {} {\bibfield  {journal} {\bibinfo  {journal} {The Astrophysical Journal}\ }\textbf {\bibinfo {volume} {487}},\ \bibinfo {pages} {1} (\bibinfo {year} {1997})}\BibitemShut {NoStop}%
\bibitem [{\citenamefont {Pajer}\ and\ \citenamefont {Zaldarriaga}(2013)}]{pajer2013renormalization}%
  \BibitemOpen
  \bibfield  {author} {\bibinfo {author} {\bibfnamefont {E.}~\bibnamefont {Pajer}}\ and\ \bibinfo {author} {\bibfnamefont {M.}~\bibnamefont {Zaldarriaga}},\ }\bibfield  {title} {\bibinfo {title} {On the renormalization of the effective field theory of large scale structures},\ }\href@noop {} {\bibfield  {journal} {\bibinfo  {journal} {Journal of Cosmology and Astroparticle Physics}\ }\textbf {\bibinfo {volume} {2013}}\bibinfo  {number} { (08)},\ \bibinfo {pages} {037}}\BibitemShut {NoStop}%
\bibitem [{\citenamefont {Senatore}\ and\ \citenamefont {Zaldarriaga}(2017)}]{senatore2017effective}%
  \BibitemOpen
\bibfield  {number} {  }\bibfield  {author} {\bibinfo {author} {\bibfnamefont {L.}~\bibnamefont {Senatore}}\ and\ \bibinfo {author} {\bibfnamefont {M.}~\bibnamefont {Zaldarriaga}},\ }\bibfield  {title} {\bibinfo {title} {The effective field theory of large-scale structure in the presence of massive neutrinos},\ }\href@noop {} {\bibfield  {journal} {\bibinfo  {journal} {arXiv preprint arXiv:1707.04698}\ } (\bibinfo {year} {2017})}\BibitemShut {NoStop}%
\bibitem [{\citenamefont {{Noriega}}\ \emph {et~al.}(2022)\citenamefont {{Noriega}} \emph {et~al.}}]{Avilesfolpnu}%
  \BibitemOpen
  \bibfield  {author} {\bibinfo {author} {\bibfnamefont {H.~E.}\ \bibnamefont {{Noriega}}} \emph {et~al.},\ }\bibfield  {title} {\bibinfo {title} {{Fast computation of non-linear power spectrum in cosmologies with massive neutrinos}},\ }\href {https://doi.org/10.1088/1475-7516/2022/11/038} {\bibfield  {journal} {\bibinfo  {journal} {\jcap}\ }\textbf {\bibinfo {volume} {2022}},\ \bibinfo {eid} {038} (\bibinfo {year} {2022})},\ \Eprint {https://arxiv.org/abs/2208.02791} {arXiv:2208.02791 [astro-ph.CO]} \BibitemShut {NoStop}%
\bibitem [{\citenamefont {Hu}\ \emph {et~al.}(1998)\citenamefont {Hu}, \citenamefont {Eisenstein},\ and\ \citenamefont {Tegmark}}]{hu1998weighing}%
  \BibitemOpen
  \bibfield  {author} {\bibinfo {author} {\bibfnamefont {W.}~\bibnamefont {Hu}}, \bibinfo {author} {\bibfnamefont {D.~J.}\ \bibnamefont {Eisenstein}},\ and\ \bibinfo {author} {\bibfnamefont {M.}~\bibnamefont {Tegmark}},\ }\bibfield  {title} {\bibinfo {title} {Weighing neutrinos with galaxy surveys},\ }\href@noop {} {\bibfield  {journal} {\bibinfo  {journal} {Physical Review Letters}\ }\textbf {\bibinfo {volume} {80}},\ \bibinfo {pages} {5255} (\bibinfo {year} {1998})}\BibitemShut {NoStop}%
\bibitem [{\citenamefont {Blas}\ \emph {et~al.}(2011)\citenamefont {Blas} \emph {et~al.}}]{Diego_Blas_2011}%
  \BibitemOpen
  \bibfield  {author} {\bibinfo {author} {\bibfnamefont {D.}~\bibnamefont {Blas}} \emph {et~al.},\ }\bibfield  {title} {\bibinfo {title} {The cosmic linear anisotropy solving system (class). part ii: Approximation schemes},\ }\href {https://doi.org/10.1088/1475-7516/2011/07/034} {\bibfield  {journal} {\bibinfo  {journal} {Journal of Cosmology and Astroparticle Physics}\ }\textbf {\bibinfo {volume} {2011}}\bibinfo  {number} { (07)},\ \bibinfo {pages} {034–034}}\BibitemShut {NoStop}%
\bibitem [{\citenamefont {Lesgourgues}(2011)}]{lesgourgues2011cosmic}%
  \BibitemOpen
\bibfield  {number} {  }\bibfield  {author} {\bibinfo {author} {\bibfnamefont {J.}~\bibnamefont {Lesgourgues}},\ }\bibfield  {title} {\bibinfo {title} {The cosmic linear anisotropy solving system (class) i: Overview},\ }\href@noop {} {\bibfield  {journal} {\bibinfo  {journal} {arXiv preprint arXiv:1104.2932}\ } (\bibinfo {year} {2011})}\BibitemShut {NoStop}%
\bibitem [{\citenamefont {Blas}\ \emph {et~al.}(2016)\citenamefont {Blas} \emph {et~al.}}]{blas2016time}%
  \BibitemOpen
  \bibfield  {author} {\bibinfo {author} {\bibfnamefont {D.}~\bibnamefont {Blas}} \emph {et~al.},\ }\bibfield  {title} {\bibinfo {title} {Time-sliced perturbation theory ii: baryon acoustic oscillations and infrared resummation},\ }\href@noop {} {\bibfield  {journal} {\bibinfo  {journal} {Journal of Cosmology and Astroparticle Physics}\ }\textbf {\bibinfo {volume} {2016}}\bibinfo  {number} { (07)},\ \bibinfo {pages} {028}}\BibitemShut {NoStop}%
\bibitem [{\citenamefont {Senatore}\ and\ \citenamefont {Trevisan}(2018)}]{senatore2018ir}%
  \BibitemOpen
\bibfield  {number} {  }\bibfield  {author} {\bibinfo {author} {\bibfnamefont {L.}~\bibnamefont {Senatore}}\ and\ \bibinfo {author} {\bibfnamefont {G.}~\bibnamefont {Trevisan}},\ }\bibfield  {title} {\bibinfo {title} {On the ir-resummation in the eftoflss},\ }\href@noop {} {\bibfield  {journal} {\bibinfo  {journal} {Journal of Cosmology and Astroparticle Physics}\ }\textbf {\bibinfo {volume} {2018}}\bibinfo  {number} { (05)},\ \bibinfo {pages} {019}}\BibitemShut {NoStop}%
\bibitem [{\citenamefont {Ivanov}\ and\ \citenamefont {Sibiryakov}(2018)}]{ivanov2018infrared}%
  \BibitemOpen
\bibfield  {number} {  }\bibfield  {author} {\bibinfo {author} {\bibfnamefont {M.~M.}\ \bibnamefont {Ivanov}}\ and\ \bibinfo {author} {\bibfnamefont {S.}~\bibnamefont {Sibiryakov}},\ }\bibfield  {title} {\bibinfo {title} {Infrared resummation for biased tracers in redshift space},\ }\href@noop {} {\bibfield  {journal} {\bibinfo  {journal} {Journal of Cosmology and Astroparticle Physics}\ }\textbf {\bibinfo {volume} {2018}}\bibinfo  {number} { (07)},\ \bibinfo {pages} {053}}\BibitemShut {NoStop}%
\bibitem [{\citenamefont {Lewandowski}\ and\ \citenamefont {Senatore}(2020)}]{lewandowski2020analytic}%
  \BibitemOpen
\bibfield  {number} {  }\bibfield  {author} {\bibinfo {author} {\bibfnamefont {M.}~\bibnamefont {Lewandowski}}\ and\ \bibinfo {author} {\bibfnamefont {L.}~\bibnamefont {Senatore}},\ }\bibfield  {title} {\bibinfo {title} {An analytic implementation of the ir-resummation for the bao peak},\ }\href@noop {} {\bibfield  {journal} {\bibinfo  {journal} {Journal of Cosmology and Astroparticle Physics}\ }\textbf {\bibinfo {volume} {2020}}\bibinfo  {number} { (03)},\ \bibinfo {pages} {018}}\BibitemShut {NoStop}%
\bibitem [{\citenamefont {Levi}\ and\ \citenamefont {Vlah}(2016)}]{levi2016massive}%
  \BibitemOpen
\bibfield  {number} {  }\bibfield  {author} {\bibinfo {author} {\bibfnamefont {M.}~\bibnamefont {Levi}}\ and\ \bibinfo {author} {\bibfnamefont {Z.}~\bibnamefont {Vlah}},\ }\bibfield  {title} {\bibinfo {title} {Massive neutrinos in nonlinear large scale structure: A consistent perturbation theory},\ }\href@noop {} {\bibfield  {journal} {\bibinfo  {journal} {arXiv preprint arXiv:1605.09417}\ } (\bibinfo {year} {2016})}\BibitemShut {NoStop}%
\bibitem [{\citenamefont {Ruggeri}\ \emph {et~al.}(2018)\citenamefont {Ruggeri} \emph {et~al.}}]{ruggeri2018demnuni}%
  \BibitemOpen
  \bibfield  {author} {\bibinfo {author} {\bibfnamefont {R.}~\bibnamefont {Ruggeri}} \emph {et~al.},\ }\bibfield  {title} {\bibinfo {title} {Demnuni: Massive neutrinos and the bispectrum of large scale structures},\ }\href@noop {} {\bibfield  {journal} {\bibinfo  {journal} {Journal of Cosmology and Astroparticle Physics}\ }\textbf {\bibinfo {volume} {2018}}\bibinfo  {number} { (03)},\ \bibinfo {pages} {003}}\BibitemShut {NoStop}%
\bibitem [{\citenamefont {Hannestad}\ \emph {et~al.}(2020)\citenamefont {Hannestad}, \citenamefont {Upadhye},\ and\ \citenamefont {Wong}}]{hannestad2020spoon}%
  \BibitemOpen
\bibfield  {number} {  }\bibfield  {author} {\bibinfo {author} {\bibfnamefont {S.}~\bibnamefont {Hannestad}}, \bibinfo {author} {\bibfnamefont {A.}~\bibnamefont {Upadhye}},\ and\ \bibinfo {author} {\bibfnamefont {Y.~Y.}\ \bibnamefont {Wong}},\ }\bibfield  {title} {\bibinfo {title} {Spoon or slide? the non-linear matter power spectrum in the presence of massive neutrinos},\ }\href@noop {} {\bibfield  {journal} {\bibinfo  {journal} {Journal of Cosmology and Astroparticle Physics}\ }\textbf {\bibinfo {volume} {2020}}\bibinfo  {number} { (11)},\ \bibinfo {pages} {062}}\BibitemShut {NoStop}%
\bibitem [{\citenamefont {{Biagetti}}\ \emph {et~al.}(2022)\citenamefont {{Biagetti}} \emph {et~al.}}]{biagetti2022covariance}%
  \BibitemOpen
\bibfield  {number} {  }\bibfield  {author} {\bibinfo {author} {\bibfnamefont {M.}~\bibnamefont {{Biagetti}}} \emph {et~al.},\ }\bibfield  {title} {\bibinfo {title} {{The covariance of squeezed bispectrum configurations}},\ }\href {https://doi.org/10.1088/1475-7516/2022/09/009} {\bibfield  {journal} {\bibinfo  {journal} {\jcap}\ }\textbf {\bibinfo {volume} {2022}},\ \bibinfo {eid} {009} (\bibinfo {year} {2022})},\ \Eprint {https://arxiv.org/abs/2111.05887} {arXiv:2111.05887 [astro-ph.CO]} \BibitemShut {NoStop}%
\bibitem [{\citenamefont {{Mohammed}}\ \emph {et~al.}(2017{\natexlab{a}})\citenamefont {{Mohammed}}, \citenamefont {{Seljak}},\ and\ \citenamefont {{Vlah}}}]{mohamed}%
  \BibitemOpen
  \bibfield  {author} {\bibinfo {author} {\bibfnamefont {I.}~\bibnamefont {{Mohammed}}}, \bibinfo {author} {\bibfnamefont {U.}~\bibnamefont {{Seljak}}},\ and\ \bibinfo {author} {\bibfnamefont {Z.}~\bibnamefont {{Vlah}}},\ }\bibfield  {title} {\bibinfo {title} {{Perturbative approach to covariance matrix of the matter power spectrum}},\ }\href {https://doi.org/10.1093/mnras/stw3196} {\bibfield  {journal} {\bibinfo  {journal} {\mnras}\ }\textbf {\bibinfo {volume} {466}},\ \bibinfo {pages} {780} (\bibinfo {year} {2017}{\natexlab{a}})},\ \Eprint {https://arxiv.org/abs/1607.00043} {arXiv:1607.00043 [astro-ph.CO]} \BibitemShut {NoStop}%
\bibitem [{\citenamefont {Novell-Masot}\ \emph {et~al.}(2024)\citenamefont {Novell-Masot} \emph {et~al.}}]{novell2024approximations}%
  \BibitemOpen
  \bibfield  {author} {\bibinfo {author} {\bibfnamefont {S.}~\bibnamefont {Novell-Masot}} \emph {et~al.},\ }\bibfield  {title} {\bibinfo {title} {On approximations of the redshift-space bispectrum and power spectrum multipoles covariance matrix},\ }\href@noop {} {\bibfield  {journal} {\bibinfo  {journal} {Journal of Cosmology and Astroparticle Physics}\ }\textbf {\bibinfo {volume} {2024}}\bibinfo  {number} { (06)},\ \bibinfo {pages} {048}}\BibitemShut {NoStop}%
\bibitem [{\citenamefont {Hamilton}\ \emph {et~al.}(2006)\citenamefont {Hamilton} \emph {et~al.}}]{hamilton2006measuring}%
  \BibitemOpen
\bibfield  {number} {  }\bibfield  {author} {\bibinfo {author} {\bibfnamefont {A.~J.}\ \bibnamefont {Hamilton}} \emph {et~al.},\ }\bibfield  {title} {\bibinfo {title} {On measuring the covariance matrix of the non-linear power spectrum from simulations},\ }\href@noop {} {\bibfield  {journal} {\bibinfo  {journal} {Monthly Notices of the Royal Astronomical Society}\ }\textbf {\bibinfo {volume} {371}},\ \bibinfo {pages} {1188} (\bibinfo {year} {2006})}\BibitemShut {NoStop}%
\bibitem [{\citenamefont {Wadekar}\ and\ \citenamefont {Scoccimarro}(2020)}]{wadekar2020galaxy}%
  \BibitemOpen
  \bibfield  {author} {\bibinfo {author} {\bibfnamefont {D.}~\bibnamefont {Wadekar}}\ and\ \bibinfo {author} {\bibfnamefont {R.}~\bibnamefont {Scoccimarro}},\ }\bibfield  {title} {\bibinfo {title} {Galaxy power spectrum multipoles covariance in perturbation theory},\ }\href@noop {} {\bibfield  {journal} {\bibinfo  {journal} {Physical Review D}\ }\textbf {\bibinfo {volume} {102}},\ \bibinfo {pages} {123517} (\bibinfo {year} {2020})}\BibitemShut {NoStop}%
\bibitem [{\citenamefont {{Hartlap}}\ \emph {et~al.}(2007)\citenamefont {{Hartlap}}, \citenamefont {{Simon}},\ and\ \citenamefont {{Schneider}}}]{hartlap}%
  \BibitemOpen
  \bibfield  {author} {\bibinfo {author} {\bibfnamefont {J.}~\bibnamefont {{Hartlap}}}, \bibinfo {author} {\bibfnamefont {P.}~\bibnamefont {{Simon}}},\ and\ \bibinfo {author} {\bibfnamefont {P.}~\bibnamefont {{Schneider}}},\ }\bibfield  {title} {\bibinfo {title} {{Why your model parameter confidences might be too optimistic. Unbiased estimation of the inverse covariance matrix}},\ }\href {https://doi.org/10.1051/0004-6361:20066170} {\bibfield  {journal} {\bibinfo  {journal} {\aap}\ }\textbf {\bibinfo {volume} {464}},\ \bibinfo {pages} {399} (\bibinfo {year} {2007})},\ \Eprint {https://arxiv.org/abs/astro-ph/0608064} {arXiv:astro-ph/0608064 [astro-ph]} \BibitemShut {NoStop}%
\bibitem [{\citenamefont {Grove}\ \emph {et~al.}(2022)\citenamefont {Grove} \emph {et~al.}}]{grove2022desi}%
  \BibitemOpen
  \bibfield  {author} {\bibinfo {author} {\bibfnamefont {C.}~\bibnamefont {Grove}} \emph {et~al.},\ }\bibfield  {title} {\bibinfo {title} {The desi n-body simulation project--i. testing the robustness of simulations for the desi dark time survey},\ }\href@noop {} {\bibfield  {journal} {\bibinfo  {journal} {Monthly Notices of the Royal Astronomical Society}\ }\textbf {\bibinfo {volume} {515}},\ \bibinfo {pages} {1854} (\bibinfo {year} {2022})}\BibitemShut {NoStop}%
\bibitem [{\citenamefont {{Desjacques}}\ \emph {et~al.}(2018)\citenamefont {{Desjacques}}, \citenamefont {{Jeong}},\ and\ \citenamefont {{Schmidt}}}]{desjacques}%
  \BibitemOpen
  \bibfield  {author} {\bibinfo {author} {\bibfnamefont {V.}~\bibnamefont {{Desjacques}}}, \bibinfo {author} {\bibfnamefont {D.}~\bibnamefont {{Jeong}}},\ and\ \bibinfo {author} {\bibfnamefont {F.}~\bibnamefont {{Schmidt}}},\ }\bibfield  {title} {\bibinfo {title} {{Large-scale galaxy bias}},\ }\href {https://doi.org/10.1016/j.physrep.2017.12.002} {\bibfield  {journal} {\bibinfo  {journal} {\physrep}\ }\textbf {\bibinfo {volume} {733}},\ \bibinfo {pages} {1} (\bibinfo {year} {2018})},\ \Eprint {https://arxiv.org/abs/1611.09787} {arXiv:1611.09787 [astro-ph.CO]} \BibitemShut {NoStop}%
\bibitem [{\citenamefont {McDonald}\ and\ \citenamefont {Roy}(2009{\natexlab{a}})}]{mcdonald2009clustering}%
  \BibitemOpen
  \bibfield  {author} {\bibinfo {author} {\bibfnamefont {P.}~\bibnamefont {McDonald}}\ and\ \bibinfo {author} {\bibfnamefont {A.}~\bibnamefont {Roy}},\ }\bibfield  {title} {\bibinfo {title} {Clustering of dark matter tracers: generalizing bias for the coming era of precision lss},\ }\href@noop {} {\bibfield  {journal} {\bibinfo  {journal} {Journal of Cosmology and Astroparticle Physics}\ }\textbf {\bibinfo {volume} {2009}}\bibinfo  {number} { (08)},\ \bibinfo {pages} {020}}\BibitemShut {NoStop}%
\bibitem [{\citenamefont {Leonardo}(2015)}]{leonardo2015bias}%
  \BibitemOpen
\bibfield  {number} {  }\bibfield  {author} {\bibinfo {author} {\bibfnamefont {S.}~\bibnamefont {Leonardo}},\ }\bibfield  {title} {\bibinfo {title} {Bias in the effective field theory of large scale structures},\ }\href@noop {} {\bibfield  {journal} {\bibinfo  {journal} {JCAP}\ }\textbf {\bibinfo {volume} {11}}\bibinfo  {number} { (007)}}\BibitemShut {NoStop}%
\bibitem [{\citenamefont {Assassi}\ \emph {et~al.}(2014)\citenamefont {Assassi}, \citenamefont {Baumann}, \citenamefont {Green},\ and\ \citenamefont {Zaldarriaga}}]{assassi2014renormalized}%
  \BibitemOpen
\bibfield  {number} {  }\bibfield  {author} {\bibinfo {author} {\bibfnamefont {V.}~\bibnamefont {Assassi}}, \bibinfo {author} {\bibfnamefont {D.}~\bibnamefont {Baumann}}, \bibinfo {author} {\bibfnamefont {D.}~\bibnamefont {Green}},\ and\ \bibinfo {author} {\bibfnamefont {M.}~\bibnamefont {Zaldarriaga}},\ }\bibfield  {title} {\bibinfo {title} {Renormalized halo bias},\ }\href@noop {} {\bibfield  {journal} {\bibinfo  {journal} {Journal of Cosmology and Astroparticle Physics}\ }\textbf {\bibinfo {volume} {2014}}\bibinfo  {number} { (08)},\ \bibinfo {pages} {056}}\BibitemShut {NoStop}%
\bibitem [{\citenamefont {Mirbabayi}\ \emph {et~al.}(2015)\citenamefont {Mirbabayi}, \citenamefont {Schmidt},\ and\ \citenamefont {Zaldarriaga}}]{mirbabayi2015biased}%
  \BibitemOpen
\bibfield  {number} {  }\bibfield  {author} {\bibinfo {author} {\bibfnamefont {M.}~\bibnamefont {Mirbabayi}}, \bibinfo {author} {\bibfnamefont {F.}~\bibnamefont {Schmidt}},\ and\ \bibinfo {author} {\bibfnamefont {M.}~\bibnamefont {Zaldarriaga}},\ }\bibfield  {title} {\bibinfo {title} {Biased tracers and time evolution},\ }\href@noop {} {\bibfield  {journal} {\bibinfo  {journal} {Journal of Cosmology and Astroparticle Physics}\ }\textbf {\bibinfo {volume} {2015}}\bibinfo  {number} { (07)},\ \bibinfo {pages} {030}}\BibitemShut {NoStop}%
\bibitem [{\citenamefont {Ortolá~Leonard}\ \emph {et~al.}(2025)\citenamefont {Ortolá~Leonard}, \citenamefont {Slepian},\ and\ \citenamefont {Hou}}]{Ortol_Leonard_2025}%
  \BibitemOpen
\bibfield  {number} {  }\bibfield  {author} {\bibinfo {author} {\bibfnamefont {W.}~\bibnamefont {Ortolá~Leonard}}, \bibinfo {author} {\bibfnamefont {Z.}~\bibnamefont {Slepian}},\ and\ \bibinfo {author} {\bibfnamefont {J.}~\bibnamefont {Hou}},\ }\bibfield  {title} {\bibinfo {title} {A model for the redshift-space galaxy 4-point correlation function},\ }\href {https://doi.org/10.1088/1475-7516/2025/01/090} {\bibfield  {journal} {\bibinfo  {journal} {Journal of Cosmology and Astroparticle Physics}\ }\textbf {\bibinfo {volume} {2025}}\bibinfo  {number} { (01)},\ \bibinfo {pages} {090}}\BibitemShut {NoStop}%
\bibitem [{\citenamefont {Ivanov}\ \emph {et~al.}(2020)\citenamefont {Ivanov}, \citenamefont {Simonovi{\'c}},\ and\ \citenamefont {Zaldarriaga}}]{ivanov2020cosmological}%
  \BibitemOpen
\bibfield  {number} {  }\bibfield  {author} {\bibinfo {author} {\bibfnamefont {M.~M.}\ \bibnamefont {Ivanov}}, \bibinfo {author} {\bibfnamefont {M.}~\bibnamefont {Simonovi{\'c}}},\ and\ \bibinfo {author} {\bibfnamefont {M.}~\bibnamefont {Zaldarriaga}},\ }\bibfield  {title} {\bibinfo {title} {Cosmological parameters from the boss galaxy power spectrum},\ }\href@noop {} {\bibfield  {journal} {\bibinfo  {journal} {Journal of Cosmology and Astroparticle Physics}\ }\textbf {\bibinfo {volume} {2020}}\bibinfo  {number} { (05)},\ \bibinfo {pages} {042}}\BibitemShut {NoStop}%
\bibitem [{\citenamefont {Carrasco}\ \emph {et~al.}(2012)\citenamefont {Carrasco}, \citenamefont {Hertzberg},\ and\ \citenamefont {Senatore}}]{carrasco2012effective}%
  \BibitemOpen
\bibfield  {number} {  }\bibfield  {author} {\bibinfo {author} {\bibfnamefont {J.~J.~M.}\ \bibnamefont {Carrasco}}, \bibinfo {author} {\bibfnamefont {M.~P.}\ \bibnamefont {Hertzberg}},\ and\ \bibinfo {author} {\bibfnamefont {L.}~\bibnamefont {Senatore}},\ }\bibfield  {title} {\bibinfo {title} {The effective field theory of cosmological large scale structures},\ }\href@noop {} {\bibfield  {journal} {\bibinfo  {journal} {Journal of High Energy Physics}\ }\textbf {\bibinfo {volume} {2012}},\ \bibinfo {pages} {1} (\bibinfo {year} {2012})}\BibitemShut {NoStop}%
\bibitem [{\citenamefont {Senatore}(2015)}]{senatore2015bias}%
  \BibitemOpen
  \bibfield  {author} {\bibinfo {author} {\bibfnamefont {L.}~\bibnamefont {Senatore}},\ }\bibfield  {title} {\bibinfo {title} {Bias in the effective field theory of large scale structures},\ }\href@noop {} {\bibfield  {journal} {\bibinfo  {journal} {Journal of Cosmology and Astroparticle Physics}\ }\textbf {\bibinfo {volume} {2015}}\bibinfo  {number} { (11)},\ \bibinfo {pages} {007}}\BibitemShut {NoStop}%
\bibitem [{\citenamefont {Feldman}\ \emph {et~al.}(2001)\citenamefont {Feldman} \emph {et~al.}}]{feldman2001constraints}%
  \BibitemOpen
\bibfield  {number} {  }\bibfield  {author} {\bibinfo {author} {\bibfnamefont {H.~A.}\ \bibnamefont {Feldman}} \emph {et~al.},\ }\bibfield  {title} {\bibinfo {title} {Constraints on galaxy bias, matter density, and primordial non-gaussianity from the psc z galaxy redshift survey},\ }\href@noop {} {\bibfield  {journal} {\bibinfo  {journal} {Physical Review Letters}\ }\textbf {\bibinfo {volume} {86}},\ \bibinfo {pages} {1434} (\bibinfo {year} {2001})}\BibitemShut {NoStop}%
\bibitem [{\citenamefont {{Guha Sarkar}}\ and\ \citenamefont {{Hazra}}(2013)}]{Guhaprimordial}%
  \BibitemOpen
  \bibfield  {author} {\bibinfo {author} {\bibfnamefont {T.}~\bibnamefont {{Guha Sarkar}}}\ and\ \bibinfo {author} {\bibfnamefont {D.~K.}\ \bibnamefont {{Hazra}}},\ }\bibfield  {title} {\bibinfo {title} {{Probing primordial non-Gaussianity: the 3D Bispectrum of Ly-{\ensuremath{\alpha}} forest and the redshifted 21-cm signal from the post reionization epoch}},\ }\href {https://doi.org/10.1088/1475-7516/2013/04/002} {\bibfield  {journal} {\bibinfo  {journal} {\jcap}\ }\textbf {\bibinfo {volume} {2013}},\ \bibinfo {eid} {002} (\bibinfo {year} {2013})},\ \Eprint {https://arxiv.org/abs/1211.4756} {arXiv:1211.4756 [astro-ph.CO]} \BibitemShut {NoStop}%
\bibitem [{\citenamefont {{Scoccimarro}}\ \emph {et~al.}(2004)\citenamefont {{Scoccimarro}}, \citenamefont {{Sefusatti}},\ and\ \citenamefont {{Zaldarriaga}}}]{bisp_png}%
  \BibitemOpen
  \bibfield  {author} {\bibinfo {author} {\bibfnamefont {R.}~\bibnamefont {{Scoccimarro}}}, \bibinfo {author} {\bibfnamefont {E.}~\bibnamefont {{Sefusatti}}},\ and\ \bibinfo {author} {\bibfnamefont {M.}~\bibnamefont {{Zaldarriaga}}},\ }\bibfield  {title} {\bibinfo {title} {{Probing primordial non-Gaussianity with large-scale structure}},\ }\href {https://doi.org/10.1103/PhysRevD.69.103513} {\bibfield  {journal} {\bibinfo  {journal} {\prd}\ }\textbf {\bibinfo {volume} {69}},\ \bibinfo {eid} {103513} (\bibinfo {year} {2004})},\ \Eprint {https://arxiv.org/abs/astro-ph/0312286} {arXiv:astro-ph/0312286 [astro-ph]} \BibitemShut {NoStop}%
\bibitem [{\citenamefont {Emiliano}\ \emph {et~al.}(2006)\citenamefont {Emiliano}, \citenamefont {Martin},\ and\ \citenamefont {Sebastian}}]{emiliano2006cosmology}%
  \BibitemOpen
  \bibfield  {author} {\bibinfo {author} {\bibfnamefont {S.}~\bibnamefont {Emiliano}}, \bibinfo {author} {\bibfnamefont {C.}~\bibnamefont {Martin}},\ and\ \bibinfo {author} {\bibfnamefont {P.}~\bibnamefont {Sebastian}},\ }\bibfield  {title} {\bibinfo {title} {Cosmology and the bispectrum},\ }\href@noop {} {\bibfield  {journal} {\bibinfo  {journal} {Phys. Rev. D}\ }\textbf {\bibinfo {volume} {74}},\ \bibinfo {pages} {023522} (\bibinfo {year} {2006})}\BibitemShut {NoStop}%
\bibitem [{\citenamefont {Scoccimarro}(2000)}]{scoccimarro2000bispectrum}%
  \BibitemOpen
  \bibfield  {author} {\bibinfo {author} {\bibfnamefont {R.}~\bibnamefont {Scoccimarro}},\ }\bibfield  {title} {\bibinfo {title} {The bispectrum: from theory to observations},\ }\href@noop {} {\bibfield  {journal} {\bibinfo  {journal} {The Astrophysical Journal}\ }\textbf {\bibinfo {volume} {544}},\ \bibinfo {pages} {597} (\bibinfo {year} {2000})}\BibitemShut {NoStop}%
\bibitem [{\citenamefont {Dizgah}\ \emph {et~al.}(2021)\citenamefont {Dizgah} \emph {et~al.}}]{dizgah2021primordial}%
  \BibitemOpen
  \bibfield  {author} {\bibinfo {author} {\bibfnamefont {A.~M.}\ \bibnamefont {Dizgah}} \emph {et~al.},\ }\bibfield  {title} {\bibinfo {title} {Primordial non-gaussianity from biased tracers: likelihood analysis of real-space power spectrum and bispectrum},\ }\href@noop {} {\bibfield  {journal} {\bibinfo  {journal} {Journal of Cosmology and Astroparticle Physics}\ }\textbf {\bibinfo {volume} {2021}}\bibinfo  {number} { (05)},\ \bibinfo {pages} {015}}\BibitemShut {NoStop}%
\bibitem [{\citenamefont {Baldauf}\ \emph {et~al.}(2011)\citenamefont {Baldauf}, \citenamefont {Seljak},\ and\ \citenamefont {Senatore}}]{baldauf2011primordial}%
  \BibitemOpen
\bibfield  {number} {  }\bibfield  {author} {\bibinfo {author} {\bibfnamefont {T.}~\bibnamefont {Baldauf}}, \bibinfo {author} {\bibfnamefont {U.}~\bibnamefont {Seljak}},\ and\ \bibinfo {author} {\bibfnamefont {L.}~\bibnamefont {Senatore}},\ }\bibfield  {title} {\bibinfo {title} {Primordial non-gaussianity in the bispectrum of the halo density field},\ }\href@noop {} {\bibfield  {journal} {\bibinfo  {journal} {Journal of Cosmology and Astroparticle Physics}\ }\textbf {\bibinfo {volume} {2011}}\bibinfo  {number} { (04)},\ \bibinfo {pages} {006}}\BibitemShut {NoStop}%
\bibitem [{\citenamefont {{D'Amico}}\ \emph {et~al.}(2025)\citenamefont {{D'Amico}}, \citenamefont {{Lewandowski}}, \citenamefont {{Senatore}},\ and\ \citenamefont {{Zhang}}}]{LewandowskiLimits}%
  \BibitemOpen
\bibfield  {number} {  }\bibfield  {author} {\bibinfo {author} {\bibfnamefont {G.}~\bibnamefont {{D'Amico}}}, \bibinfo {author} {\bibfnamefont {M.}~\bibnamefont {{Lewandowski}}}, \bibinfo {author} {\bibfnamefont {L.}~\bibnamefont {{Senatore}}},\ and\ \bibinfo {author} {\bibfnamefont {P.}~\bibnamefont {{Zhang}}},\ }\bibfield  {title} {\bibinfo {title} {{Limits on primordial non-Gaussianities from BOSS galaxy-clustering data}},\ }\href {https://doi.org/10.1103/PhysRevD.111.063514} {\bibfield  {journal} {\bibinfo  {journal} {\prd}\ }\textbf {\bibinfo {volume} {111}},\ \bibinfo {eid} {063514} (\bibinfo {year} {2025})},\ \Eprint {https://arxiv.org/abs/2201.11518} {arXiv:2201.11518 [astro-ph.CO]} \BibitemShut {NoStop}%
\bibitem [{\citenamefont {{Shirasaki}}(2021)}]{ShirasakiConstraining}%
  \BibitemOpen
  \bibfield  {author} {\bibinfo {author} {\bibfnamefont {M.~o.}\ \bibnamefont {{Shirasaki}}},\ }\bibfield  {title} {\bibinfo {title} {{Constraining primordial non-Gaussianity with postreconstructed galaxy bispectrum in redshift space}},\ }\href {https://doi.org/10.1103/PhysRevD.103.023506} {\bibfield  {journal} {\bibinfo  {journal} {\prd}\ }\textbf {\bibinfo {volume} {103}},\ \bibinfo {eid} {023506} (\bibinfo {year} {2021})},\ \Eprint {https://arxiv.org/abs/2010.04567} {arXiv:2010.04567 [astro-ph.CO]} \BibitemShut {NoStop}%
\bibitem [{\citenamefont {{Gualdi}}\ \emph {et~al.}(2019)\citenamefont {{Gualdi}}, \citenamefont {{Gil-Mar{\'\i}n}}, \citenamefont {{Schuhmann}}, \citenamefont {{Manera}}, \citenamefont {{Joachimi}},\ and\ \citenamefont {{Lahav}}}]{GualdiEnhancing}%
  \BibitemOpen
  \bibfield  {author} {\bibinfo {author} {\bibfnamefont {D.}~\bibnamefont {{Gualdi}}}, \bibinfo {author} {\bibfnamefont {H.}~\bibnamefont {{Gil-Mar{\'\i}n}}}, \bibinfo {author} {\bibfnamefont {R.~L.}\ \bibnamefont {{Schuhmann}}}, \bibinfo {author} {\bibfnamefont {M.}~\bibnamefont {{Manera}}}, \bibinfo {author} {\bibfnamefont {B.}~\bibnamefont {{Joachimi}}},\ and\ \bibinfo {author} {\bibfnamefont {O.}~\bibnamefont {{Lahav}}},\ }\bibfield  {title} {\bibinfo {title} {{Enhancing BOSS bispectrum cosmological constraints with maximal compression}},\ }\href {https://doi.org/10.1093/mnras/stz051} {\bibfield  {journal} {\bibinfo  {journal} {\mnras}\ }\textbf {\bibinfo {volume} {484}},\ \bibinfo {pages} {3713} (\bibinfo {year} {2019})},\ \Eprint {https://arxiv.org/abs/1806.02853} {arXiv:1806.02853 [astro-ph.CO]} \BibitemShut {NoStop}%
\bibitem [{\citenamefont {{Ivanov}}\ \emph {et~al.}(2022)\citenamefont {{Ivanov}} \emph {et~al.}}]{IvanovPrecision}%
  \BibitemOpen
  \bibfield  {author} {\bibinfo {author} {\bibfnamefont {M.~M.}\ \bibnamefont {{Ivanov}}} \emph {et~al.},\ }\bibfield  {title} {\bibinfo {title} {{Precision analysis of the redshift-space galaxy bispectrum}},\ }\href {https://doi.org/10.1103/PhysRevD.105.063512} {\bibfield  {journal} {\bibinfo  {journal} {\prd}\ }\textbf {\bibinfo {volume} {105}},\ \bibinfo {eid} {063512} (\bibinfo {year} {2022})},\ \Eprint {https://arxiv.org/abs/2110.10161} {arXiv:2110.10161 [astro-ph.CO]} \BibitemShut {NoStop}%
\bibitem [{\citenamefont {{Philcox}}\ \emph {et~al.}(2022)\citenamefont {{Philcox}} \emph {et~al.}}]{PhilcoxCosmology}%
  \BibitemOpen
  \bibfield  {author} {\bibinfo {author} {\bibfnamefont {O.~H.~E.}\ \bibnamefont {{Philcox}}} \emph {et~al.},\ }\bibfield  {title} {\bibinfo {title} {{Cosmology with the redshift-space galaxy bispectrum monopole at one-loop order}},\ }\href {https://doi.org/10.1103/PhysRevD.106.043530} {\bibfield  {journal} {\bibinfo  {journal} {\prd}\ }\textbf {\bibinfo {volume} {106}},\ \bibinfo {eid} {043530} (\bibinfo {year} {2022})},\ \Eprint {https://arxiv.org/abs/2206.02800} {arXiv:2206.02800 [astro-ph.CO]} \BibitemShut {NoStop}%
\bibitem [{\citenamefont {{Bragan{\c{c}}a}}\ \emph {et~al.}(2023)\citenamefont {{Bragan{\c{c}}a}}, \citenamefont {{Donath}}, \citenamefont {{Senatore}},\ and\ \citenamefont {{Zheng}}}]{senatoreforecast}%
  \BibitemOpen
  \bibfield  {author} {\bibinfo {author} {\bibfnamefont {D.}~\bibnamefont {{Bragan{\c{c}}a}}}, \bibinfo {author} {\bibfnamefont {Y.}~\bibnamefont {{Donath}}}, \bibinfo {author} {\bibfnamefont {L.}~\bibnamefont {{Senatore}}},\ and\ \bibinfo {author} {\bibfnamefont {H.}~\bibnamefont {{Zheng}}},\ }\bibfield  {title} {\bibinfo {title} {{Peeking into the next decade in Large-Scale Structure Cosmology with its Effective Field Theory}},\ }\href {https://doi.org/10.48550/arXiv.2307.04992} {\bibfield  {journal} {\bibinfo  {journal} {arXiv e-prints}\ ,\ \bibinfo {eid} {arXiv:2307.04992}} (\bibinfo {year} {2023})},\ \Eprint {https://arxiv.org/abs/2307.04992} {arXiv:2307.04992 [astro-ph.CO]} \BibitemShut {NoStop}%
\bibitem [{\citenamefont {{Hahn}}\ \emph {et~al.}(2024)\citenamefont {{Hahn}} \emph {et~al.}}]{simbigchang}%
  \BibitemOpen
  \bibfield  {author} {\bibinfo {author} {\bibfnamefont {C.}~\bibnamefont {{Hahn}}} \emph {et~al.},\ }\bibfield  {title} {\bibinfo {title} {{Cosmological constraints from the nonlinear galaxy bispectrum}},\ }\href {https://doi.org/10.1103/PhysRevD.109.083534} {\bibfield  {journal} {\bibinfo  {journal} {\prd}\ }\textbf {\bibinfo {volume} {109}},\ \bibinfo {eid} {083534} (\bibinfo {year} {2024})},\ \Eprint {https://arxiv.org/abs/2310.15243} {arXiv:2310.15243 [astro-ph.CO]} \BibitemShut {NoStop}%
\bibitem [{\citenamefont {Hahn}\ \emph {et~al.}(2023)\citenamefont {Hahn} \emph {et~al.}}]{hahn2023rmsscriptsizeimbigcosmological}%
  \BibitemOpen
  \bibfield  {author} {\bibinfo {author} {\bibfnamefont {C.}~\bibnamefont {Hahn}} \emph {et~al.},\ }\href {https://arxiv.org/abs/2310.15243} {\bibinfo {title} {\textsc{SimBIG}: The first cosmological constraints from the non-linear galaxy bispectrum}} (\bibinfo {year} {2023}),\ \Eprint {https://arxiv.org/abs/2310.15243} {arXiv:2310.15243 [astro-ph.CO]} \BibitemShut {NoStop}%
\bibitem [{\citenamefont {{Fry}}\ and\ \citenamefont {{Gaztanaga}}(1993)}]{fry_1993}%
  \BibitemOpen
  \bibfield  {author} {\bibinfo {author} {\bibfnamefont {J.~N.}\ \bibnamefont {{Fry}}}\ and\ \bibinfo {author} {\bibfnamefont {E.}~\bibnamefont {{Gaztanaga}}},\ }\bibfield  {title} {\bibinfo {title} {{Biasing and Hierarchical Statistics in Large-Scale Structure}},\ }\href {https://doi.org/10.1086/173015} {\bibfield  {journal} {\bibinfo  {journal} {\apj}\ }\textbf {\bibinfo {volume} {413}},\ \bibinfo {pages} {447} (\bibinfo {year} {1993})},\ \Eprint {https://arxiv.org/abs/astro-ph/9302009} {arXiv:astro-ph/9302009 [astro-ph]} \BibitemShut {NoStop}%
\bibitem [{\citenamefont {Yankelevich}\ and\ \citenamefont {Porciani}(2019)}]{yankelevich2019cosmological}%
  \BibitemOpen
  \bibfield  {author} {\bibinfo {author} {\bibfnamefont {V.}~\bibnamefont {Yankelevich}}\ and\ \bibinfo {author} {\bibfnamefont {C.}~\bibnamefont {Porciani}},\ }\bibfield  {title} {\bibinfo {title} {Cosmological information in the redshift-space bispectrum},\ }\href@noop {} {\bibfield  {journal} {\bibinfo  {journal} {Monthly Notices of the Royal Astronomical Society}\ }\textbf {\bibinfo {volume} {483}},\ \bibinfo {pages} {2078} (\bibinfo {year} {2019})}\BibitemShut {NoStop}%
\bibitem [{\citenamefont {Hahn}\ \emph {et~al.}(2020)\citenamefont {Hahn} \emph {et~al.}}]{Hahn_2020}%
  \BibitemOpen
  \bibfield  {author} {\bibinfo {author} {\bibfnamefont {C.}~\bibnamefont {Hahn}} \emph {et~al.},\ }\bibfield  {title} {\bibinfo {title} {Constraining m$\nu$ with the bispectrum. part i. breaking parameter degeneracies},\ }\href {https://doi.org/10.1088/1475-7516/2020/03/040} {\bibfield  {journal} {\bibinfo  {journal} {Journal of Cosmology and Astroparticle Physics}\ }\textbf {\bibinfo {volume} {2020}}\bibinfo  {number} { (03)},\ \bibinfo {pages} {040–040}}\BibitemShut {NoStop}%
\bibitem [{\citenamefont {{Behera}}\ \emph {et~al.}(2024)\citenamefont {{Behera}}, \citenamefont {{Rezaie}}, \citenamefont {{Samushia}},\ and\ \citenamefont {{Ereza}}}]{rezaei_mahdi_BAO}%
  \BibitemOpen
\bibfield  {number} {  }\bibfield  {author} {\bibinfo {author} {\bibfnamefont {J.}~\bibnamefont {{Behera}}}, \bibinfo {author} {\bibfnamefont {M.}~\bibnamefont {{Rezaie}}}, \bibinfo {author} {\bibfnamefont {L.}~\bibnamefont {{Samushia}}},\ and\ \bibinfo {author} {\bibfnamefont {J.}~\bibnamefont {{Ereza}}},\ }\bibfield  {title} {\bibinfo {title} {{Modelling the BAO feature in bispectrum}},\ }\href {https://doi.org/10.1093/mnras/stae1161} {\bibfield  {journal} {\bibinfo  {journal} {\mnras}\ }\textbf {\bibinfo {volume} {531}},\ \bibinfo {pages} {3326} (\bibinfo {year} {2024})},\ \Eprint {https://arxiv.org/abs/2312.05942} {arXiv:2312.05942 [astro-ph.CO]} \BibitemShut {NoStop}%
\bibitem [{\citenamefont {{Gil-Mar{\'\i}n}}\ \emph {et~al.}(2017)\citenamefont {{Gil-Mar{\'\i}n}} \emph {et~al.}}]{Gilclustering}%
  \BibitemOpen
  \bibfield  {author} {\bibinfo {author} {\bibfnamefont {H.}~\bibnamefont {{Gil-Mar{\'\i}n}}} \emph {et~al.},\ }\bibfield  {title} {\bibinfo {title} {{The clustering of galaxies in the SDSS-III Baryon Oscillation Spectroscopic Survey: RSD measurement from the power spectrum and bispectrum of the DR12 BOSS galaxies}},\ }\href {https://doi.org/10.1093/mnras/stw2679} {\bibfield  {journal} {\bibinfo  {journal} {\mnras}\ }\textbf {\bibinfo {volume} {465}},\ \bibinfo {pages} {1757} (\bibinfo {year} {2017})},\ \Eprint {https://arxiv.org/abs/1606.00439} {arXiv:1606.00439 [astro-ph.CO]} \BibitemShut {NoStop}%
\bibitem [{\citenamefont {{Gil-Mar{\'\i}n}}\ \emph {et~al.}(2015)\citenamefont {{Gil-Mar{\'\i}n}} \emph {et~al.}}]{Gilinterpretation}%
  \BibitemOpen
  \bibfield  {author} {\bibinfo {author} {\bibfnamefont {H.}~\bibnamefont {{Gil-Mar{\'\i}n}}} \emph {et~al.},\ }\bibfield  {title} {\bibinfo {title} {{The power spectrum and bispectrum of SDSS DR11 BOSS galaxies - II. Cosmological interpretation}},\ }\href {https://doi.org/10.1093/mnras/stv1359} {\bibfield  {journal} {\bibinfo  {journal} {\mnras}\ }\textbf {\bibinfo {volume} {452}},\ \bibinfo {pages} {1914} (\bibinfo {year} {2015})},\ \Eprint {https://arxiv.org/abs/1408.0027} {arXiv:1408.0027 [astro-ph.CO]} \BibitemShut {NoStop}%
\bibitem [{\citenamefont {{Child}}\ \emph {et~al.}(2018)\citenamefont {{Child}} \emph {et~al.}}]{ChildBispectrum}%
  \BibitemOpen
  \bibfield  {author} {\bibinfo {author} {\bibfnamefont {H.~L.}\ \bibnamefont {{Child}}} \emph {et~al.},\ }\bibfield  {title} {\bibinfo {title} {{Bispectrum as baryon acoustic oscillation interferometer}},\ }\href {https://doi.org/10.1103/PhysRevD.98.123521} {\bibfield  {journal} {\bibinfo  {journal} {\prd}\ }\textbf {\bibinfo {volume} {98}},\ \bibinfo {eid} {123521} (\bibinfo {year} {2018})},\ \Eprint {https://arxiv.org/abs/1806.11147} {arXiv:1806.11147 [astro-ph.CO]} \BibitemShut {NoStop}%
\bibitem [{\citenamefont {{Pearson}}\ and\ \citenamefont {{Samushia}}(2018)}]{PearsonDetection}%
  \BibitemOpen
  \bibfield  {author} {\bibinfo {author} {\bibfnamefont {D.~W.}\ \bibnamefont {{Pearson}}}\ and\ \bibinfo {author} {\bibfnamefont {L.}~\bibnamefont {{Samushia}}},\ }\bibfield  {title} {\bibinfo {title} {{A Detection of the Baryon Acoustic Oscillation features in the SDSS BOSS DR12 Galaxy Bispectrum}},\ }\href {https://doi.org/10.1093/mnras/sty1266} {\bibfield  {journal} {\bibinfo  {journal} {\mnras}\ }\textbf {\bibinfo {volume} {478}},\ \bibinfo {pages} {4500} (\bibinfo {year} {2018})},\ \Eprint {https://arxiv.org/abs/1712.04970} {arXiv:1712.04970 [astro-ph.CO]} \BibitemShut {NoStop}%
\bibitem [{\citenamefont {{Slepian}}\ \emph {et~al.}(2017)\citenamefont {{Slepian}} \emph {et~al.}}]{SlepianDetection}%
  \BibitemOpen
  \bibfield  {author} {\bibinfo {author} {\bibfnamefont {Z.}~\bibnamefont {{Slepian}}} \emph {et~al.},\ }\bibfield  {title} {\bibinfo {title} {{Detection of baryon acoustic oscillation features in the large-scale three-point correlation function of SDSS BOSS DR12 CMASS galaxies}},\ }\href {https://doi.org/10.1093/mnras/stx488} {\bibfield  {journal} {\bibinfo  {journal} {\mnras}\ }\textbf {\bibinfo {volume} {469}},\ \bibinfo {pages} {1738} (\bibinfo {year} {2017})},\ \Eprint {https://arxiv.org/abs/1607.06097} {arXiv:1607.06097 [astro-ph.CO]} \BibitemShut {NoStop}%
\bibitem [{\citenamefont {{Cabass}}\ \emph {et~al.}(2022)\citenamefont {{Cabass}} \emph {et~al.}}]{CabassConstraints}%
  \BibitemOpen
  \bibfield  {author} {\bibinfo {author} {\bibfnamefont {G.}~\bibnamefont {{Cabass}}} \emph {et~al.},\ }\bibfield  {title} {\bibinfo {title} {{Constraints on multifield inflation from the BOSS galaxy survey}},\ }\href {https://doi.org/10.1103/PhysRevD.106.043506} {\bibfield  {journal} {\bibinfo  {journal} {\prd}\ }\textbf {\bibinfo {volume} {106}},\ \bibinfo {eid} {043506} (\bibinfo {year} {2022})},\ \Eprint {https://arxiv.org/abs/2204.01781} {arXiv:2204.01781 [astro-ph.CO]} \BibitemShut {NoStop}%
\bibitem [{\citenamefont {{Spaar}}\ and\ \citenamefont {{Zhang}}(2025)}]{SpaarCosmological}%
  \BibitemOpen
  \bibfield  {author} {\bibinfo {author} {\bibfnamefont {S.}~\bibnamefont {{Spaar}}}\ and\ \bibinfo {author} {\bibfnamefont {P.}~\bibnamefont {{Zhang}}},\ }\bibfield  {title} {\bibinfo {title} {{Cosmological constraints from combined probes with the three-point statistics of galaxies at one-loop precision}},\ }\href {https://doi.org/10.1103/PhysRevD.111.023529} {\bibfield  {journal} {\bibinfo  {journal} {\prd}\ }\textbf {\bibinfo {volume} {111}},\ \bibinfo {eid} {023529} (\bibinfo {year} {2025})},\ \Eprint {https://arxiv.org/abs/2312.15164} {arXiv:2312.15164 [astro-ph.CO]} \BibitemShut {NoStop}%
\bibitem [{\citenamefont {Gil-Mar{\'\i}n}\ \emph {et~al.}(2016)\citenamefont {Gil-Mar{\'\i}n} \emph {et~al.}}]{gil2016clustering}%
  \BibitemOpen
  \bibfield  {author} {\bibinfo {author} {\bibfnamefont {H.}~\bibnamefont {Gil-Mar{\'\i}n}} \emph {et~al.},\ }\bibfield  {title} {\bibinfo {title} {The clustering of galaxies in the sdss-iii baryon oscillation spectroscopic survey: Rsd measurement from the los-dependent power spectrum of dr12 boss galaxies},\ }\href@noop {} {\bibfield  {journal} {\bibinfo  {journal} {Monthly Notices of the Royal Astronomical Society}\ }\textbf {\bibinfo {volume} {460}},\ \bibinfo {pages} {4188} (\bibinfo {year} {2016})}\BibitemShut {NoStop}%
\bibitem [{\citenamefont {D’Amico}\ \emph {et~al.}(2024)\citenamefont {D’Amico} \emph {et~al.}}]{D_Amico_2024}%
  \BibitemOpen
  \bibfield  {author} {\bibinfo {author} {\bibfnamefont {G.}~\bibnamefont {D’Amico}} \emph {et~al.},\ }\bibfield  {title} {\bibinfo {title} {The boss bispectrum analysis at one loop from the effective field theory of large-scale structure},\ }\href {https://doi.org/10.1088/1475-7516/2024/05/059} {\bibfield  {journal} {\bibinfo  {journal} {Journal of Cosmology and Astroparticle Physics}\ }\textbf {\bibinfo {volume} {2024}}\bibinfo  {number} { (05)},\ \bibinfo {pages} {059}}\BibitemShut {NoStop}%
\bibitem [{\citenamefont {Ivanov}\ \emph {et~al.}(2022)\citenamefont {Ivanov} \emph {et~al.}}]{Ivanov_2022}%
  \BibitemOpen
\bibfield  {number} {  }\bibfield  {author} {\bibinfo {author} {\bibfnamefont {M.~M.}\ \bibnamefont {Ivanov}} \emph {et~al.},\ }\bibfield  {title} {\bibinfo {title} {Precision analysis of the redshift-space galaxy bispectrum},\ }\bibfield  {journal} {\bibinfo  {journal} {Physical Review D}\ }\textbf {\bibinfo {volume} {105}},\ \href {https://doi.org/10.1103/physrevd.105.063512} {10.1103/physrevd.105.063512} (\bibinfo {year} {2022})\BibitemShut {NoStop}%
\bibitem [{\citenamefont {Philcox}\ and\ \citenamefont {Ivanov}(2022)}]{philcox2022boss}%
  \BibitemOpen
  \bibfield  {author} {\bibinfo {author} {\bibfnamefont {O.~H.}\ \bibnamefont {Philcox}}\ and\ \bibinfo {author} {\bibfnamefont {M.~M.}\ \bibnamefont {Ivanov}},\ }\bibfield  {title} {\bibinfo {title} {Boss dr12 full-shape cosmology: $\lambda$ cdm constraints from the large-scale galaxy power spectrum and bispectrum monopole},\ }\href@noop {} {\bibfield  {journal} {\bibinfo  {journal} {Physical Review D}\ }\textbf {\bibinfo {volume} {105}},\ \bibinfo {pages} {043517} (\bibinfo {year} {2022})}\BibitemShut {NoStop}%
\bibitem [{\citenamefont {{Lu}}\ \emph {et~al.}(2025)\citenamefont {{Lu}}, \citenamefont {{Simon}},\ and\ \citenamefont {{Zhang}}}]{LuPreference}%
  \BibitemOpen
  \bibfield  {author} {\bibinfo {author} {\bibfnamefont {Z.}~\bibnamefont {{Lu}}}, \bibinfo {author} {\bibfnamefont {T.}~\bibnamefont {{Simon}}},\ and\ \bibinfo {author} {\bibfnamefont {P.}~\bibnamefont {{Zhang}}},\ }\bibfield  {title} {\bibinfo {title} {{Preference for evolving dark energy in light of the galaxy bispectrum}},\ }\href {https://doi.org/10.48550/arXiv.2503.04602} {\bibfield  {journal} {\bibinfo  {journal} {arXiv e-prints}\ ,\ \bibinfo {eid} {arXiv:2503.04602}} (\bibinfo {year} {2025})},\ \Eprint {https://arxiv.org/abs/2503.04602} {arXiv:2503.04602 [astro-ph.CO]} \BibitemShut {NoStop}%
\bibitem [{\citenamefont {{Ivanov}}\ \emph {et~al.}(2023)\citenamefont {{Ivanov}} \emph {et~al.}}]{Ivanovoptimal}%
  \BibitemOpen
  \bibfield  {author} {\bibinfo {author} {\bibfnamefont {M.~M.}\ \bibnamefont {{Ivanov}}} \emph {et~al.},\ }\bibfield  {title} {\bibinfo {title} {{Cosmology with the galaxy bispectrum multipoles: Optimal estimation and application to BOSS data}},\ }\href {https://doi.org/10.1103/PhysRevD.107.083515} {\bibfield  {journal} {\bibinfo  {journal} {\prd}\ }\textbf {\bibinfo {volume} {107}},\ \bibinfo {eid} {083515} (\bibinfo {year} {2023})},\ \Eprint {https://arxiv.org/abs/2302.04414} {arXiv:2302.04414 [astro-ph.CO]} \BibitemShut {NoStop}%
\bibitem [{\citenamefont {{Pearson}}\ and\ \citenamefont {{Samushia}}(2019)}]{PearsonErratum}%
  \BibitemOpen
  \bibfield  {author} {\bibinfo {author} {\bibfnamefont {D.~W.}\ \bibnamefont {{Pearson}}}\ and\ \bibinfo {author} {\bibfnamefont {L.}~\bibnamefont {{Samushia}}},\ }\bibfield  {title} {\bibinfo {title} {{Erratum: A Detection of the Baryon Acoustic Oscillation Features in the SDSS BOSS DR12 Galaxy Bispectrum}},\ }\href {https://doi.org/10.1093/mnras/sty3173} {\bibfield  {journal} {\bibinfo  {journal} {\mnras}\ }\textbf {\bibinfo {volume} {483}},\ \bibinfo {pages} {915} (\bibinfo {year} {2019})}\BibitemShut {NoStop}%
\bibitem [{\citenamefont {Scoccimarro}\ \emph {et~al.}(1998)\citenamefont {Scoccimarro} \emph {et~al.}}]{scoccimarro1998nonlinear}%
  \BibitemOpen
  \bibfield  {author} {\bibinfo {author} {\bibfnamefont {R.}~\bibnamefont {Scoccimarro}} \emph {et~al.},\ }\bibfield  {title} {\bibinfo {title} {Nonlinear evolution of the bispectrum of cosmological perturbations},\ }\href@noop {} {\bibfield  {journal} {\bibinfo  {journal} {The Astrophysical Journal}\ }\textbf {\bibinfo {volume} {496}},\ \bibinfo {pages} {586} (\bibinfo {year} {1998})}\BibitemShut {NoStop}%
\bibitem [{\citenamefont {Scoccimarro}\ \emph {et~al.}(1999)\citenamefont {Scoccimarro}, \citenamefont {Couchman},\ and\ \citenamefont {Frieman}}]{scoccimarro1999bispectrum}%
  \BibitemOpen
  \bibfield  {author} {\bibinfo {author} {\bibfnamefont {R.}~\bibnamefont {Scoccimarro}}, \bibinfo {author} {\bibfnamefont {H.}~\bibnamefont {Couchman}},\ and\ \bibinfo {author} {\bibfnamefont {J.~A.}\ \bibnamefont {Frieman}},\ }\bibfield  {title} {\bibinfo {title} {The bispectrum as a signature of gravitational instability in redshift space},\ }\href@noop {} {\bibfield  {journal} {\bibinfo  {journal} {The Astrophysical Journal}\ }\textbf {\bibinfo {volume} {517}},\ \bibinfo {pages} {531} (\bibinfo {year} {1999})}\BibitemShut {NoStop}%
\bibitem [{\citenamefont {Tegmark}\ \emph {et~al.}(1998)\citenamefont {Tegmark} \emph {et~al.}}]{Tegmark_1998}%
  \BibitemOpen
  \bibfield  {author} {\bibinfo {author} {\bibfnamefont {M.}~\bibnamefont {Tegmark}} \emph {et~al.},\ }\bibfield  {title} {\bibinfo {title} {Measuring the galaxy power spectrum with future redshift surveys},\ }\href {https://doi.org/10.1086/305663} {\bibfield  {journal} {\bibinfo  {journal} {The Astrophysical Journal}\ }\textbf {\bibinfo {volume} {499}},\ \bibinfo {pages} {555–576} (\bibinfo {year} {1998})}\BibitemShut {NoStop}%
\bibitem [{\citenamefont {Noriega}\ \emph {et~al.}(2022)\citenamefont {Noriega}, \citenamefont {Aviles}, \citenamefont {Fromenteau},\ and\ \citenamefont {Vargas-Maga{\~n}a}}]{noriega2022fast}%
  \BibitemOpen
  \bibfield  {author} {\bibinfo {author} {\bibfnamefont {H.~E.}\ \bibnamefont {Noriega}}, \bibinfo {author} {\bibfnamefont {A.}~\bibnamefont {Aviles}}, \bibinfo {author} {\bibfnamefont {S.}~\bibnamefont {Fromenteau}},\ and\ \bibinfo {author} {\bibfnamefont {M.}~\bibnamefont {Vargas-Maga{\~n}a}},\ }\bibfield  {title} {\bibinfo {title} {Fast computation of non-linear power spectrum in cosmologies with massive neutrinos},\ }\href@noop {} {\bibfield  {journal} {\bibinfo  {journal} {Journal of Cosmology and Astroparticle Physics}\ }\textbf {\bibinfo {volume} {2022}}\bibinfo  {number} { (11)},\ \bibinfo {pages} {038}}\BibitemShut {NoStop}%
\bibitem [{\citenamefont {{Fumagalli}}\ \emph {et~al.}(2022)\citenamefont {{Fumagalli}} \emph {et~al.}}]{FumagalliFitting}%
  \BibitemOpen
\bibfield  {number} {  }\bibfield  {author} {\bibinfo {author} {\bibfnamefont {A.}~\bibnamefont {{Fumagalli}}} \emph {et~al.},\ }\bibfield  {title} {\bibinfo {title} {{Fitting covariance matrix models to simulations}},\ }\href {https://doi.org/10.1088/1475-7516/2022/12/022} {\bibfield  {journal} {\bibinfo  {journal} {\jcap}\ }\textbf {\bibinfo {volume} {2022}},\ \bibinfo {eid} {022} (\bibinfo {year} {2022})},\ \Eprint {https://arxiv.org/abs/2206.05191} {arXiv:2206.05191 [astro-ph.CO]} \BibitemShut {NoStop}%
\bibitem [{\citenamefont {{Mohammed}}\ \emph {et~al.}(2017{\natexlab{b}})\citenamefont {{Mohammed}}, \citenamefont {{Seljak}},\ and\ \citenamefont {{Vlah}}}]{sugi_covar}%
  \BibitemOpen
  \bibfield  {author} {\bibinfo {author} {\bibfnamefont {I.}~\bibnamefont {{Mohammed}}}, \bibinfo {author} {\bibfnamefont {U.}~\bibnamefont {{Seljak}}},\ and\ \bibinfo {author} {\bibfnamefont {Z.}~\bibnamefont {{Vlah}}},\ }\bibfield  {title} {\bibinfo {title} {{Perturbative approach to covariance matrix of the matter power spectrum}},\ }\href {https://doi.org/10.1093/mnras/stw3196} {\bibfield  {journal} {\bibinfo  {journal} {\mnras}\ }\textbf {\bibinfo {volume} {466}},\ \bibinfo {pages} {780} (\bibinfo {year} {2017}{\natexlab{b}})},\ \Eprint {https://arxiv.org/abs/1607.00043} {arXiv:1607.00043 [astro-ph.CO]} \BibitemShut {NoStop}%
\bibitem [{\citenamefont {{Oddo}}\ \emph {et~al.}(2021)\citenamefont {{Oddo}} \emph {et~al.}}]{Oddolikelihood}%
  \BibitemOpen
  \bibfield  {author} {\bibinfo {author} {\bibfnamefont {A.}~\bibnamefont {{Oddo}}} \emph {et~al.},\ }\bibfield  {title} {\bibinfo {title} {{Cosmological parameters from the likelihood analysis of the galaxy power spectrum and bispectrum in real space}},\ }\href {https://doi.org/10.1088/1475-7516/2021/11/038} {\bibfield  {journal} {\bibinfo  {journal} {\jcap}\ }\textbf {\bibinfo {volume} {2021}},\ \bibinfo {eid} {038} (\bibinfo {year} {2021})},\ \Eprint {https://arxiv.org/abs/2108.03204} {arXiv:2108.03204 [astro-ph.CO]} \BibitemShut {NoStop}%
\bibitem [{\citenamefont {{Salvalaggio}}\ \emph {et~al.}(2024)\citenamefont {{Salvalaggio}} \emph {et~al.}}]{Salvalaggiocovariance}%
  \BibitemOpen
  \bibfield  {author} {\bibinfo {author} {\bibfnamefont {J.}~\bibnamefont {{Salvalaggio}}} \emph {et~al.},\ }\bibfield  {title} {\bibinfo {title} {{Bispectrum non-Gaussian covariance in redshift space}},\ }\href {https://doi.org/10.1088/1475-7516/2024/08/046} {\bibfield  {journal} {\bibinfo  {journal} {\jcap}\ }\textbf {\bibinfo {volume} {2024}},\ \bibinfo {eid} {046} (\bibinfo {year} {2024})},\ \Eprint {https://arxiv.org/abs/2403.08634} {arXiv:2403.08634 [astro-ph.CO]} \BibitemShut {NoStop}%
\bibitem [{\citenamefont {Gualdi}\ and\ \citenamefont {Verde}(2020)}]{gualdi2020galaxy}%
  \BibitemOpen
  \bibfield  {author} {\bibinfo {author} {\bibfnamefont {D.}~\bibnamefont {Gualdi}}\ and\ \bibinfo {author} {\bibfnamefont {L.}~\bibnamefont {Verde}},\ }\bibfield  {title} {\bibinfo {title} {Galaxy redshift-space bispectrum: the importance of being anisotropic},\ }\href@noop {} {\bibfield  {journal} {\bibinfo  {journal} {Journal of Cosmology and Astroparticle Physics}\ }\textbf {\bibinfo {volume} {2020}}\bibinfo  {number} { (06)},\ \bibinfo {pages} {041}}\BibitemShut {NoStop}%
\bibitem [{\citenamefont {McDonald}\ and\ \citenamefont {Roy}(2009{\natexlab{b}})}]{McDonald_2009}%
  \BibitemOpen
\bibfield  {number} {  }\bibfield  {author} {\bibinfo {author} {\bibfnamefont {P.}~\bibnamefont {McDonald}}\ and\ \bibinfo {author} {\bibfnamefont {A.}~\bibnamefont {Roy}},\ }\bibfield  {title} {\bibinfo {title} {Clustering of dark matter tracers: generalizing bias for the coming era of precision lss},\ }\href {https://doi.org/10.1088/1475-7516/2009/08/020} {\bibfield  {journal} {\bibinfo  {journal} {Journal of Cosmology and Astroparticle Physics}\ }\textbf {\bibinfo {volume} {2009}}\bibinfo  {number} { (08)},\ \bibinfo {pages} {020–020}}\BibitemShut {NoStop}%
\bibitem [{\citenamefont {Saito}\ \emph {et~al.}(2014)\citenamefont {Saito} \emph {et~al.}}]{Saito_2014}%
  \BibitemOpen
\bibfield  {number} {  }\bibfield  {author} {\bibinfo {author} {\bibfnamefont {S.}~\bibnamefont {Saito}} \emph {et~al.},\ }\bibfield  {title} {\bibinfo {title} {Understanding higher-order nonlocal halo bias at large scales by combining the power spectrum with the bispectrum},\ }\bibfield  {journal} {\bibinfo  {journal} {Physical Review D}\ }\textbf {\bibinfo {volume} {90}},\ \href {https://doi.org/10.1103/physrevd.90.123522} {10.1103/physrevd.90.123522} (\bibinfo {year} {2014})\BibitemShut {NoStop}%
\bibitem [{\citenamefont {Lewis}(2019)}]{lewis2019getdistpythonpackageanalysing}%
  \BibitemOpen
  \bibfield  {author} {\bibinfo {author} {\bibfnamefont {A.}~\bibnamefont {Lewis}},\ }\href {https://arxiv.org/abs/1910.13970} {\bibinfo {title} {Getdist: a python package for analysing monte carlo samples}} (\bibinfo {year} {2019}),\ \Eprint {https://arxiv.org/abs/1910.13970} {arXiv:1910.13970 [astro-ph.IM]} \BibitemShut {NoStop}%
\bibitem [{\citenamefont {{Noriega}}\ \emph {et~al.}(2025)\citenamefont {{Noriega}} \emph {et~al.}}]{NoriegaComparing}%
  \BibitemOpen
  \bibfield  {author} {\bibinfo {author} {\bibfnamefont {H.~E.}\ \bibnamefont {{Noriega}}} \emph {et~al.},\ }\bibfield  {title} {\bibinfo {title} {{Comparing Compressed and Full-Modeling analyses with FOLPS: implications for DESI 2024 and beyond}},\ }\href {https://doi.org/10.1088/1475-7516/2025/01/136} {\bibfield  {journal} {\bibinfo  {journal} {\jcap}\ }\textbf {\bibinfo {volume} {2025}},\ \bibinfo {eid} {136} (\bibinfo {year} {2025})},\ \Eprint {https://arxiv.org/abs/2404.07269} {arXiv:2404.07269 [astro-ph.CO]} \BibitemShut {NoStop}%
\end{thebibliography}%

\end{document}